\documentclass[structabstract]{aa}  
\usepackage{graphicx}
\usepackage[comma,authoryear]{natbib}
\usepackage{amsmath, amssymb} 
\usepackage{pdflscape}

\usepackage{multirow}

\begin{document}
\title{Search for starless clumps in the ATLASGAL survey}
% \titlerunning{J. Tackenberg: A catalog of starless clumps}

% \subtitle{A classification of peaks found in the ATLASGAL survey using GLIMPSE and MIPSGAL images as well as their catalogs.}
% \subtitle{Peaks in the ATLASGAL survey; a classification of starless clumps incorporating GLIMPSE and MIPSGAL images as well as their catalogs.}
% \subtitle{A search for starless peaks in the ATLASGAL survey using GLIMPSE and MIPSGAL}
\author{J. Tackenberg\inst{1},
  H. Beuther\inst{1},
  T. Henning\inst{1},
  F. Schuller\inst{2},
  M. Wienen\inst{2},
  F. Motte\inst{7}
  F. Wyrowski\inst{2},
  S. Bontemps\inst{3},
  L. Bronfman\inst{4},
  K. Menten\inst{2},
  L. Testi\inst{5},
  B. Lefloch\inst{6}
}

\institute{Max-Planck-Institut f\"ur Astronomie (MPIA), K\"onigstuhl 17, 69117 Heidelberg, Germany\\
  \email{last-name@mpia.de}
  % \and
  % Max-Planck Institut fuer Astronomie (MPIA), Koenigstuhl 17, 69117 Heidelberg\\
  \and
  Max-Planck-Institut f\"ur Radioastronomie (MPIfR), Auf dem H\"ugel 69, 53121 Bonn, Germany
  \and
  OASU/LAB-UMR5804, CNRS/INSU, Universit\'{e} Bordeaux 1,  2 rue de l'Observatoire, 33270 Floirac, France %%Laboratoire d'Astrophysique de Bordeaus, CNRS/INSU, UMR 5804
  \and
  Departamento de Astronom\'ia, Universidad de Chile, Casilla 36-D, Santiago, Chile
  \and
  ESO, Karl-Schwarzschild Str. 2, 85748 Garching, Germany
  \and
  Laboratoire d'Astrophysique de Grenoble, UMR 5571-CNRS, Universit\'{e} Joseph Fourier, Grenoble, France
  \and
  Laboratoire AIM, CEA/IRFU - CNRS/INSU - Universit\'{e} Paris Diderot, CEA-Saclay, 91191 Gif-sur-Yvette Cedex, France
}

\date{Received June 6, 2011; accepted January 16, 2012}

\abstract
% context heading (optional)
% {} leave it empty if necessary  
{Understanding massive star formation requires comprehensive knowledge about the initial conditions of this process. The cradles of massive stars are believed to be located in dense and massive molecular clumps.}
% aims heading (mandatory)
{In this study, we present an unbiased sample of the earliest stages of massive star formation across 20 deg$^2$ of the sky.}
% methods heading (mandatory)
{Within the region 10$^{\circ}$ $<$ l $<$ 20$^{\circ}$ and $|$b$|$ $<$ 1$^{\circ}$, we search the ATLASGAL survey at 870 $\mu$m for dense gas
  condensations. These clumps are carefully examined for indications of ongoing star formation using YSOs from the GLIMPSE source catalog as
  well as sources in the 24 $\mu$m MIPSGAL images, to search for starless clumps. We calculate the column densities as well as the kinematic distances and masses for sources where the v$_{lsr}$ is known from spectroscopic observations.}
% results heading (mandatory)
{Within the given region, we identify 210 starless clumps\thanks{The catalog (Table \ref{tab:list}) is available in electronic form at the CDS via anonymous ftp to cdsarc.u-strasbg.fr (130.79.128.5) or via http://cdsweb.u-strasbg.fr/cgi-bin/qcat?J/A+A/.} with peak column densities $>$ 1 $\times$ 10$^{23}$ cm$^{-2}$. In particular, we identify
  potential starless clumps on the other side of the Galaxy. The sizes of the clumps range between 0.1 pc and 3 pc with masses between a few tens of
  M$_{\odot}$ up to several ten thousands of M$_{\odot}$.
  Most of them may form massive
  stars, but in the 20 deg$^2$ area we only find 14 regions massive enough to form stars more massive than 20 M$_{\odot}$ and 3 regions with the potential
  to form stars more massive than 40 M$_{\odot}$. 
  The slope of the high-mass tail of the clump mass function for clumps on the near side of the Galaxy is $\alpha$ = 2.2 and, therefore, Salpeter-like. We estimate the lifetime of the most massive starless clumps to be (6 $\pm$ 5) $\times$ 10$^4$ yr.}
% conclusions heading (optional), leave it empty if necessary 
{The sample offers a uniform selection of starless clumps. In the large area surveyed, we only find a few potential precursors of stars in
  the excess of 40 M$_{\odot}$. It appears that the lifetime of these clumps is somewhat shorter than their free-fall times, although both values agree within the errors. 
In addition, these are ideal objects for detailed studies and follow-up observations.
}

\authorrunning{J. Tackenberg}
\titlerunning{Starless clumps in ATLASGAL}
\maketitle

\section{Introduction}
\label{sec:intro}
The understanding of high-mass star formation has made tremendous progress within the last decade (cf. reviews by \citealt{Zinnecker2007, Beuther2007}). Nevertheless, there is still no consistent scenario describing how massive stars form, nor is the impact of massive stars on their environment fully understood. They undoubtedly play a crucial role throughout their whole lifetime on the physical and chemical structure of the interstellar medium, the evolution of stellar clusters, and even of entire galaxies. Therefore, massive star formation needs to be understood if we wish to make progress in our understanding of galaxy evolution.

Independent of the actual high-mass star formation scenario \citep[e.g. ][]{keto2003,mckee2003,bonnell2005,commercon2011}, there is agreement that the most massive stars form in clusters. Therefore, it is probable that we can detect an initial stage of a high-mass gas clump without any star formation signatures.

The discovery of infrared-dark clouds (IRDCs) with ISO, MSX, and Spitzer data provided an interesting sample of objects with which to characterize the earliest stages of massive star formation (e.g. \citealt{Perault1996, Carey1998}). The systematic study of IRDCs provided potential precursors of massive stars and allowed the characterization of their physical and chemical parameters (e.g. \citealt{Simon2006, Peretto2009, Vasyunina2009, Vasyunina2010}). Although \citet{Peretto2009} reported that more than 30 \% of the IRDCs have no IR counterparts at 24 $\mu$m, follow-up studies of IRDCs often revealed signs of ongoing star formation \citep{Beuther2007a, Cyganowski2008, Vasyunina2010}.

All IRDC studies are biased by the variation of the background, foreground confusion, and extinction caused by variations in the dust properties. Longward of 200 $\mu$m, the thermal emission from
dust grains in IRDCs is optically thin and can be measured at mm and sub-mm wavelengths \citep{Hildebrand1983}. This can be used to obtain
extinction-independent mass measurements of the cold gas inside these objects.

Until recently, there were no available systematic surveys of the Galactic plane directly tracing the cold dust associated with molecular clumps. Now, the
Bolocam Galactic Plane Survey (BGPS, \citealt{Aguirre2011}) at 1.1 mm, as well as the APEX Telescope Large Area Survey of the GALaxy (ATLASGAL,
\citealt{Schuller2009}) at 870 $\mu$m, offer surveys of the Galactic plane's cold dust. Only ATLASGAL covers the full range l = -60$^{\circ}$ to 60$^{\circ}$ of the inner Galactic plane at 19$\arcsec$ resolution.

In this paper, we present a compilation of clumps of high column density, located in a region of Galactic longitude 10$^{\circ}$ $<$ l $<$ 20$^{\circ}$ and latitude -1$^{\circ}$ $<$ b $<$ 1$^{\circ}$, showing no signs of star formation. To confirm their starless nature, we carefully examined each clump for GLIMPSE \citep{Benjamin2003} and MIPSGAL 24 $\mu$m \citep{Carey2009} sources, either of which would indicate that star formation had already started.
The column density threshold we imposed on our survey is 1 $\times$ 10$^{23}$ cm$^{-2}$.

After a short description of the surveys we employed (Sec. \ref{data}), we describe the individual steps of the classification and its limitations in Sec. \ref{root_classification}. In Sec. \ref{results}, we present both the results of the classification and the direct clump properties and discuss the clump column densities. Using the ammonia velocities given in Wienen et al. (submitted), we derived distances to $\sim$ 71 \% of the clumps (Sec. \ref{sec:dist_col_mass}). Sec. \ref{sec:masses_CMF} presents the clump masses and the clump mass function. Next we estimated the lifetimes of starless clumps (Sec. \ref{sec:lifetimes}). In Sec. \ref{sec:discussion}, we discuss our results and compare them to other surveys (Sec. \ref{sec:dunham_etc}). Our conclusions, in Sec. \ref{sec:conclusion_outlook}, summarizes the main results of this work and provides an outlook to future work.
% \textit{In this study we try to identify starless clumps within 20 deg$^2$ of the Galactic plane. We carefully check for stellar emission using GLIMPSE and MIPSGAL 24 $\mu$m.}

\section{Employed data}
\label{data}
All data for this study were taken from large surveys, most of them publicly available. Clumps were identified by searching for continuum peaks at 870 $\mu$m in the ATLASGAL survey \citep{Schuller2009} and then classified using both the GLIMPSE point source catalog and MIPSGAL 24 $\mu$m images.

%% \subsection{Incorporated Data}
In contrast to most other searches for massive prestellar clumps of high column density using extinction maps \citep{Simon2006, Peretto2009, Kainulainen2011}, we used emission at 870 $\mu$m as a tracer of cold dust. The APEX telescope large area survey of the Galaxy (ATLASGAL, \citealt{Schuller2009}) is a systematic survey of the Galactic plane at 870 $\mu$m with LABOCA \citep{Siringo2009}. Its beam size is 19.2$\arcsec$, the pixel size in the maps is 6$\arcsec$, and the average rms noise of the selected maps is below 50 mJy. To obtain a statistically meaningful sample, we covered 20 deg$^2$ on the sky, the region of Galactic longitude 10$^{\circ}$ $<$ l $<$ 20$^{\circ}$ and Galactic latitude $|$b$|$ $<$ 1$^{\circ}$. 

As described in Sec. \ref{sec:ident_starless}, we extracted young stellar objects from the GLIMPSE {\sc i} Spring '07 catalog \citep{Benjamin2003}. Among other criteria, the GLIMPSE source catalog requires a minimum flux of 0.6 mJy, 0.4 mJy, 2 mJy, or 10 mJy in either the 3.8, 4.5, 5.8, or 8.0 $\mu$m
band, respectively, for a source to be taken into account. For the region given, we investigated more than 5.6 million GLIMPSE sources. The IRAC/SPITZER pointing
accuracy is better than 1$\arcsec$ and the pixel resolution is 0.6$\arcsec$. As an additional tracer of ongoing star formation, we used the MIPSGAL 24 $\mu$m survey \citep{Carey2009}. The rms noise of the MIPSGAL images is $\sim$ 0.67mJy and its resolution is 6$\arcsec$. 

\section{Classification}
\label{root_classification}
The naming of clumps in the literature refers to various physical objects. %% Especially confusing, 
Research groups working on high- and low-mass star formation have different naming schemes for the objects named clumps and cores, including the sub-categories starless and prestellar \citep{Enoch2008}. An often used
nomenclature tries to bind things to physical properties, denoting gravitationally bound objects ``cores'' and unbound objects ``clumps''
\citep{Chabrier2010}. In this paper, we use the term clumps for all emission peaks revealed by the CLUMPFIND algorithm
\citep{Williams1994}. Typically, these are massive and large enough to form massive clusters. As
shown in Fig. \ref{fig:angular_effective_radius}, typical sizes derived from the effective radii of these clumps range from 18$\arcsec$ to
70$\arcsec$. These can either be bound or unbound systems, but are assumed to be coherent in lbv space (Galactic longitude and latitude, and radial velocity, thus distance, \citealt{Williams2000}). 
In the remainder of this paper, clumps are called starless if they host no mid-IR tracers of ongoing star formation. Nevertheless, as mentioned in Sec. \ref{sec:intro} and Sec. \ref{biases:24micron}, many IRDCs not hosting 24 $\mu$m sources have been shown to host star formation using other tracers such as SiO emission. In this study, we cannot completely rule out the presence of star formation, but only present starless clump candidates. In this context, the MALT90 survey \citep{Foster2011} will improve future classifications.

\subsection{Clump extraction}
\label{sec:clump_identification}
\begin{figure}
  \includegraphics[width=0.5\textwidth]{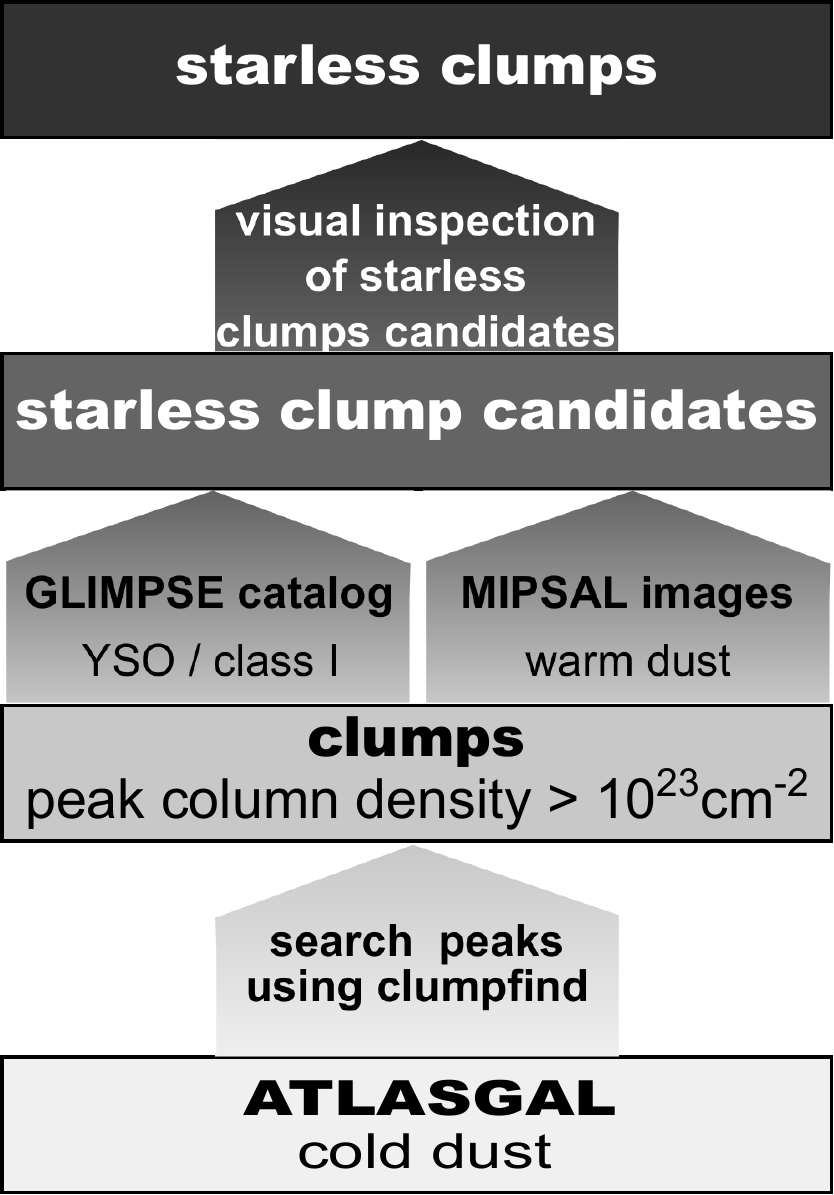}
  \caption{Schematic visualization of classification.}
  \label{fig:flowchart}
\end{figure}
\begin{figure*}[tb]
  \includegraphics[width=\textwidth]{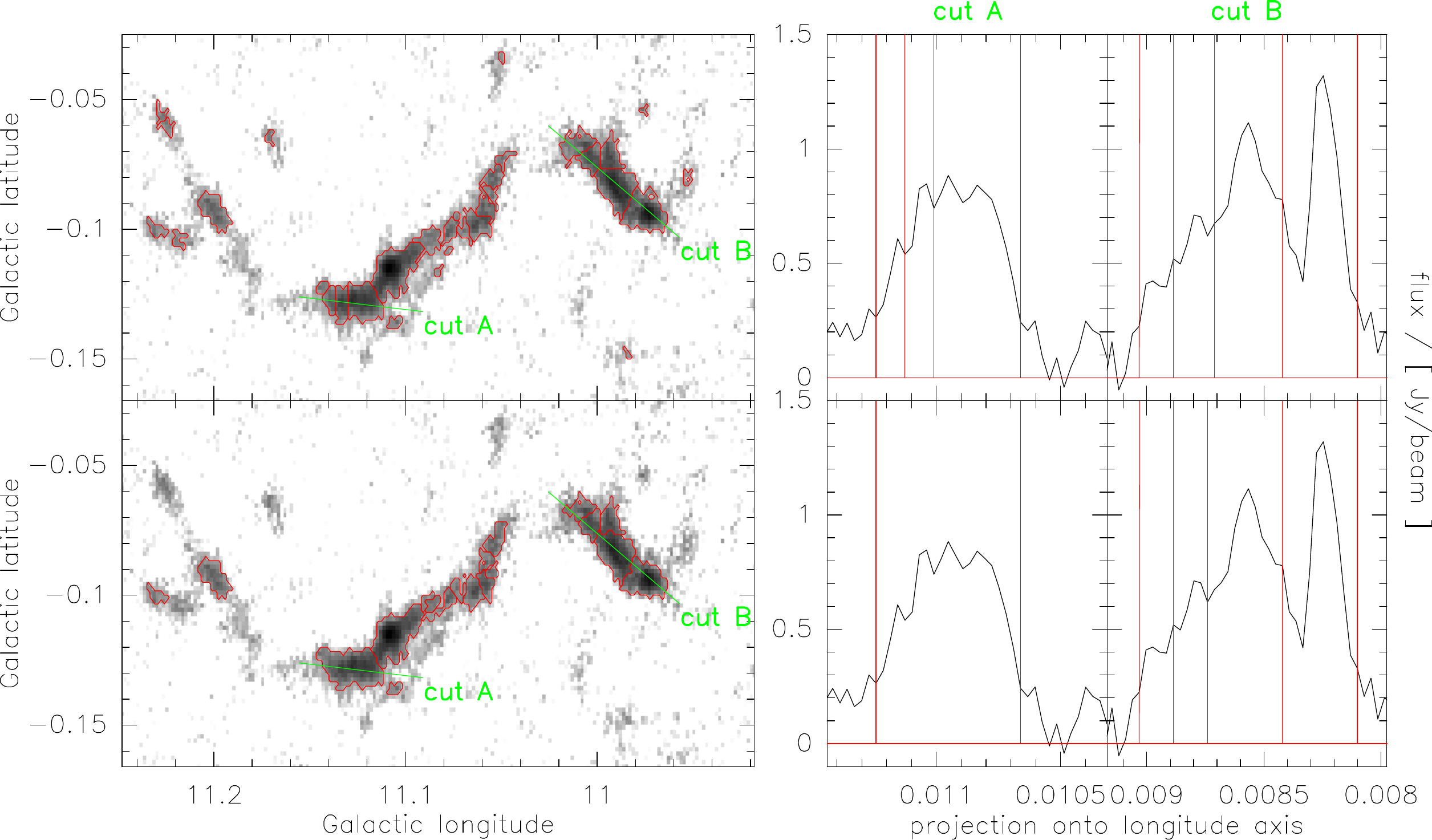}
  \caption{The figures show the clump definition as used in this paper (bottom row), compared to 'classical' 3$\sigma$ spacing contour levels (top row). While the left most column shows the ATLASGAL image of G11.11 with the clump definitions (red), the other columns show two profiles along the lines shown in the left panel ('cut A' and 'cut B'). The red lines indicate the clump borders.}
  \label{fig:clump_profile}
\end{figure*}
To identify starless clumps, we first employed the CLUMPFIND algorithm by \citet{Williams1994} to search for dust condensations. 
It has been reported that CLUMPFIND is less reliable in very crowded regions \citep{Kainulainen2009} and that the extracted clump parameters strongly depend on the distance \citep{Smith2008}. The second point is unavoidable in observed data and is discussed further in Sec. \ref{sec:far_field}. Nevertheless, we are interested in column-density peak positions and the associated fluxes/masses, which CLUMPFIND can extract reliably. \citet{Pineda2009} demonstrated that the exponent of the derived mass function is not very sensitive to the chosen step size.

We set the lowest detection level to 6$\sigma$, or 0.3 Jy. The additional thresholds, 0.3, 0.4, 0.5, 0.7, 0.9, 1.3, 1.8, 2.5, 4, and 7 Jy, were chosen (1) to account for the degree of variation relative to the actual emission level and (2) to trace the structures recognized by observers. The use of non-constant intervals for the various emission levels in CLUMPFIND prevents bright clumps from being artificially sub-divided because of brightness changes that are very small relative to the flux level of the clump. % We believe that, different from cores or compact sources, clumps do not need to have a strictly monotonically decreasing radial profile, since they will fragment further and may even form clusters.

To test the robustness of the chosen thresholds, we compared the integrated clump fluxes of our clump extraction to classical 3 $\sigma$ spaced thresholds as proposed in \citet{Kainulainen2009}.
The two right columns of Fig. \ref{fig:clump_profile} compare the flux distributions along the lines plotted in the left column. The upper row shows our clump definition, the lower row shows the clumps of a pure 3$\sigma$ spacing of the contour levels. While some regions do not differ at all, e.g. as shown by the right-most column of Fig. \ref{fig:clump_profile}, the additional contours in the evenly spaced situation can subdivide large clumps into a number of smaller ones, shown by the middle column of Fig. \ref{fig:clump_profile}.

While flux level spacings that are not bound to some objective criteria introduce subjectivity, Fig. \ref{fig:clump_profile} shows that the chosen levels trace structures that we call clumps. 
In addition, we compared the fluxes of our clumps to the fluxes of the 3 $\sigma$ extraction and could not find an excess of bright clumps, as might have been expected. Furthermore, we compared our CLUMPFIND sources to sources found by Contreras (priv. communication) using SExtractor as described in \citet{Schuller2009}. For fluxes above our thresholds, almost all sources identified by Contreras had a counterpart within our clumps with matching peak fluxes. In addition to this, the comparison shows that we identified smaller fragments of lower mass, and the integrated fluxes in our catalog are lower than the integrated fluxes of the corresponding SExtractor clumps. This assures us that we do not produce sources with artificially high fluxes.

As we aim to study sites of high-mass star formation, clumps with a peak flux of less than 0.5 Jy, corresponding to a column density of 1 $\times$ 10$^{23}$ cm$^{-2}$, are ignored in the following. (For a further discussion of the derivation of column densities, we refer to Sec. \ref{sec:column_densities}.)

To motivate this threshold, we used the Orion nebula cluster (ONC). Its stellar mass is about 1800 M$_{\odot}$ \citep{Hillenbrand1998}. To be consistent
with estimates carried out in Sec. \ref{sec:most_massive_star}, we assumed a star formation efficiency of 30\% and, therefore, estimated the initial gas
mass of the ONC to be 6000 M$_{\odot}$. As the cluster has dispersed over its lifetime, we set its initial radius to the typical radius we found for
our most massive clumps, hence 0.7 pc (cf. Sec. \ref{sec:most_massive_star}). With the assumptions made and a spherically symmetric mass distribution, the
initial peak column density in the ONC has been 1.8 $\times$ 10$^{23}$ cm$^{-2}$, or 0.6 g cm$^{-3}$. %  If the cluster is gravitationally bound,
% estimates of its current gas mass require even higher initial gas masses (C. Olczak, priv. communication), and therefore higher column densities. 
This agrees with the theoretical values found by \citet{Krumholz2008}. To avoid fragmentation in high-mass star formation, they require
column densities of 1 g cm$^{-2}$, or 3$\times$10$^{23}$cm$^{-2}$. 

\subsection{Identification of starless clumps}
\label{sec:ident_starless}
Although it is unclear whether high- or low-mass stars form first, starless clumps should not host young stellar objects (class {\sc i} sources, YSOs). To identify clumps hosting YSOs, we searched the GLIMPSE source catalog for stars with colors similar to known YSOs and compared those to our clumps. To do so, we followed the classification given by \citet{Gutermuth2008}. We used additional color criteria, given in \citet{Gutermuth2008} as well, to reject contaminating extragalactic sources and AGNs `that masquerade as bona fide YSOs'. Afterwards we selected objects obeying the following IRAC criteria:
%% \begin{center}
\begin{align*}
  [4.5] - [5.8] &> 1.0 \text{ OR}\\
  ( [4.5]-[5.8] &> 0.7 \text{ AND } [3.6]-[4.5] > 0.7 ) \text{.}
\end{align*}
%% \end{center}
In addition, we required a source to be detected at 8 $\mu$m. The identified YSOs were
directly compared to the clumps and their extensions according to CLUMPFIND
using the CLUMPFIND maps produced. If a YSO is located on a clump (in
projection), the clump was considered as star forming and is ignored in
the following. Nevertheless, at the onset of star formation, sources  may be too cold to be detectable in the GLIMPSE bands, but show weak 24 $\mu$m emission. Unfortunately, no MIPSGAL 24 $\mu$m point source catalog has been published. Therefore, we used the STARFINDER algorithm by \citet{Diolaiti2000} to
search the $24 \mu$m MIPSGAL images for point sources. To avoid mis-identifications we only
extracted stars with a detection better than $7\sigma$. Again, clumps with a 24 $\mu$m source were assumed to host stellar activity.
In a last step, all remaining clumps were classified by visual inspection. Here
the main focus was on 24 $\mu$m objects that had not been identified by
STARFINDER. % (NOTE: Since we expect that all young stellar objects should still
% show emission at 24micron.) 
%% Therefore, the identified YSOs from GLIMPSE and the 24 $\mu$m detections
%% are overplotted on the 24 $\mu$m MIPSGAL maps as well as on the ATLASGAL maps and CLUMPFIND maps. Now all
%% clumps that show no signs of ongoing star formation are examined individually
%% and are than classified by hand. 
A schematic summary of the classification is given in Fig. \ref{fig:flowchart}.
Parts of M17, in which MIPSGAL is saturated due to extended emission, were omitted as well as a few additional regions. Exact positions of omitted
regions are listed in the Appendix, Table \ref{tab:omitted}. 

\subsection{Limitations and observational biases}
\label{limitations}
Although the visual verification of the classification ensures a maximum reliability, technical limitations of the data sets impose various biases. To point out the limitations of this study, next we carefully discuss the biases. 

\subsubsection{ATLASGAL and clump finding limitations}
The spatial limitations of ATLASGAL vary with the distance and are discussed in detail in Sec. \ref{sec:far_field}. The flux threshold for the clump
extraction was chosen to be $\sim$ 6$\sigma$ or 0.3 Jy, and the higher thresholds were chosen to trace clearly recognizable structures. These threshold spacings are larger than the estimated rms. Contours in steps of the noise level are less biased and would generate more substructure, hence clumps. However, noise would, more likely, generate artificial clumps as studied by \citet{Reid2010}. The chosen peak flux threshold of  0.5 Jy/beam corresponds to 1 $\times$ 10$^{23}$ cm$^{-2}$. In the context of massive star formation, this provides a rough lower limit for potential regions of massive star formation (see Sec. \ref{sec:clump_identification}). %  Nevertheless, no clump consisting only of 3 pixels was identified, and only 2 clumps with 4 pixels. Effective clump sizes measured with CLUMPFIND which are smaller than the beam size have been set to the beam size. \\
% %%We note again, that the 'clumps' are not necessarily gravitationally bound structures and their physical extension varies over two orders of magnitude.
Sources for which the integrated flux is less than its peak flux are considered as artificial and 28 out of 929 sources were rejected.

\subsubsection{GLIMPSE catalog limitations}
The detection thresholds of GLIMPSE and MIPSGAL as well as the point source extraction from the GLIMPSE catalog and the color-color criteria for young sources described in Sec. \ref{sec:ident_starless} have a major impact on the classification. \citet{Gutermuth2008} included criteria to reject contaminating extragalactic sources, but AGB stars have similar colors to YSOs and obey the color-color criteria Gutermuth used to identify YSOs. Their contribution to the list of YSOs may be as high as $\sim$ 30\% \citep{Robitaille2008}, rejecting potential starless clumps. Nevertheless, their likelihood of being projected onto a clump is significantly lower. Since we expect embedded YSOs to have detectable 24 $\mu$m flux for which we will check again, the given sensitivity limits of the GLIMPSE source catalog do not influence the population of starless clumps. Additional sources in the list of YSOs as well as chance alignments could lead to an artificial rejection of clumps, but will not produce artificial starless clumps.

\subsubsection{MIPSGAL 24 $\mu$m limitations}
\label{biases:24micron}
The situation is different for the 24 $\mu$m MIPSGAL images. Here the sensitivity limit is the key parameter and sources hidden in the rms can lead to misidentifications of starless clumps. The brightness of the faintest sources still detectable varies over the images with respect to their surroundings, but for the visual inspection method we estimated it to be $\sim$ 1 mJy. This is slightly smaller than the 2 mJy level for a 3$\sigma$ detection given in \citet{Carey2009}.

For sources hidden in the dust, one may assume that all flux gets re-emitted by the dust producing a black body spectrum. From this, one can estimate the integrated luminosity of the internal source. 
% {\bf We took a well measured but faint 24 $\mu$m source as reference (18223-3 at 3.6 kpc, \citealt{Beuther2010}) and set its 24 $\mu$m to 2 mJy. Using near- and mid-IR dust opacities from \citet{Draine1984} we fitted two black body spectra to the SED, one for the cold dust component, one for the hotter 24 $\mu$m. The luminosity of the warm component then becomes 17 L$_{\odot}$, which corresponds to a main sequence star of $\sim$ 1.9 M$_{\odot}$ \citep{Siess2000}. For sources further away this becomes significantly larger. 17 L$_{\odot}$ at 3.6 kpc become $\sim$ 300 L$_{\odot}$ at 15 kpc, which allows main sequence stars of up to 4 M$_{\odot}$ \citep{Siess2000} to be hidden in the dust.}
Since the faint 24 $\mu$m sources in question are not detected at GLIMPSE wavelengths, we used both the GLIMPSE and MIPSGAL detection limits to construct a SED of three data points at 3.8 $\mu$m, 8.0 $\mu$m, and 24 $\mu$m with 0.6 mJy, 10.0 mJy, and 2.0 mJy, respectively. Using near- and mid-IR dust opacities from \citet{Draine1984}, we fitted a black body spectrum to the SED, with an integrated luminosity of 1.1 L$_{\odot}$ at 3 kpc, or 27 L$_{\odot}$ at 15 kpc. These luminosities correspond to main sequence stars of 1.1 M$_{\odot}$ or 2.1 M$_{\odot}$ \citep{Siess2000}.
\citet{Krumholz2007} found that accretion luminosities in massive star formation reach several hundred solar luminosities very early on, excluding that a massive collapsing core could be hidden in the dust.
Furthermore, there have been observations of objects that may form high-mass stars, but are not yet that luminous \citep[][Ragan et al., in prep.]{Beuther2007b, Bontemps2010a, Motte2010}. Nevertheless, their luminosities are still higher (on the order of several 10 L$_{\odot}$) than our detection limit for 24 $\mu$m fluxes on the far side of the Galaxy. Therefore, only low-mass objects can be hidden.

Observations using the Herschel satellite have shown that some 24 $\mu$m dark regions, hence starless clumps already show 70 $\mu$m emission
\citep{Beuther2010, Wilcock2011}. As discussed in \citet{Henning2010}, these sources may be either starless or protostellar. Similarly, \citet{Motte2007} and \citet{Russeil2010} find MSX and 24 $\mu$m dark cores, driving SiO outflows. Although \citet{Motte2007} and \citet{Russeil2010} are less sensitive at 24 $\mu$m, future studies will need to disentangle this situation.

\subsection{Verification of classification by comparison to other tracers and studies}
\label{biases:verification}
\begin{figure}
  \includegraphics[angle=-90., width=0.5\textwidth]{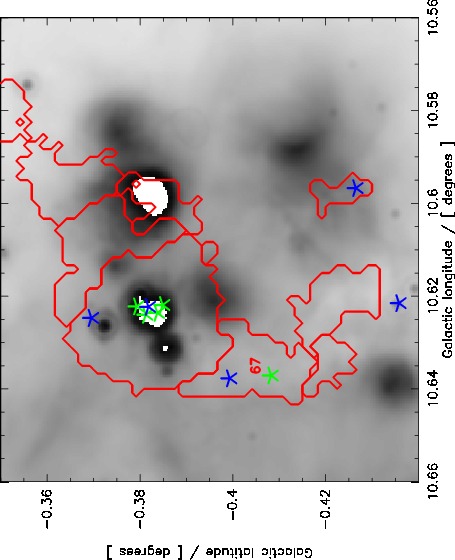}
  \caption{Clump 67 overplotted on a MIPSGAL 24 $\mu$m image. The clump definition is in red and its peak position is marked by a red asterisk. Overplotted in green are H{\sc II} regions \citep{Purcell2010}, and in blue Red Sources found in GLIMPSE by \citet{Robitaille2008}.}
  \label{fig:cl67}
\end{figure}
To test the classification, a comparison with other tracers and catalogs is helpful.

Similar to \citet{Gutermuth2008}, \citet{Robitaille2008} identified Intrinsically Red Sources (R08 in the following) by applying color-color criteria to the Spitzer GLIMPSE catalog. Both sets of color criteria differ and, in addition to a large number of common sources, both catalogs also identify different sources. We take these different identifications as statistical variations that set the approximate uncertainties in the different catalogs. If we now compare the population of starless clumps to the Red Sources given in RO8, this gives us a feeling for the classification statistics. As it turns out, only two clumps that we identified as starless have a SPITZER Red Source. With knowledge of its position, we have been able to associate the R08 source in clump 67 with a peak in the 24 $\mu$m image (see Fig. \ref{fig:cl67}).  The other Red Source is supposed to be in clump 1216, at a ridge of bright continuous emission. This hampers the identification and we cannot identify a 24 $\mu$m counter part. Therefore, it is unclear to us whether this source is still very young.

Another test of our classification is to check the clumps for additional tracers of star formation. H{\sc ii} regions are a well-accepted tracer of ongoing massive-star formation and several surveys have searched the Galactic plane systematically. In this context, CORNISH \citep{Purcell2010}, a Galactic plane survey at 5 GHz with the VLA in B configuration, identified more than 600 H{\sc ii} regions in our region of study. With their high spatial resolution, matches can be made unambiguously. We found H{\sc ii} regions on only three of our clumps. In clump 67 (shown in Fig. \ref{fig:cl67}) and clump 87, no 24 $\mu$m source is in the vicinity of the cm emission peak. This suggests that star formation is already taking place, but so embedded that (almost) no light can escape. However, clump 505 has a very bright 24 $\mu$m source at the edge of the clump. This might power the H{\sc ii} region, which is offset by ~11$\arcsec$ towards the emission peak of clump 505.

A comparison to the Green Bank Telescope H{\sc ii} Region Survey \citep[GBT HRDS, ][]{Bania2010} and the Red MSX Source Survey \citep[RMS, ][]{Hoare2004,Mottram2011} did not discover any matches.

In summary, since only three incorrect classifications have been found, all tests confirm our classification and establish its credibility.
For consistency reasons, we flagged clump 67 as star forming, but kept the other two sources in our sample. %%Nevertheless, we marked those sources in our tables.

% The HERSCHEL key program of the Earliest Phases Of Star-formation (EPOS, Krause in prep.) studies 45 high-mass regions selected from IRDC catalogs. The high sensitivity at the various far-IR wavelengths of HERSCHEL allows us a more detailed study of these objects. Especially at 70 $\mu$m the
% initial stages of massive star formation will be uncovered. Modeling the full SED will give further insights. Several objects studied
% by EPOS lie within the area of this survey (e.g. 18223 \citep{Beuther2010} and G11.11 \citep{Henning2010}, both within 10$^{\circ}$ $<$ l $<$ 20$^{\circ}$, and have already been published).\\
Furthermore, both pointed HERSCHEL observations (e.g. EPOS, Krause et al., in prep., Ragan et al., in prep.) and the HERSCHEL Galactic plane survey HiGal \citep{Molinari2010} revealed a new population of very young sources, detectable at 70 $\mu$m, but yet dark at 24 $\mu$m. 
Comparing the embedded protostars found in \citet{Henning2010} to the starless clumps we found in G11.11, we conclude that one out of six starless clumps harbors an embedded source, that is invisible at 24 $\mu$m.
This suggests that not all clumps presented here will be starless at 70 $\mu$m. %%Here, the HI-GAL/HERSCHEL survey \citep{Molinari2010} will allow a further distinction of our clumps. 
Future studies will need to clarify the HERSCHEL view of our starless clumps.
% The Herschel EPOS survey covers some of the regions even deeper in all HERSCHEL bands and allows detailed studies of these regions. E.g. \citet{ Beuther2010, Henning2010} discovered new cores in some of the starless clumps discussed here, which only become visible at 70 $\mu$m.

\section{Distance independent results}
\label{results}
\begin{figure*}[tbp]
  \includegraphics[width=\textwidth]{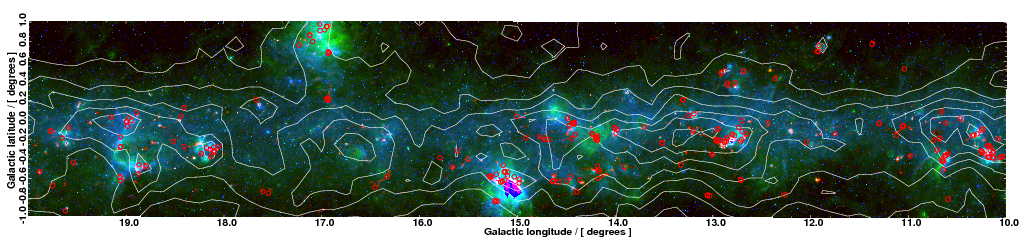}
  \caption{RGB image of the Galactic plane with Galactic latitude l = 10$^{\circ}$ to 20$^{\circ}$ using GLIMPSE 8 $\mu$m, MIPSGAL 24 $\mu$m, and ATLASGAL 870 $\mu$m, respectively. Overplotted are CO contours from \citet{Dame2001}. Starless cores are indicated as circles.}
  \label{fig:whole_region}
\end{figure*}
\onlfig{1}{
  \begin{figure*}[tbp]
    \includegraphics[angle=0,width=1.\textwidth]{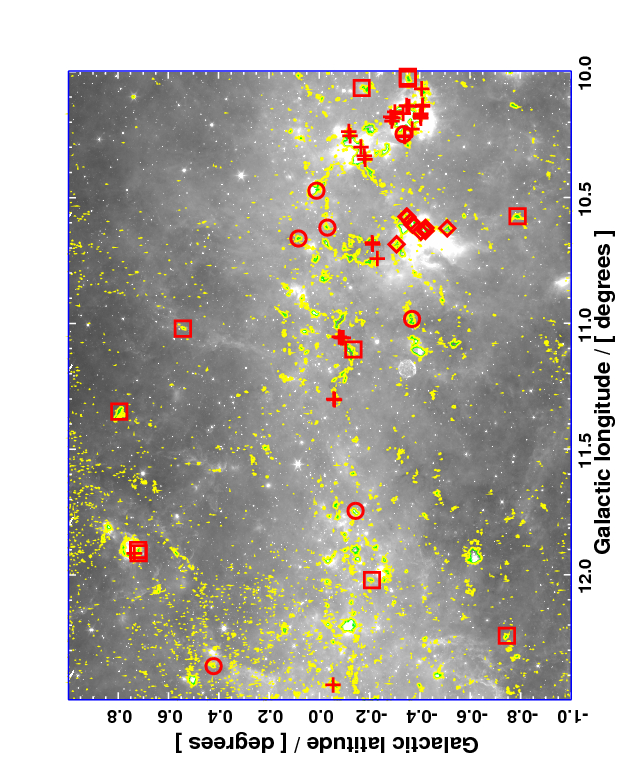}
    \caption{ATLASGAL 3 and 6 $\sigma$ contours in yellow and green, respectively, on top of a 24 $\mu$m MIPSGAL image in logarithmic scale, with clumps plotted in overlay. Plus signs represent clumps for which the near solutions is assumed while circles are clumps with far solution assumed. For clumps plotted with a diamond only the far solution exists, while for sources with a box no velocity information is present. White stripes at the edges are artifacts from the MIPSGAL coverage.}
    \label{fig:overview1}
  \end{figure*}
}
\onlfig{2}{
  \begin{figure*}[tbp]
    \includegraphics[angle=0.,width=1.\textwidth]{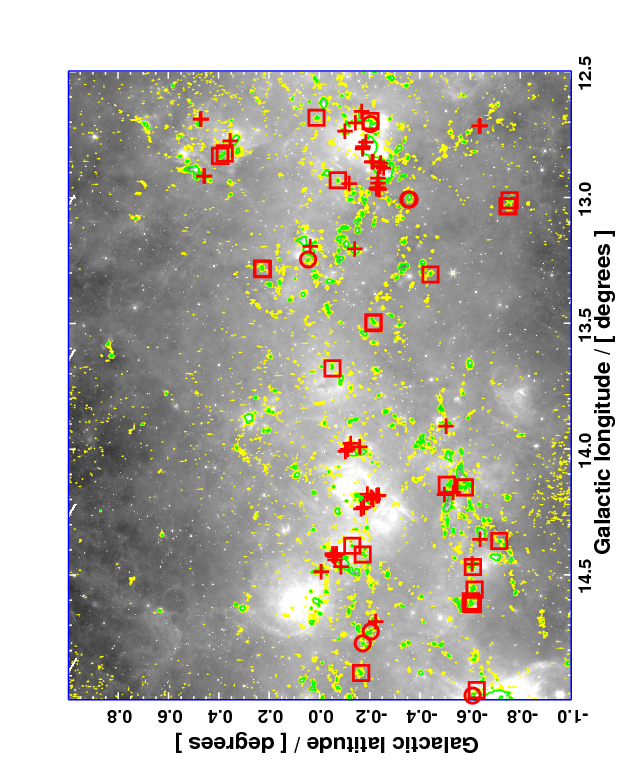}
    \caption{continued.}
    \label{fig:overview2}
  \end{figure*}
}
\onlfig{3}{
\begin{figure*}[tbp]
  \includegraphics[angle=0., width=1.\textwidth]{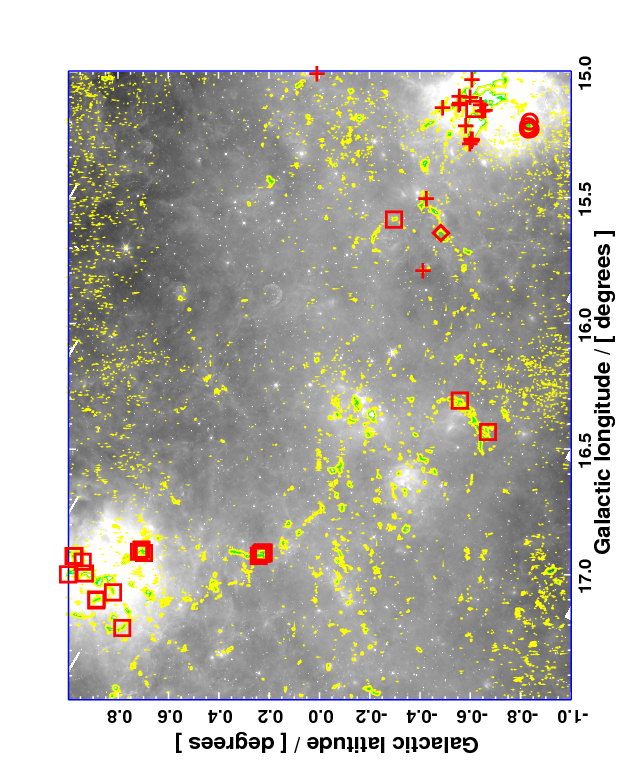}
  \caption{continued.}
   \label{fig:overview3}
\end{figure*}
}
\onlfig{4}{
\begin{figure*}[tbp]
  \includegraphics[angle=0., width=1.\textwidth]{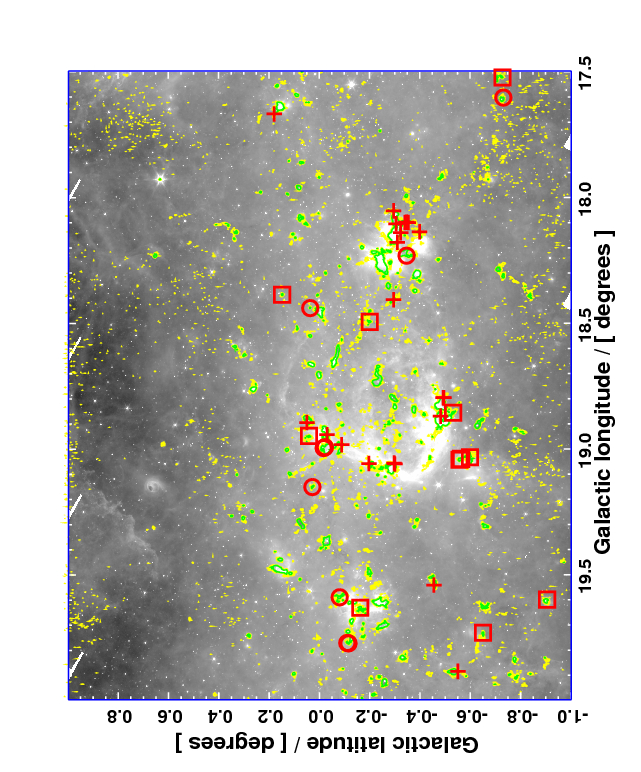}
  \caption{continued.}
   \label{fig:overview4}
\end{figure*}
}
Using CLUMPFIND, we therefore extracted 901 clumps with peak column densities above our threshold of 1 $\times$ 10$^{23}$ cm$^2$. We found that 291 clumps have a Spitzer counterpart classified as YSO using the Gutermuth criteria. For 238 objects, STARFINDER identified a 24 $\mu$m (point) source in the MIPSGAL images, which had no YSO inside. During the visual inspection of the remaining 372 clumps, 103 additional 24 $\mu$m sources were found, while 59 clumps were found to be partially or completely saturated in the MIPSGAL 24 $\mu$m images. Therefore, 210 clumps, or $\sim$ 23$\%$, show no signs of a heating source with the data employed. These can be considered as starless.

The large number of visually identified 24 $\mu$m sources show that this step was crucial for a reliable source catalog. Unfortunately, 24 $\mu$m point sources are often hidden in the unsteady background emission, hence algorithmic point source extraction is unable to distinguish the weakest sources. % This view is encouraged by the fact that still no MIPSGAL 24 $\mu$m point source catalog could be extracted for the whole galactic plane.
The positions of the starless clumps are shown in a three color image, Fig. \ref{fig:whole_region}, and full details are given in Table \ref{tab:list}. These clumps build a sample of potential starless clumps. As we show, most of them can be considered as massive.

\subsection{Results based on the classification}
\label{sec:direct_results}
It can be seen from Fig. \ref{fig:overview1} in the online appendix that almost all of the clumps are embedded in larger structures (cf. \citealt{Schuller2009}). These form filaments with lower density gas, indicated by the 3 $\sigma$ contour. Only a few clumps seem to be isolated.

The majority of the gas is concentrated towards the Galactic plane, with a small offset towards negative latitudes. Enhanced concentrations of
gas/clumps are visible towards known regions, mainly W31, W33, M17, M16, and W39 (from west to east).
% Due to additional observational coverage, improvements in the flux calibration and sky subtraction, ATLASGAL maps used in this study differ from the maps
% presented in \citet{Schuller2009}. Therefore, only one of the clumps presented from \citet{Schuller2009} can be matched unambiguously to a starless clump found in our study.

A study to identify IRDCs, solely based on SPITZER 8 $\mu$m extinction, was conducted by \citet{Peretto2009}. They found the fraction of starless IRDCs
to be 32\%. This is similar to the fraction of 23\% found within this work. Nevertheless, as pointed out in \citet{Peretto2010}, the detection of
column densities via SPITZER 8 $\mu$m extinction breaks down at column densities larger than $\sim$ 1 $\times$ 10$^{23}$ cm$^{-2}$, which we require as minimum column density in our study. In addition,
extinction is very unlikely to be observed on the far side of the Galaxy. Therefore, more than a quarter of the complexes have no \citet{Peretto2009}
IRDC close by and are a completely new sample, which is likely at the far side of the Galaxy. 

% \subsection{Effective angular radius}
\label{sec:angular_effective_radius}
\begin{figure}[tbp]
  \includegraphics[width=.5\textwidth]{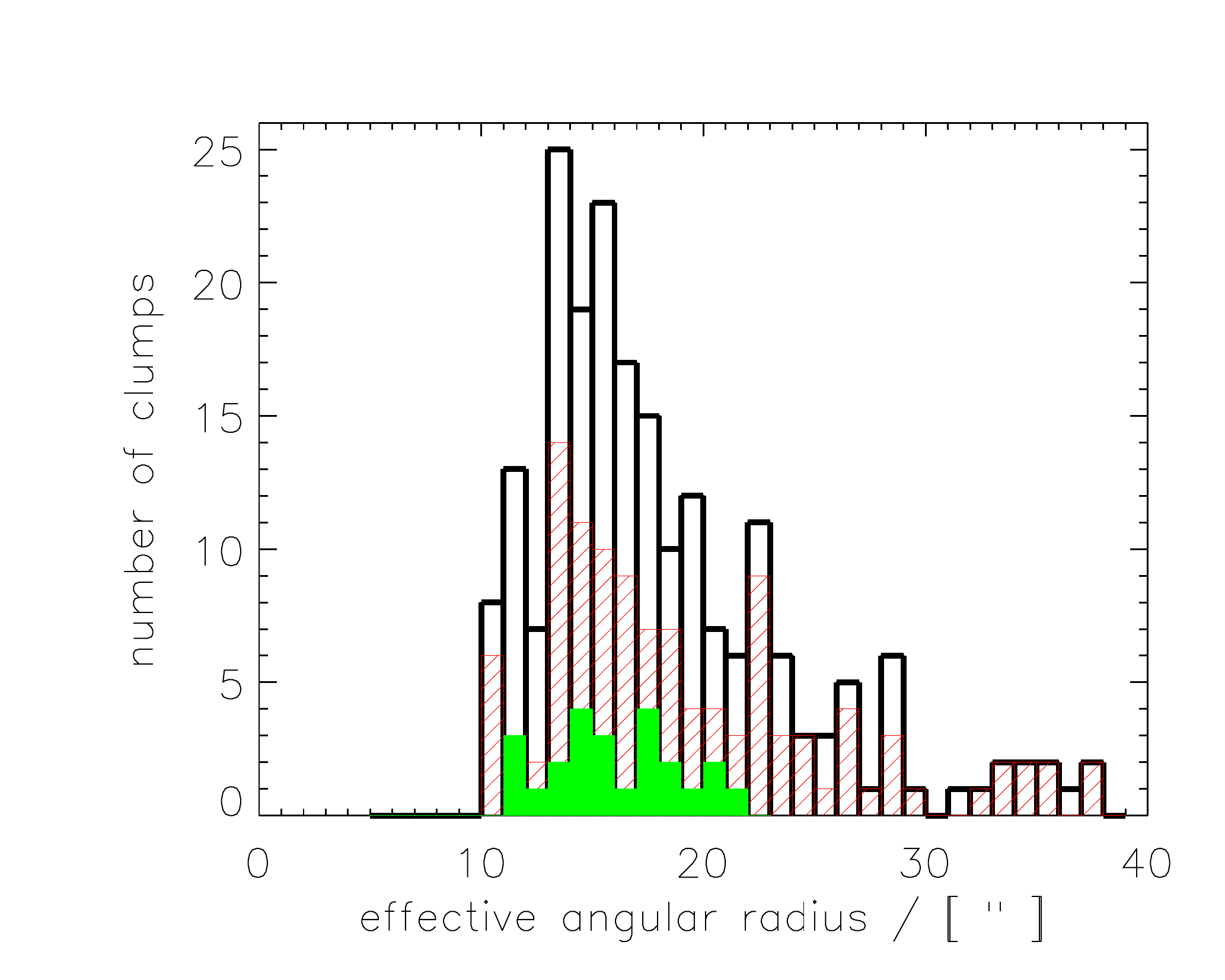}
  \caption{Histogram of the effective radius derived by CLUMPFIND for all clumps (black), for clumps with IRDC connection (hatched red area), only, and for clumps without any IRDC indication (solid green area).}
  \label{fig:angular_effective_radius}
\end{figure}
The CLUMPFIND algorithm calculates the effective radius of the clumps by % estimating their large and small half-axis and requiring the same area for the ellipses and a circle.  -> wrong!!!
equating the area of a theoretical circular clump to the sum of the pixels. 
Results for the clump radii are shown in black in Fig. \ref{fig:angular_effective_radius}. The clump radii range between 10\arcsec and 40\arcsec, with an average radius of 18\arcsec.

\subsection{Column densities}
\label{sec:column_densities}
The fluxes at the peak positions of the clumps can be used to derive a beam-averaged peak column density. Within their large NH$_3$ survey, Wienen et al. (submitted) measured the rotation temperature of 15 of our starless clumps directly. Both the mean and median temperature are T = 15 K at these
peak positions. This is in agreement with temperature estimates for IRDCs \citep{Sridharan2005, Pillai2006, Vasyunina2010, Peretto2010a}. Since we required all clumps to be devoid of 24 $\mu$m emission, we assumed that all our clumps have similar temperatures. We calculated the column density of the gas via
\begin{equation}
  N_{H_2} = \frac{ R F_{\lambda} }{ B_{\lambda}(\lambda,T) m_{H_2} \kappa \Omega}
\end{equation}
for a gas-to-dust ratio of R = 100, where F$_{\lambda}$ is the flux at the given wavelength, B($\lambda$,T) the blackbody radiation as a function of wavelength and
temperature, m$_{H_2}$ the mass per H$_2$ molecule, and $\Omega$ the beam size. The mass absorption coefficient $\kappa$ = 0.77 cm$^2$ g$^{-1}$ is
adopted from the values given in \citet{Hildebrand1983} using a dust emissivity index $\beta$ = 2 and an emissivity at 250 $\mu$m of 3.75 $\times$
10$^{-4}$. This is consistent with the value for the diffuse ISM in \citet{Ossenkopf1994}, a frequently used value for dark clouds. The calculated column
densities for the starless clumps are given in table \ref{tab:list}. For intermediate volume densities of 10$^5$ cm$^{-3}$ and thin ice mantles, one
can extrapolate $\kappa$ = 1.85 cm$^2$ g$^{-1}$ from \citet{Ossenkopf1994},  as, e.g., used by \citet{Schuller2009}. Including their different assumption of the mean
molecular weight for the ISM, column density estimates in \citet{Schuller2009} would be smaller by a factor of three. 

\begin{figure}[tbp]
  \includegraphics[width=0.5\textwidth]{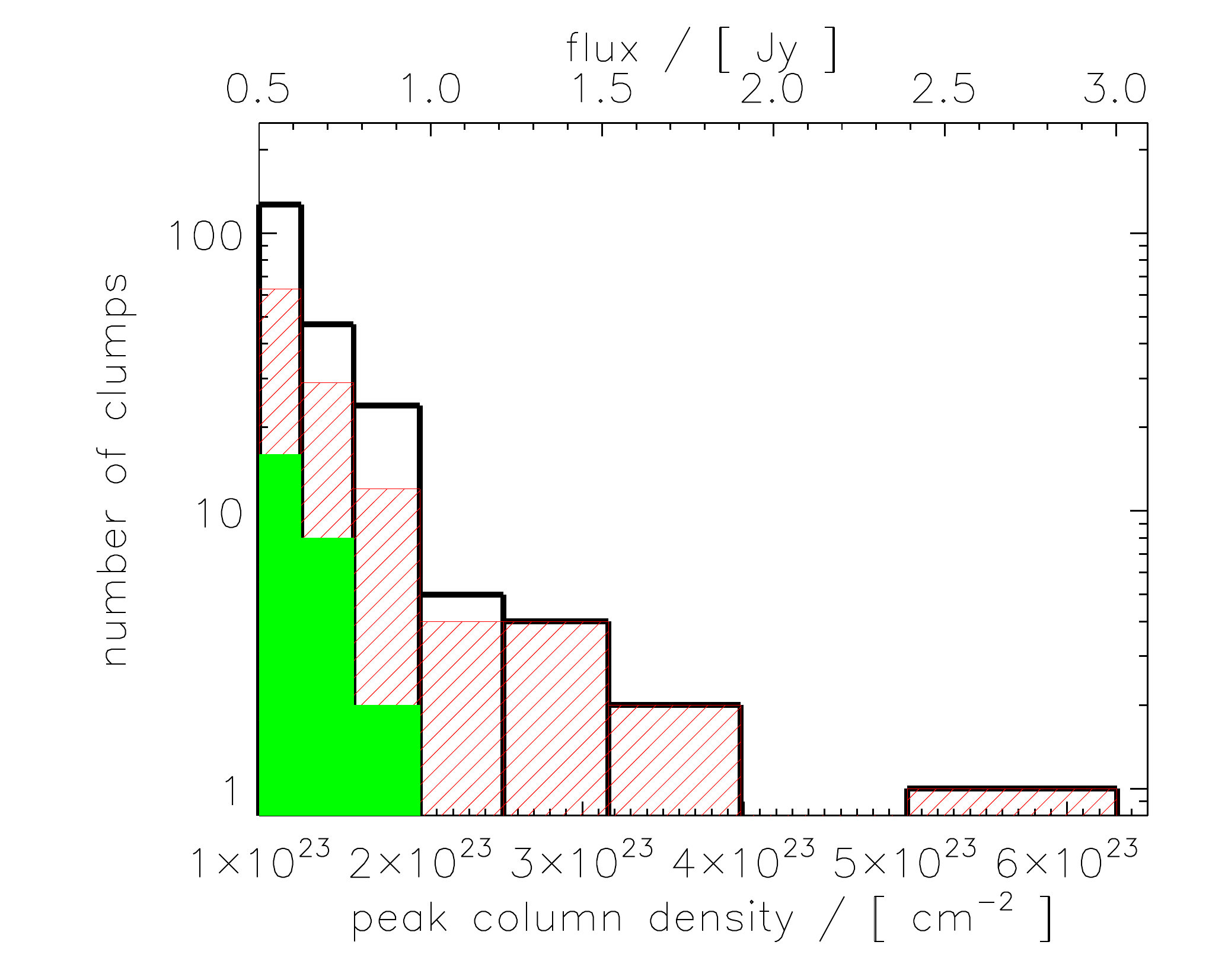}
  \caption{Logarithmic histogram plot of the column density (lower x-axis) and flux (upper axis). While the black histogram represents the full sample of starless clumps, the hatched red and filled green histograms correspond to the near and far sample, respectively. (For details, cf. Sec. \ref{sec:distances}.)%   94\% of the starless clumps have column densities below 2 $\times$ 10$^{23}$ cm$^{-2}$.
  }
  \label{fig:cd_histogram}
\end{figure}
The peak column densities vary only slightly. As shown in Fig. \ref{fig:cd_histogram}, 94$\%$ of the starless clumps have column densities in the
range between our survey threshold 1 $\times$ 10$^{23}$ cm$^{-2}$ and 2 $\times$ 10$^{23}$ cm$^{-2}$. Only 5 clumps, or 2$\%$, have a peak column density larger than 3
$\times$ 10$^{23}$ cm$^{-2}$, which corresponds to 1 g cm$^{-2}$. Nevertheless, the beam at a distance of 3 kpc corresponds to 0.26 pc, hence is too large to
resolve individual cores ($\sim$ 0.01 pc - 0.1 pc). The given column densities are beam-averaged over large spatial scales and the actual peak column densities could be
considerably larger. This effect preferentially reduces the column densities of clumps further away more significantly than those of nearby clumps, introducing an artificial difference between the clumps on this side and on the far side of the Galaxy. This difference is clearly illustrated by the red and green histograms in Fig. \ref{fig:cd_histogram} for the near and far clumps, respectively (see Sec. \ref{sec:dist_col_mass}).

% Assuming a r$^{-1}$ density profile for starless cores 
To get a feeling for the small-scale peak column densities, \citet{Vasyunina2009} studied the effects of distance and telescope resolution onto the peak column density. First, they produced an artificial r$^{-1}$ density distribution grid of 2000\,AU, or 0.01\,pc, resolution. Secondly, they smoothed the grid with different Gaussian kernels to imitate observations with a 24$\arcsec$ beam at different distances. They next compared the obtained column densities to the unsmoothed peak column densities and calculated from these ratios correction factors for different distances, which resemble the peak column density seen with a linear spatial resolution of 2000\,AU. The correction factors applied for a distance of $\sim$ 2 kpc start at around $\sim$ 17, and go up to $\sim$ 40 for distances around 4.5
kpc. 
% In order to get a feeling for the true peak column densities, \citet{Vasyunina2009} smoothed an artificial r$^{-1}$ density distribution with a synthetic 24$\arcsec$ beam and placed the synthetic cloud at different distances. {\bf In their model, each pixel corresponds to 2000\,AU or 0.01\,pc. After smoothing, they then compared
% the obtained column densities to the unsmoothed peak column densities and calculated a correction factor for different distances from these
% ratios, which resembles the peak column density as seen with a linear spatial resolution of 2000\,AU.} The correction factors applied for a distance of $\sim$ 2 kpc start at around $\sim$ 17, and go up to $\sim$ 40 for distances around 4.5
% kpc.
Even assuming a minimal correction factor of $\sim$ 10 and applying it to our sample, all clumps should contain smaller subregions of higher column densities, larger than 3$\times$10$^{23}$cm$^{-2}$, following the Krumholz criterion for high-mass star formation. Nevertheless, this procedure cannot be applied to clumps at all distances. For clumps at the far side of the Galaxy in particular, the beam averages over several/many clumps and projection effects become more likely.%  Assuming a similar column density distribution for both populations, one could try to statistically disentangle the number of clumps per beam on the far side. 
% But due to the various problems explained in Sec. \ref{sec:far_field} the statistical disentanglement is not trivial.
%
%% \subsection{Results of classification}
%% \label{dis:class}

\section{Distances}
\label{sec:dist_col_mass}
\begin{table*}[htbp]
%%\begin{tabular}{|p{4.cm}|c|c|c|c|}
\begin{tabular}{|c|c|c|c|c|}
\hline
 Property Reviewed & Near & Far & Only Far & Estimated Error Near / Far / Only Far\\
\hline
number of clumps & 115 & 26 &9 &\\
distance / [kpc] & 3.1 & 12.8 & 16.9& 0.5 \\
average effective radius / [pc] & 0.3 & 1.0 & 2.2& 0.28/1.2/1.6 \\
average mass / [M$_{\odot}$] & 620 &  5560 & 26400& factor of 4 \\
median mass / [M$_{\odot}$] & 320 & 4600 & 21400&  factor of 4 \\
particle density / [cm$^{-3}$] & 1.0$\times$ 10$^5$ & 2.3$\times$ 10$^4$ & 1.3$\times$ 10$^4$ & factor of 2 \\
\hline
\end{tabular}
\caption{Overview of typical clump properties for near and far clumps. The origin of the uncertainties is explained in Appendix \ref{uncertainties}.}
\label{tab:comparison}
\end{table*}

To determine additional physical parameters, in particular the size and the mass of the clumps, the distance is a major parameter. A Galactic rotation
curve was utilized to determine distances from the clumps' radial velocities. In the following, only clumps with a distance estimate are discussed.

\label{sec:distances}
\begin{figure}[tbp]
  \includegraphics[width=0.5\textwidth]{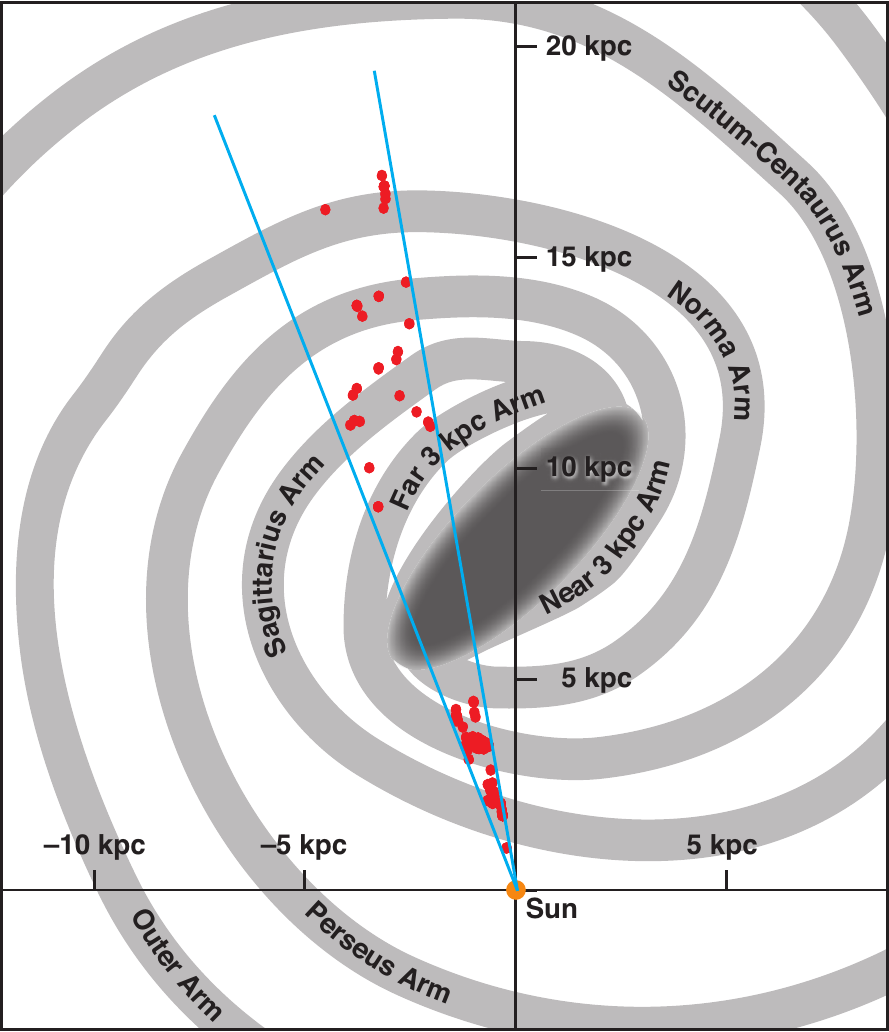}
  \caption{Artist impression of face-on view of the Milky Way by R. Hurt (SSC-Caltech) / MPIA graphic. Plotted on top are the starless clumps presented here with the distance according to the distance flag in Table \ref{tab:list}.}
  \label{fig:reid_galaxy}
\end{figure}
As the idea of this study was an unbiased survey of a large area of the sky with continuum data, a priori we have no information about the distances to the clumps found. To tackle this problem, we employed the Galactic rotation curve given in \citet{Reid2009}. The necessary velocities are provided by Wienen et al. (submitted). Wienen et al. (submitted) conducted spectroscopic follow-up observations of NH$_3$ towards bright peaks in the ATLASGAL survey. If no counterpart was found in Wienen et al. (submitted), we used the HCO$^+$ survey of BGPS sources by \citet{Schlingman2011}.

To maximize the number of clumps to which we could assign a velocity, all clumps that lie within the same lowest significant contour were assumed to be connected. With this assumption, we were able to assign to connected clumps the same velocities as their neighbors.
Incorporating all information at hand, the velocities of 150 starless clumps, or 71\%, are known.
The uncertainties in the velocities can be estimated by comparing the NH$_3$ and HCO$^+$ velocities of clumps that have both measurements. We note that 134 of all clumps (not only starless clumps presented here) were observed by both Wienen et al (submitted) and \citet{Schlingman2011}. The average difference between both velocity measurements is 0.5 km/s, while their median difference is 0.3 km/s, with 2.3 km/s being the largest difference. Therefore, we estimated the velocity uncertainties to be 0.5 km/s.

Owing to the rotational structure and symmetry of the Galaxy, a Galactic rotation curve usually yields two distance solutions for a given direction and velocity. To solve this distance ambiguity, additional information or assumptions are required. While in studies of IRDCs it has often been argued that all sources lie at the near solution because their identification requires a bright mid-IR background, this argument could not be adopted here. The optically thin dust emission at 870 $\mu$m instead allowed us to identify clumps across the entire Galaxy. Nevertheless, coincidence with an IRDC favors the near solution and we used the catalog of IRDCs given in \citet{Peretto2009} to identify nearby objects within our sample. Although they cover a different column density range (for details see section \ref{sec:direct_results}), considerable overlap can still be expected. During the visual inspection of the 24 $\mu$m emission, additional dark patches connected to our clumps were identified and noted as IRDC. In the following, all clumps with an associated IRDC were assumed to be on the near side.

For 9 sources with velocities between -5 km/s and -1 km/s, the rotation curve only allows the far solution, because their near solution is meaningless (it places the source in the outer Galaxy, while we looked in the opposite direction towards the inner Galaxy). The far solution places them in, or close to, the Norma arm at $\sim$ 17 kpc. However, as discussed in \citet{Dame2008} and \citet{Green2011}, the velocities could also place them in the near 3 kpc arm at $\sim$ 5.2 kpc distance. For consistency with the adopted Galactic rotation model, we prefer the Norma solution. %%At the same time, their velocities and longitudes are consistent with the CO emission of the 3 kpc arm \citep{Green2011}. Nevertheless, since we cannot decide which solution is correct we accept the result of the rotation curve for consistency.}\\
Future studies of these clumps could also use H{\sc i} self-absorption or $^{13}$CO associations with well-known regions \citep{Liszt1981} to better solve the distance ambiguity.

In summary, out of the 160 sources with velocity measurements 115 clumps are likely on the near side and 35 clumps are on the far side of the Galaxy (cf. Table \ref{tab:comparison} and Sec. \ref{sec:far_field}). Only few starless clumps on the far side of the Galaxy have been known previously \citep{Battersby2011}, thus about a quarter of the sources are newly identified. Fig. \ref{fig:reid_galaxy} shows the locations of the starless clumps within the Milky Way Galaxy. One notes a clear gap between 5 kpc and 11 kpc in the source distribution, which can be explained in several ways: (1) The elliptical orbits in the bulge of the Milky Way randomize its clouds' velocities and the rotation curve places them at random distances. (2) Circular orbits close to the tangent point have very large $\frac{d (dist)}{dv}$, hence small errors in the velocities propagate into large distance offsets. (3) The majority of the cold gas is homogeneously distributed in a molecular ring around the Galactic center with 4 kpc $<$ R$_{GC}$ $<$ 8 kpc \citep{Solomon1989}. Therefore, no clumps are expected outside that region.

\section{Masses and clump mass function}
\label{sec:masses_CMF}
\subsection{Masses}
\label{sec:masses}
\begin{figure}[tb]
  \includegraphics[width=0.5\textwidth]{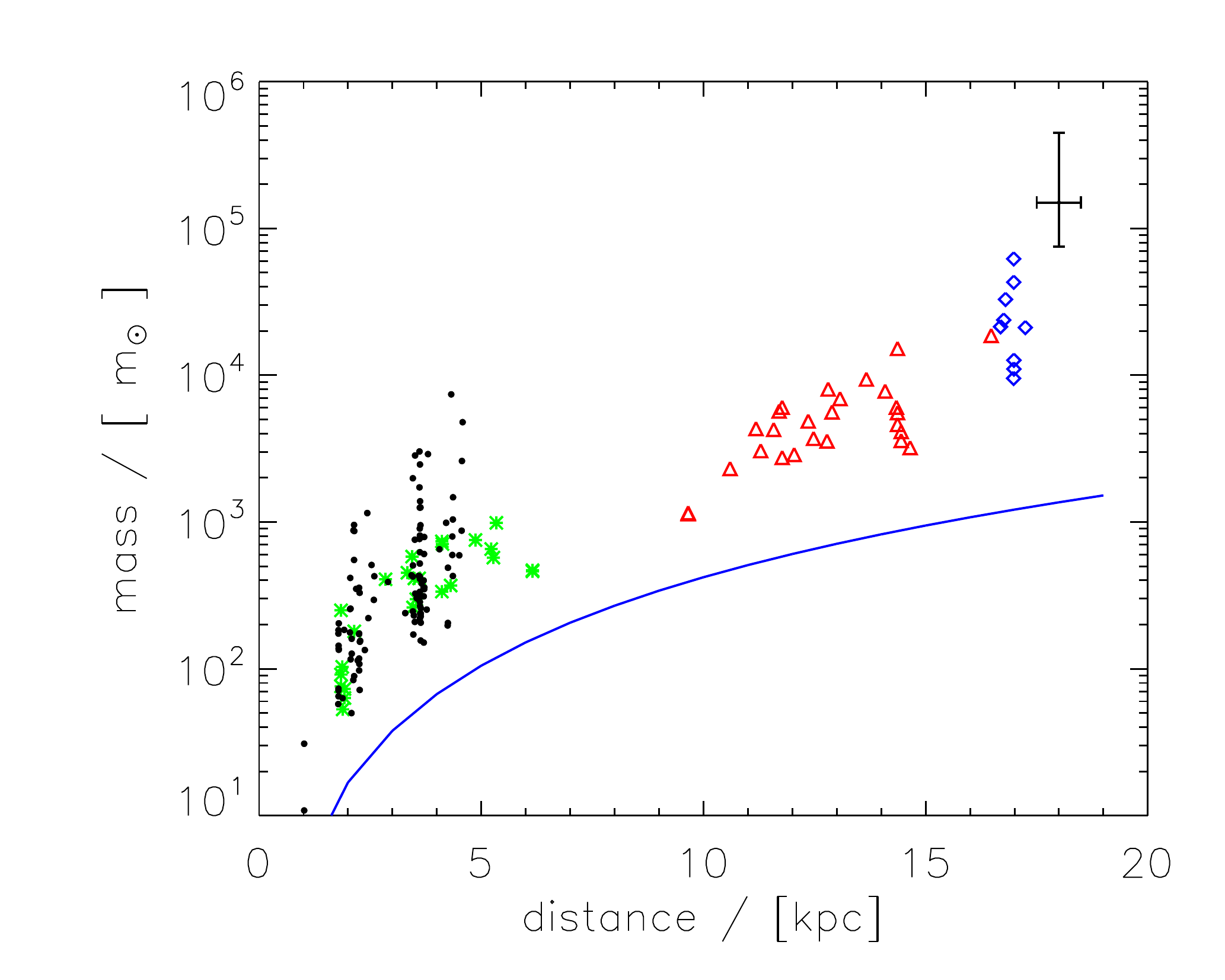}
  \caption{%% Left: Plotted is the column density over the distance. The solid line indicates the column density threshold. The black filled dots represent sources with IRDC association, therefore plotted is the near distance. The red triangles represent the far solution of sources without an IRDC for which both a far and a near solution can be calculated; complementary the green asterisks are their near solution. Blue diamonds show sources with such low velocities that only a single solution can be found in the given direction. The error bar in the top right corner indicates uncertainties of a factor 1.5 for the column densities, while the distance uncertainty is 0.5 kpc.  
    The mass in solar masses is plotted over the distance in kpc.  The solid line indicates the sensitivity/completeness limit of our clump extraction, which depends on
    the distance. The black filled dots represent sources with IRDC associations, and, therefore, the near distance is plotted. The red triangles
    represent the far solution of sources without an IRDC for which both a far and a near solution can be calculated; in addition, the green asterisks represent the corresponding near solutions. Blue diamonds show sources with such low velocities that only a single solution can be found in the given direction. The error bar in the top right corner indicates uncertainties of a factor of two for the masses and 0.5 kpc for the distance.}
  \label{fig:mass_sensitivity}
\end{figure}

Assuming optically thin emission, the mass of these clumps can be calculated from the dust continuum emission via 
\begin{equation}
  M_{gas} = \frac{  R d^2 F_{\lambda} }{ B_{\lambda}(\lambda,T) \kappa} \text{,}
  \label{eqn:mass}
\end{equation}
where most of the parameters are the same as defined in Sec. \ref{sec:column_densities}, and d is the distance. Therefore, the mass can only
be calculated for sources with distance measurements. 
For completeness, the mass is calculated for both the near and far solutions produced by the distance ambiguity and listed in Table \ref{tab:list}. % As explained in
% Sec. \ref{sec:distances}, a first indication for the solution of the distance is set by a connection to an IRDC.
Fig. \ref{fig:mass_sensitivity} shows the calculated masses for both the near and far solutions. The solid line indicates the theoretical sensitivity limit of our source extraction. The uncertainties in the masses are discussed in Appendix \ref{uncertainties}. % Although the plot exhibits a minor gap between the theoretical cut-off and the masses found for each distance, the shape is nicely reproduced.

We expect the amount of dense gas per volume to be similar on both sides of the Galaxy. We can therefore conduct a consistency check by comparing the mass of the near and far population relative to the volume covered.

%%As a consistency check to the near/far indication given in Sec. \ref{sec:distances}, we compare the mass of both populations identified that way. 
All clumps on the near side of the Galaxy have masses below 10$^4$ M$_\odot$ (cf. Fig. \ref{fig:mass_sensitivity}, black dots), while the maximum mass
within the far-clumps is a factor of about three higher than the most massive clump on the near side. In addition, when looking at
Fig. \ref{fig:mass_sensitivity}, the total number of near objects is clearly higher than the number of far objects. To make a quantitative
comparison of both populations, we estimated the common sensitivity limit for distances between 10 kpc and 15 kpc to be 1 $\times$ 10$^{3}$
M$_{\odot}$. We next calculated the volumes that we cover on the near and the far side using the scale height of $\sim$40 pc given in
\citet{Bronfman2000}. Adding up all masses above the far detection threshold for the near range, 0 kpc to 5 kpc, the total mass of clumps on the near side
is $\sim$37000 M$_{\odot}$. Doing the same for the far range 10 kpc to 15 kpc and normalizing it to the volume covered on the near side the total mass
becomes 38000 M$_{\odot}$. Both masses agree given this crude estimate, supporting the idea that, statistically, the allocation is reliable. 

\subsection{Observational biases for clumps on the far side}
\label{sec:far_field}
% [Dear Co-authors,
% for the referee it seems not to be clear that every extraction algorithm can only extract structures larger than the resolution power of the telescope. In special, working with 2D data, no algorithm can know the distance of the signal it is looking at and adjust the resolution or the extraction parameters. I agree with most of you that said this is very obvious and should not be discussed again. Unfortunately the is of different opinion. I don't know how to phrase the following passage without making it sound ridiculous. If you can help me with it, please make suggestions...
% Here is the referee's comment:
% > Sec 5.4.  It is not clear that the physical radius of sources should
% > be correlated with distance. The authors should discuss whether
% > CLUMPFIND extracting similar angular sized structures for all clouds,
% > irrespective of distance, suggests that the extracted structures do
% > not in fact represent physical entities.  Is this in fact evidence
% > that CLUMPFIND is excessively fragmenting the more nearby clouds?  The
% > authors should show and discuss a plot of the clump size (effective
% > radius) versus distance.]
%
\begin{figure}[tbp]
  \includegraphics[width=0.5\textwidth]{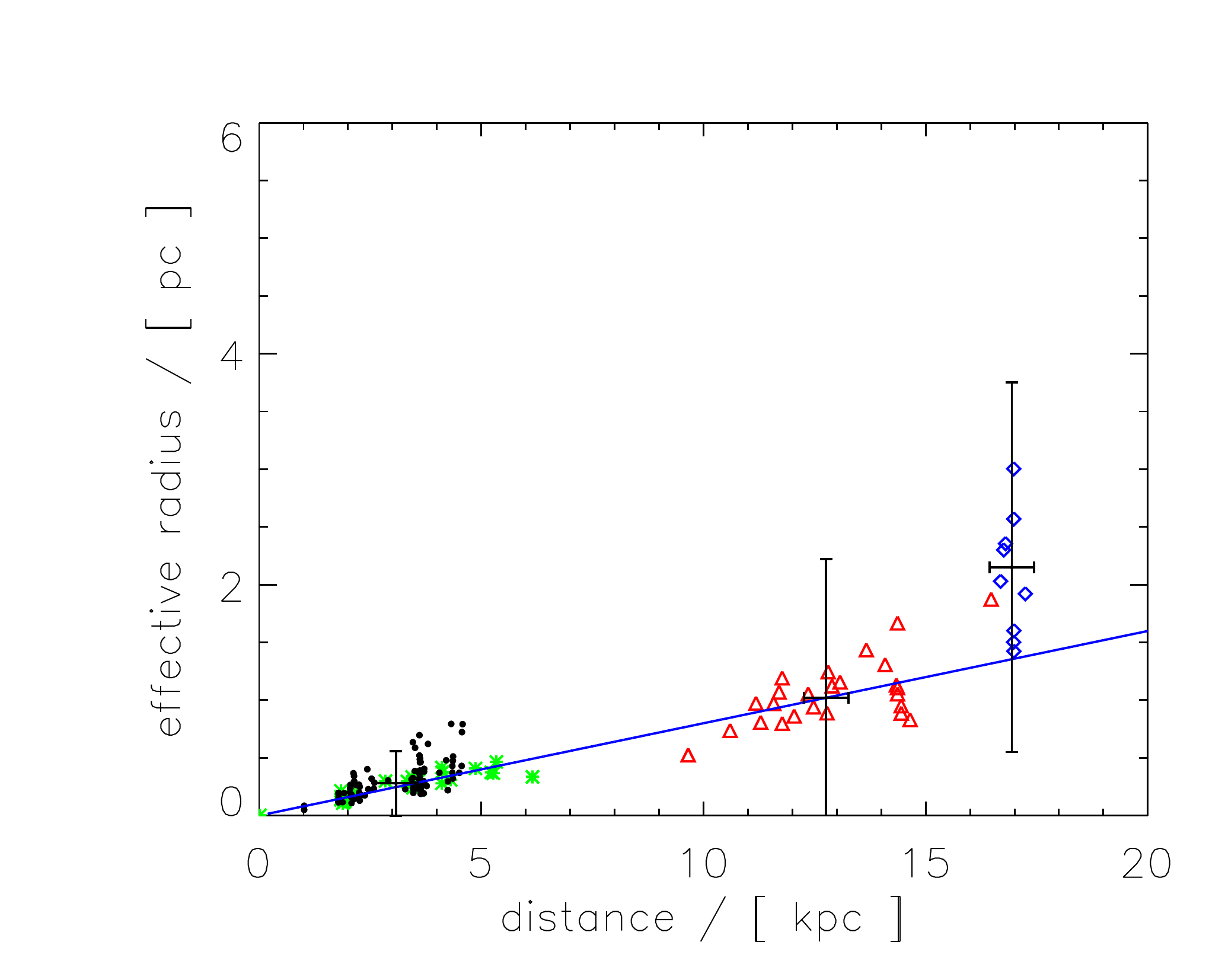}
  \caption{Effective radius in pc over the distance. Colors and symbols are as in Fig. \ref{fig:mass_sensitivity}. % Again, black dots represent near solutions of clumps connected to an IRDC. Red triangles and green asterisks are clumps' with distance ambiguity far and near solution, respectively. Blue diamonds have a far solution only.
  }
  \label{fig:size_vs_dist}
\end{figure}
The change in resolution over the survey's depth affects the sensitivity and the recognition of substructure significantly. The 19.2\arcsec beam corresponds to 0.28 pc linear spatial resolution at a distance of 3 kpc, and in contrast corresponds to 1.4 pc at a distance of 15 kpc.
% While the beam of 19.2$\arcsec$ corresponds to 0.28 pc linear spatial resolution at 3 kpc distance, it corresponds to 1.4 pc at 15 kpc distance. \\
% As the ATLASGAL continuum maps are 2D data, all source extraction algorithms ( clumpfind or any other source extraction algorithm working on 2D data [to be removed]) can solely operate on variations in the intensity distribution. Due to spatial resolution, these variations can only be on the order of the beam size or larger. In addition, distance information can only be added once structures are identified. [This might be an too ambitious statement. With plenty of crude assumptions, one could do it earlier.]
While the distances vary by more than an order of magnitude, the angular sizes of the extracted clumps vary by only a factor 2-3 and show no correlation with distance (cf Fig. \ref{fig:angular_effective_radius}). This results in an almost linear correlation between the physical size and the distance, which is shown in Fig. \ref{fig:size_vs_dist}. In addition, a single, unresolved source would be 25 times fainter at 15 kpc than at 3 kpc. As indicated by the solid line in Fig. \ref{fig:mass_sensitivity}, the completeness limit changes with distance. Both effects are studied in detail in the following.

Taking an ATLASGAL map of 3$^\circ \times$2$^\circ$, we re-extracted all clumps with CLUMPFIND using the same thresholds as explained above and calculated their masses assuming a generic distance of 3 kpc. In addition, we convolved the same map with a Gaussian profile, emulating a resolution of 96$\arcsec$, reducing the resolution by a factor of 5. This resembles the appearance of the same structure as seen at 15 kpc. Once again CLUMPFIND was used to search for clumps using the same parameters, but assuming a distance of 15 kpc when determining the mass.

While 90\% of the total mass was recovered in the lower resolution maps, the number of clumps extracted differed significantly. In the full resolution map, 328 clumps were extracted, whereas in the lower resolution map, only 20 clumps were found. This implies that structures, which can be resolved into several clumps on the near side, cannot be resolved on the far side and that their fluxes then add up.\\
The volumes covered at the two distances differ by a factor of about five. This would add to the probability of chance alignment and since the dust emission is optically thin, several faint clumps within the same beam may add up and may detectable.

Therefore, one should keep in mind that clumps discovered on the far side are slightly different types of objects.

\subsection{Clump-mass function (CMF)}
\label{sec:clump-mass_function}
\begin{figure}[tbp]
  \includegraphics[width=0.5\textwidth]{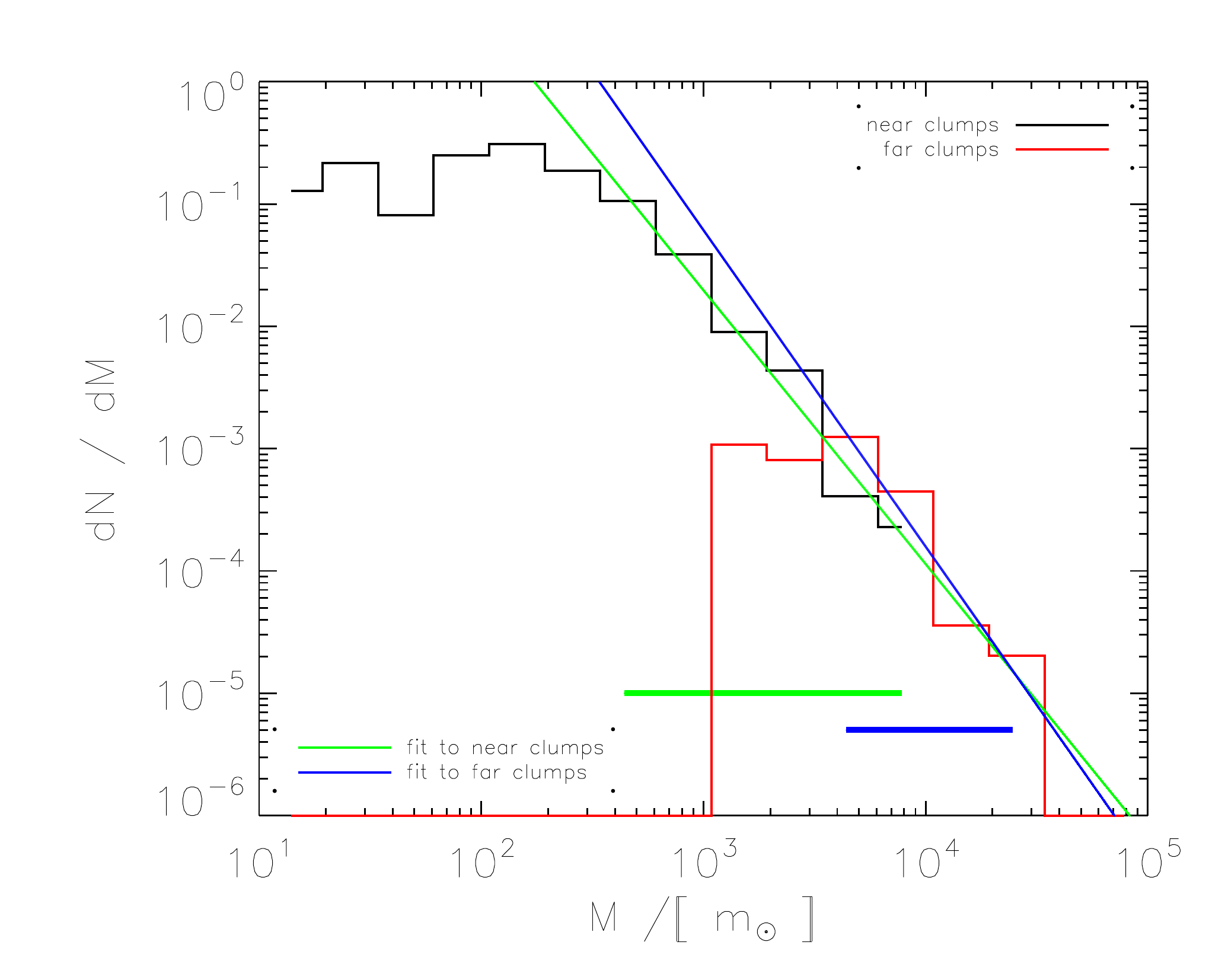}
  \caption{Clump mass function with the near population plotted in black, the far population plotted in red. Both populations' high-mass tail have been fitted, with the fitting range indicated by the thick horizontal bars at the bottom. The power law indices are fitted to -2.2 and -2.6 for the near and far population, respectively.}
  \label{fig:clump_mass_function}
\end{figure}
Mass distributions connected to stellar populations or star formation are very often compared to a power-law distribution, as first stated by
\citet{Salpeter1955} and discussed since then. The more recent study of \citet{Reid2010} even found that statistical errors allow the slope of (m)any sort(s) of astrophysical random distribution(s) to be fitted by a power-law with spectral
indices similar to the Salpeter value of 2.35 within the errors. Nevertheless, a clump-mass function (CMF) is of great interest in connecting clumps to
core- and star-formation efficiencies. To ensure that our data has sufficiently high quality statistics, we considered all 125 near clumps for a `local' CMF and the 27 far
clumps for a CMF on the other side of the Galaxy. As shown in Fig. \ref{fig:mass_sensitivity}, the completeness limit for the near and far populations is lower than 100 $M_{\odot}$ and 1000 $M_{\odot}$, respectively. The black histogram in Fig. \ref{fig:clump_mass_function} represents the number distribution of the near clumps dN/dM. %for logarithmic bins, normalized to the bin width. This implies that the spectral index is comparable to a linear version of this plot with the better statistics of logarithmic bins.\\
Owing to the completeness limits, we cannot rely on the low-mass end of the CMF below 100 $M_{\odot}$, nor do we have sufficient statistics to determine the existence of a broken power-law distribution. We instead only consider the high-mass tail. The green line indicates a fit to the high-mass tail of the near-clump mass function, with a logarithmic slope of $\alpha$ = -2.2. The red histogram is the far-clump mass function, fitted by the blue line of slope $\alpha$ = -2.6. Neglecting the uncertainties on the data points, the uncertainties on the slopes calculated by the IDL routine 'LINFIT' are 0.2 and 0.4 for the near- and far-clump mass functions, respectively. The thick bars at the bottom indicate the fitting range. If we expand the fitting range of the near population and include lower mass bins, the slope becomes immediately shallower. If we reduce the fitting range, the slope remains constant within the uncertainties. Therefore, we believe that the fitting range is reliable and the uncertainty is reasonable. For the far range, the situation is more difficult because of the smaller number of bins. An enlargement of the fitting range to smaller bins again reduces the slope dramatically. Including the next mass bin does not change the result significantly, but including additional bins of lower mass would steepen the slope to -3.1. This number is almost within the errors and can be entirely explained by the lower quality statistics. Nevertheless, the fit to the far mass distribution should be interpreted with caution.

% \subsection{Lifetime of clumps}
% Despite the naming problem occurring regularly when discussing about clouds, clumps and cores, the discussion about star-formation efficiencies from
% clumps to stars has not agreed on a common value (cf. \citealt{Chabrier2010} and references therein), and high-mass star formation has gotten even less
% attention. An estimate of the star formation efficiencies of dust clumps for massive stars was given in \citet{Johnston2009}. They found large
% variations, ranging from few percent to 25 \% with an average of 7 \%. The masses of the starless clumps identified in the present study range from
% few tens to few ten-thousands of solar masses. As discussed in Sec. \ref{sec:far_field}, clumps seen at the far side are most likely a mixture of dust
% seen in chance-alignment or groupings that we would identify as several structures on the near side. In the following, to be more consistent with
% values given in the literature, we will only consider starless clumps which are most likely located on the near side. When assuming the power-law
% M$_{\star}$ = M$^{0.5}$ found in \citet{Johnston2009}, to form a cluster with
% the most massive star to be 20 M$_{\odot}$ a clump of 400 M$_{\odot}$ is required, and a 2500 M$_{\odot}$
% clump is required for a cluster containing a 50 M$_{\odot}$ star. With these numbers, about 38 \% of our near clumps can form stars up to 20 M$_{\odot}$, while 5 \% of the
% near clumps can even form stars with masses up to 50 M$_{\odot}$. In absolute numbers, 6 out of 125 clumps may form massive stars of M > 50 M$_{\odot}$.\\
\section{Lifetimes}
\label{sec:lifetimes}
\begin{table*}[htbp]
  %% \begin{tabular}{|p{4.cm}|c|c|c|c|}
  \begin{tabular}{|c|c|c|c|c|}
    \hline
%%    \multicolumn{5}{|c|}{Lifetime}\\
    SFR & \multicolumn{2}{|c|}{H83, diffuse ISM opacity} & \multicolumn{2}{|c|}{OH94, cold dense opacities} \\
    Mass of Potential Star & 20 M$_{\odot}$ & 40 M$_{\odot}$ & 20 M$_{\odot}$ & 40 M$_{\odot}$ \\
    \hline
    Clump-mass Threshold & 1065 M$_{\odot}$ & 2960 M$_{\odot}$ & 1065 M$_{\odot}$ & 2960 M$_{\odot}$ \\
    Number of Clumps Above Threshold & 14 & 3 & 6 & 1 \\
    \hline
    SFR of 1 M$_{\odot}$ / yr & 3.3 $\times$ 10$^5$ yr & 2.0 $\times$ 10$^5$ yr & 1.4 $\times$ 10$^5$ yr & 6.6 $\times$ 10$^4$ yr \\
    SFR of 3 M$_{\odot}$ / yr & 1.1 $\times$ 10$^5$ yr & 6.6 $\times$ 10$^4$ yr & 4.8 $\times$ 10$^4$ yr & 2.2 $\times$ 10$^4$ yr \\
    SFR of 6 M$_{\odot}$ / yr & 5.5 $\times$ 10$^4$ yr & 3.3 $\times$ 10$^4$ yr & 2.4 $\times$ 10$^4$ yr & 1.1 $\times$ 10$^4$ yr \\
      % distance / [kpc] & 3.1 & 12.8 & 16.9& 0.5 \\
      % average effective radius / [pc] & 0.3 & 1.0 & 2.2& 0.28/1.2/1.6 \\
      % average mass / [M$_{\odot}$] & 580 &  5400 & 29000& factor of 4 \\
      % median mass / [M$_{\odot}$] & 290 & 4300 & 23800&  factor of 4 \\
      % particle density / [cm$^{-3}$] & 1.1$\times$ 10$^5$ & 2.3$\times$ 10$^4$ & 1.3$\times$ 10$^4$ & factor of 2 \\
      \hline
    \end{tabular}
    \caption{Lifetimes of starless clumps calculated for different sets of parameters. Estimates are calculated using different Milky Way star formation
      efficiencies for opacities for both the cold ISM (\citealt{Hildebrand1983}, H83) and dense but cold regions (\citealt{Ossenkopf1994}, OH94).}
    \label{tab:lifetime}
  \end{table*}
  
As discussed in Sec. \ref{sec:far_field}, clumps seen at the far side of the Galaxy are a mixture of clumps seen in chance-alignment or groupings that we would
identify as several structures on the near side. To form a more consistent sample, in the following we only consider starless clumps that are identified on the near side.

\subsection{Mass of the most massive star}
\label{sec:most_massive_star}
To place constraints on the lifetime of starless clumps, we first need to estimate what clump mass is required so that the final cluster can house at least one massive star.

% Despite the naming problem occurring regularly when discussing clouds, clumps and cores, the discussion about 
Star-formation efficiencies on scales from
clumps to stars do not have a common value but a number of studies estimate that it is 23\% - 50\% (\citealt{Chabrier2010} and references therein). For high-mass star formation, the numbers are even more weakly constrained (5\% - 50\%, cf. \citealt{Krumholz2007}, \citealt{Kuiper2010}).

Following the definition given in \citet{Williams2000} and \citet{Beuther2007}, these clumps will most likely form
entire clusters instead of single stars. To estimate the required clump mass to form a star of given mass, one must assume a gas-to-star
formation efficiency and the initial mass function (IMF) of the cluster produced.

We assume the IMF of \citet{Kroupa2001} and normalize it to the probability that at least one star with a mass higher than 20 M$_{\odot}$ is formed. Integrating the normalized IMF over the expected mass range of stars, from 0.08 to 150 M$_{\odot}$, the stellar mass of this cluster % a cluster containing a 20 M$_{\odot}$
is on the order of 320 M$_{\odot}$. With a star formation efficiency (SFE) of 30\% (for details see next paragraph), we estimate the mass of a clump with the potential to form at least one
star more massive than 20 M$_{\odot}$ to be 10$^3$ M$_{\odot}$. A 3 $\times$ 10$^3$ M$_{\odot}$ clump is required to form a 40 M$_{\odot}$ star (see also Table \ref{tab:lifetime}).

Since the mass distribution follows a power law, the number of clumps with masses higher than a given threshold is very sensitive to that threshold. 
In the picture in which these clumps form entire clusters following the IMF, the estimate of the stellar cluster mass relative to the most massive star seems quite reliable. Nevertheless, the estimates of the SFE vary over a wide range \citep{Lada2003, Alves2007, Johnston2009, Bontemps2010}. Here, we used the SFE given in \citet{Lada2003}, \citet{Alves2007}, and \citet{Bontemps2010} of 30\%.

% Nevertheless, the star formation efficiency of 30 \% seems to be quite high at first. But while estimates in ??? [ here I don't want to cite Johnston because the different values within that one paper confused the referee for a good reason. By now I even believe that this relates only to a single, massive star, not to a cluster or total stellar mass. That would make sense because the low mass part will not produce any ionizing flux] suggest lower star formation efficiencies and therefore higher clump masses for a given stellar mass, \citet{Johnston2009} gives as relation between the most massive star within a clump and the clump's mass: M$_{*}$ = 1.0 $\times$ M$_{clump}^{0.5}$. This relation proposes significantly smaller clumps as we require here. Therefore we will stick to the intermediate values, since we are lacking better statistics.}\\
%%To be more consistent with values given in the literature, we will only consider starless clumps which are most likely located on the near side.
In the near sample derived across 20 deg$^2$ of the sky, with this estimate only 14 starless clumps have the potential to form stars more massive than 20
M$_{\odot}$, and only 3 have the potential to form a 40 M$_{\odot}$ star.

\begin{figure*}[tbp]
  \includegraphics[width=0.33\textwidth]{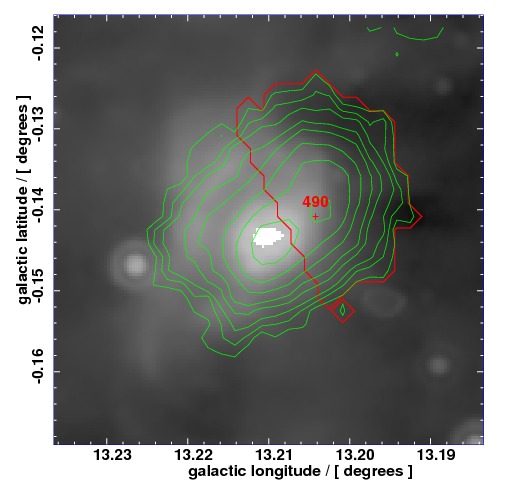}
  \includegraphics[width=0.33\textwidth]{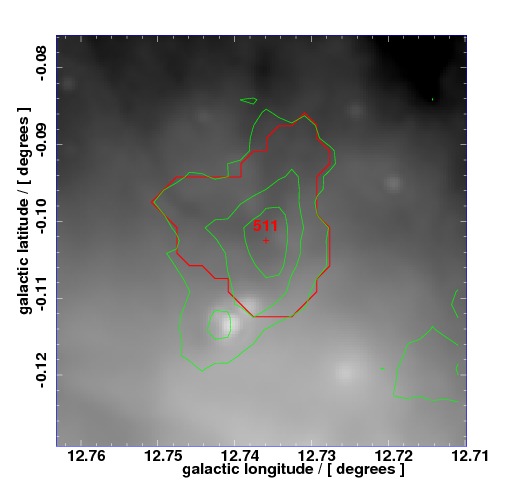}
  \includegraphics[width=0.33\textwidth]{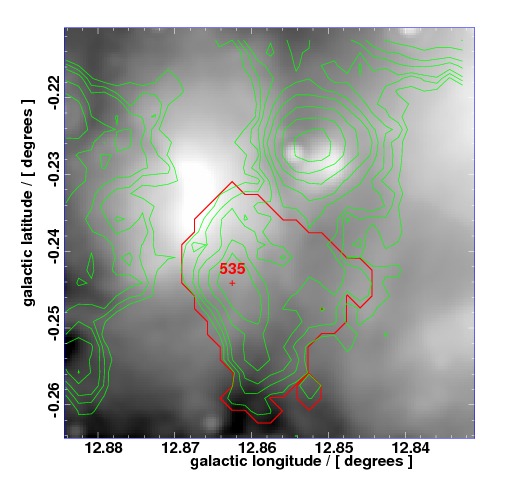}
  \caption{MIPSGAL images of the three most massive starless clumps found in this survey, with (from left to right ) 7400 M$_{\odot}$, 4700 M$_{\odot}$, and 3000
    M$_{\odot}$. Green contours are from ATLASGAL, red lines mark the boarders of the clumps identified by CLUMPFIND. }
  \label{fig:most_massive_clumps}
\end{figure*}
The MIPSGAL images of the 3 most massive clumps are shown in Fig. \ref{fig:most_massive_clumps}. As can be seen, none of the regions are isolated but
all are connected to regions already containing 24 $\mu$m sources. At the distances to these objects varying between 3.6 kpc and 4.6 kpc, their effective radii become between 0.7
pc and 0.8 pc. Therefore, their particle densities are not among the highest of our sample at only 4 $\times$ 10$^4$ cm$^{-3}$ to 8 $\times$ 10$^4$
cm$^{-3}$. Nevertheless, these are very interesting objects and very promising in the context of massive star formation.

\subsection{Lifetime of clumps}
\label{sec:lifetimeclumps}
Since there is no reason to believe that the starless clumps have ages that are correlated with those of other objects in this sample, the sample should span the entire age range expected for these clumps. If this is the case, the oldest clumps will start forming stars as new and similar clumps appear. Therefore, their lifetime can be calculated by comparing the number of these clumps identified to the number of massive stars formed.

To do so, we need to \textbf{1)} extrapolate the number of starless clumps we would find in the entire Milky Way galaxy, and \textbf{2)} estimate the number/fraction of massive stars formed every year:\\
\textbf{1)} We assume that most star-forming gas is distributed in a ring around the Galactic center, between 4 kpc and 8 kpc Galactocentric distance \citep{Solomon1989} with a scale height of 40 pc \citep{Bronfman2000}. Taking into account that we only consider clumps up to a distance of 5 kpc from the Sun within the direction of the survey, and above the clump mass thresholds for 20 M$_{\odot}$ and 40 M$_{\odot}$ stars, % or 1 $\times$ 10$^3$ M$_{\odot}$ and 3 $\times$ 10$^3$ M$_{\odot}$, respectively, 
for the whole Milky Way Galaxy, we expect to identify 1043 and 223 clumps, respectively.\\
\textbf{2)} To calculate the fraction of massive stars formed every year, one needs to assume a star-formation rate (SFR) for the Milky Way. The most recent publications suggest a star formation rate of around 1 - 2 M$_{\odot}$/yr \citep{Robitaille2010}.

% For a SFR of 1 M$_{\odot}$/yr the lifetime of massive starless clumps become (2 $\pm$ 1) $\times$ 10$^5$ yr and (1 $\pm$ 1) $\times$ 10$^5$ yr
% calculated for all clumps more massive than 10$^3$ M$_{\odot}$, and 3 $\times$ 10$^3$ M$_{\odot}$, respectively. For larger SFRs, the lifetimes become shorter in a linear fashion. Lifetimes for different parameters are summarized in Table \ref{tab:lifetime}. 
For a SFR of 1 M$_{\odot}$/yr, the lifetimes of massive starless clumps become 3 $\times$ 10$^5$ yr and 2 $\times$ 10$^5$ yr
calculated for all clumps more massive than 10$^3$ M$_{\odot}$, and 3 $\times$ 10$^3$ M$_{\odot}$, respectively. For higher SFRs, the lifetimes become shorter in a linear fashion, e.g. for 6 M$_{\odot}$/yr the lifetimes become 5.5 $\times$ 10$^4$ yr and 3.3 $\times$ 10$^4$ yr, respectively. Lifetimes for different parameters are summarized in Table \ref{tab:lifetime}. As we discuss in Sec. \ref{discussion of lifetimes}, we estimate the lifetime of these objects to be (6 $\pm$ 5) $\times$ 10$^4$ yr.

\section{Discussion}
\label{sec:discussion}

\subsection{Discussion of the clump-mass function}
% Both exponents of the clump-mass function found in Sec. \ref{sec:clump-mass_function} are consistent with the Salpeter value \citep{Salpeter1955} for the IMF.
The slope of the CMF for starless clumps found in this study, $\alpha$ = -2.2, is similar to the value of the Salpeter
IMF. Nonetheless, as we study clumps that will most likely host small clusters rather than individual stars, we do not propose a one-to-one mapping to the IMF. We emphasize that not all clumps will eventually form clusters or even be transient objects.
Another set of massive clumps was presented in \citet{Peretto2009}, including a subset of clumps without MIPS 24 $\mu$m emission. They found the mass function of IRDCs to be similar to the CO clump mass distribution \citep{Simon2006, Peretto2010} with $\alpha$ = -1.7. In addition, they used a derivative of CLUMPFIND to search their extinction maps for structures within the IRDCs. These fragments have a rather Salpeter-like slope \citep{Rathborne2006, Peretto2010} which is then similar to our result. Although the extinction method used by \citet{Peretto2009} to calculate column densities of IRDCs is sensitive to a lower column density range than that of the starless clumps we present, the 870 $\mu$m emission identifies objects similar to fragments \citet{Peretto2009} find in IRDCs.

The studies of \citet{Williams2004}, \citet{Reid2005}, and \citet{Beltran2006} all targeted the more evolved high-mass protostellar objects. They found a common break in the CMF at 100 M$_{\odot}$ and fit power laws to their high-mass end with exponents between -2.0 and -2.32. Although we did not attempt to fit the exact break, a break point of 100 M$_{\odot}$ or just above seems to be in good agreement with the starless CMF, but that clearly might be biased by the completeness limit close to 100 M$_{\odot}$. However, the exponent of the near CMF agrees with all values within the errors. Thus, comparing our results to earlier studies of more evolved clumps shows that there is no evidence that the CMF of starless clumps is different from a CMF at later evolutionary stages.

Thus, we found that the exponent of the CMF for clumps at the far side of the Galaxy is larger than most other values. Although the uncertainty is larger and the difference could be explained by the errors, this trend might equally represent a general scale-dependent trend. 
\citet{Beltran2006} distinguished between the populations at d $<$ 2 kpc and d $<$ 6 kpc and found that the exponent for the more clearly resolved population $<$ 2 kpc is shallower than for d $<$ 6 kpc. Our far population is even more distant than their sample and the far CMF's exponent would continue the trend to steeper slopes. This could be interpreted as a general scale-dependent trend and rather a matter of resolution than of true structure.

Studies of the core mass function for low-mass star formation target significantly smaller objects. To have sufficient spatial resolution, they are typically chosen to be nearby. Nevertheless, whether they present starless cores or more evolved objects, many studies have measured power-law slopes close to the Salpeter value \citep[][]{Motte1998, Johnstone2000, Alves2007, Enoch2008}.%  [what's the conclusion of that? I guess that all this is at most self similarity, or really just coincidence.]

\subsection{Discussion of the lifetimes}
\label{discussion of lifetimes}
The estimates of the lifetimes of starless clumps are based on the SFR of a given mass range and the number of clumps above a corresponding clump mass. Both estimates involve several assumptions, which may introduce errors.

Lifetimes are inversely proportional to the SFR. The SFR varies from 1 M$_{\odot}$/yr \citep{Robitaille2010} to over 4 M$_{\odot}$/yr \citep{Diehl2006} to even larger values, but the most recent publications favor the smaller values. Nevertheless, our own survey looks at a region in the vicinity of the Galactic center for which both \citet{Rosolowsky2010} and \citet{Beuther2011} found that more dense gas is located within the inner l $<$ 30$^{\circ}$ of the Galactic plane. While most gas is concentrated within a molecular ring around the Galactic center at 4 kpc $<$ R $<$ 8 kpc \citep{Solomon1989}, the ring does not seem homogeneous and the outer regions of that ring seem to contain less gas. This might indicate that the extrapolated numbers for the entire Galaxy might be higher than average and therefore require a higher SFR. This would reduce the lifetime estimate.%  [The referee misunderstood us here at all. His point was: we cannot form more stars in the little region of our survey than the entire Milky Way produces overall. Does it now become clear?]

The lifetimes are also proportional to the number of clumps of given mass and when we use the extrapolated dust opacities from \citet{Ossenkopf1994} our clump masses decrease and the number of clumps above our threshold is lower. This, in turn, reduces the lifetimes. In addition, one should keep in mind that small number statistics are involved. If one of these objects turns out to already be star-forming, the estimated lifetimes are reduced, while correcting chance alignments of mid-IR sources has the opposite effect.

Another factor contributing to the uncertainties are the star-formation efficiencies, which directly influence the mass thresholds themselves. As explained in Sec. \ref{sec:most_massive_star}, the number distribution decays as a power of the mass, which makes the lifetime estimates very sensitive to the clump mass thresholds. On the basis of the range of possible lifetimes shown in Table \ref{tab:lifetime}, we estimated the uncertainties to be one order of magnitude.

In summary, most effects seem to reduce the lifetimes. Based on these arguments, we estimate the lifetime of starless clumps to be on the order of (6 $\pm$ 5) $\times$ 10$^4$ yr.

For volume-averaged particle densities of 10$^5$ cm$^{-3}$, the free-fall time becomes $\sim$ 1.6 $\times$ 10$^5$ yr, and lower densities increase this number. Therefore, the free-fall time is about a factor of two longer than the lifetime we found for starless clumps, but both agree within the errors.

In good agreement with previous studies \citep{Motte2007, Hatchell2008, Motte2010}, Table \ref{tab:lifetime} shows that the estimated lifetimes of more massive clumps are smaller. 
Accordingly, it was expected that the lifetimes we found for starless dense clumps are shorter than the 3 $\times$ 10$^5$ yr found by \citet{Kirk2005} for low-mass cores.

An adequate comparison to the lifetime of high-mass starless clumps is difficult because only a few studies exist. Most deal with slightly different kinds of objects. For example, \citet{Motte2007} studied the nearby Cygnus X region and found cores that cover the same volume densities as our clumps, while their masses are significantly lower (see also \citealt{Motte2010}). However, they found clumps of similar mass to our objects but with lower volume densities. Using SiO as an additional tracer of star formation, \citet{Motte2007} did not find a single starless massive dense core in their sample. Therefore they proposed that the low-density starless clumps dynamically evolve into star-forming massive dense cores. Their lifetime estimate of starless massive dense cores becomes $<$ 10$^3$ yr.

A similar study covering more similar entities both in density and mass by \citet{Russeil2010} found one starless clump, hence they derived a statistical lifetime of $\sim$ 1 $\times$ 10$^4$ yr. This agrees with our estimate, but at the lower edge of the error. The difference could also be caused by their inclusion of SiO as a tracer of star formation, which could reduce the number of starless clumps we found.

\subsection{Comparison to other surveys}
\label{sec:dunham_etc}
\subsubsection{Comparison to the Bolocam Galactic Plane Survey (BGPS)}
The Bolocam Galactic Plane Survey (BGPS) performed a systematic study of the inner Galactic plane at 1.1 mm. Although their rms across the region 10$^{\circ}$ $<$ l $<$ 20$^{\circ}$ is $\sim$ 25 mJy/beam, their overall 5 $\sigma$ level is 0.4 Jy, at which they have a completeness level of 99\%. Their 0.4 Jy completeness threshold converts into a column density threshold of $\sim$ 6 $\times$ 10$^{22}$ cm$^{-2}$ over their 31\arcsec beam when using the same assumptions as for our data, while their sensitivity translates into a rms of 64 mJy/beam at 870 $\mu$m and their beam width of 31$\arcsec$. With this column density, they found 1211 sources on the same region as we studied here.

For that data set, \citet{Dunham2011} used different point source catalogs to search within all BGPS
sources for mid-infrared tracers of star formation activity. As resources they employed the Red Sources \citep[R08,][]{Robitaille2008}, the RMS catalog, the catalog of extended green objects (EGOs, \citealt{Cyganowski2008}), and the full GLIMPSE catalog. The EGOs have no direct counterpart in this study, but trace shocked gas. To be truly starless, starless clumps should not contain EGOs.

Reviewing the \citet{Dunham2011} source catalog for the same region as surveyed in this study, we found that for $\sim$ 70 \% of the sources they found neither RMS sources, R08 sources, nor EGOs.

Following a similar approach to \citet{Dunham2011} and using the identical mid-IR tracers, the R08, RMS, and EGO catalogs, for the ATLASGAL clumps, we found R08 counterparts within 163 clumps, 39 RMS counterparts within clumps without R08 sources, and 9 matches among EGOs and clumps without the previous tracers. This leaves 719, or 77\%, of our ATLASGAL clumps starless. Both fractions of infrared quiet clumps are significantly larger than those of \citet{Peretto2009} (32\%) or we found (23\%) including all tracers. Even without visual inspection but including MIPSGAL 24 $\mu$m sources found by STARFINDER, we would have found only 42\% of the clumps to be starless. These differences in our statistics from both of the latter studies can be explained by the superior sensitivity of the ``by eye'' source confirmation and the higher spatial resolution of the MIPSGAL survey over the MSX images.

\subsubsection{Comparison to PLANCK's Early Cold Cores}
That Planck's Early Cold Core Source List (ECC) \citep{PlanckCollaboration2011} does not contain any cold sources in the region of the sky we
surveyed here is apparently remarkable. However, the ECC contains only sources colder than T $<$ 14 K averaged over PLANCK's beam sizes of several arc minutes. Owing to the high gas density and ongoing star formation as well as confusion, no sources are expected to be found within the Galactic plane \citep{PlanckCollaboration2011}.

% \subsection{General discussion and comparison to earlier work}
% {\bf We cover similar volume densities as in studies of typical low-mass star-forming regions \citep{Enoch2008}, but select only the highest column densities. The comparable column densities are a clear bias due to the larger distance. If one corrects the peak column densities for the larger beams at these distances using the correction from \citet{Vasyunina2009}, the peak column densities become significantly larger than in studies of low-mass star formation (\citealt{Enoch2008, Chabrier2010} and references therein).}

\section{Conclusion and outlook} 
\label{sec:conclusion_outlook}
\subsection{Conclusion}
\label{sec:conclusion}
For the first time, we have presented a complete and unbiased sample of high-mass starless clumps on 20 deg$^2$ of the sky.
To concentrate on the actual potential precursors of massive stars, we imposed a minimum peak column density of 1 $\times$ 10$^{23}$ cm$^{-2}$. From ATLASGAL, we extracted 901 clumps across the region 10$^{\circ}$ $<$ l $<$ 20$^{\circ}$ of the Galactic plane. Using the GLIMPSE source catalog and MIPSGAL 24 $\mu$m images, we compared clumps found at 870 $\mu$m to near- and mid-infrared tracers of ongoing star formation.

Neglecting clumps that are saturated at MIPSGAL 24 $\mu$m, we identified 210 clumps, or 23\%, to be starless. Their effective radii range from 10\arcsec to 40\arcsec, and most of the beam-averaged peak column densities are in the range 1 - 2 $\times$ 10$^{23}$ cm$^{-2}$. Correcting the single-dish peak column densities to true peak column densities as discussed in \citet{Vasyunina2009} even suggests that all clumps should have peak column densities above the fragmentation threshold for massive star formation of 1 cm$^{-2}$ proposed by \citet{Krumholz2008}.

For $\sim$ 71\% of the starless clumps, we had the velocities and were able to calculate their distance. If a clump is connected to an IRDC, then we assumed a
distance on the near side. We found that about a quarter of the starless clumps lie on the far side of the Galaxy and were therefore previously unknown.

We found that the masses of starless clumps on the near side range from 10 M$_{\odot}$ to 7500 M$_{\odot}$, and that objects on the far side have masses between
1000 M$_{\odot}$ to several 10000 M$_{\odot}$. The different mass regimes are a consequence of our limited spatial resolution, biasing us towards detecting larger structures as a single clump on the far side, which we would resolve into several clumps on the near side.
The mass distributions of the near and far populations could be fitted by power laws with slopes $\alpha$ = -2.2 and $\alpha$ = -2.6, respectively, and agree within the errors. This shows that the mass distribution of clump populations on the near and far side of the Galaxy do not differ significantly.

Drawn from the population of clumps on the near side, we found that only 14 objects are massive enough to form clusters with stars more massive than 20
M$_{\odot}$. Only 3 starless clumps have the potential to form stars even more massive than 40 M$_{\odot}$. We estimate the minimum clump mass
required to form a cluster with a 20 M$_{\odot}$ or 40 M$_{\odot}$ star to be 10$^3$ M$_{\odot}$ or 3 $\times$ 10$^3$ M$_{\odot}$, respectively. Since
the star-formation efficiency used is an upper limit, these numbers are lower limits.

Extrapolating the numbers of massive starless clumps from our survey volume to the Milky Way Galaxy, we estimated the lifetime of the most massive
starless clumps to be on the order of (6 $\pm$ 5) $\times$ 10$^4$ yr. To do so, we assumed a star formation rate of 1 M$_{\odot}$ to 3 M$_{\odot}$ per
year for the entire Galaxy. We also discussed a possibly enhanced star formation activity within the surveyed volume and its implications for the assumed global star formation rate.

\subsection{Outlook}
In the future, we wish to extend our classification of the clumps to more evolved stages. To identify massive young stellar objects in particular, the Red
MSX Source (RMS) survey will be used. Incorporating CORNISH (see Sec. \ref{biases:verification}) and RMS, in addition to GLIMPSE and MIPSGAL, we plan to establish an evolutionary sequence of the
early stages of massive star formation. With this classification, we will be able to derive the relative timescales for the full sequence. Using the
absolute timescale established in Sec. \ref{sec:lifetimeclumps}, we will be able to translate the relative into absolute timescales and compare these to earlier studies. 

In addition, the objects that we have presented herein represent an ideal and unique sample for follow-up studies of the early stages of massive star formation. The upcoming HI-GAL/HERSCHEL survey \citep{Molinari2010} will help to provide additional constraints on the nature of these clumps, but deep integrations with the HERSCHEL 100 $\mu$m band will be needed to uncover all embedded sources. In addition, the high resolution of ALMA will allow detailed studies of star-forming regions on the other side of the Galaxy.

\section{Acknowledgments}
We wish to thank the anonymous referee for her/his careful reading and comments, which substantially enhanced both the science and appearance of the paper.
This publication is partially based on data acquired with the Atacama
Pathfinder Experiment (APEX). APEX is a collaboration between the
Max-Planck-Institut f\"ur Radioastronomie, the European Southern Observatory,
and the Onsala Space Observatory. This work is based, in part, on observations made with the Spitzer Space Telescope, which is operated by the Jet Propulsion Laboratory, California
Institute of Technology under a contract with NASA. This research has made use of the NASA/ IPAC Infrared Science Archive, which is operated by the Jet Propulsion Laboratory, California Institute of Technology, under contract with the National Aeronautics and Space Administration.
L. B. acknowledges support from CONICYT projects  FONDAP 15010003 and Basal PFB-06.
J. T. is supported by the International Max Planck Research School (IMPRS) for Astronomy and Cosmic Physics.
%%
% \section{Things to do}
% - nochmal besser EPOS diskutieren
% - einzelne clumps eine association zuordnen (z.B. M17 oder W31 o. ae.)
% - Planck ECC discuss Fig 10, 11 
% - explain filtering issue
% - conclusion
%%
\begin{landscape} 
  \centering
  %% \onecolumn
  \begin{table}[tbp]
  \caption{Properties of starless clumps as described in the paper. The full table is available in the online version of this paper. }
  \begin{tabular}{ p{.7cm} *{4}{l} *{3}{p{.7cm}} p{1.3cm} p{.6cm} p{1.3cm} p{.6cm} *{2}{p{.9cm}} p{1.2cm} *{2}{p{1.1cm}} *{1}{p{.6cm}}}
    Global Identifier & Gal Lon & Gal Lat & Ra & Dec & Peak Flux & R & Total Flux & NH$_3$ Velocity & NH$_3$ Flag & HCO$^+$ Velocity & HCO$^+$ Flag & Near Distance & Far Distance & Peak Column Density & Near Mass & Far Mass & Near Flag\\
    & [ $^{\circ}$ ] & [ $^{\circ}$ ] & [ $^{\circ}$ ] & [ $^{\circ}$ ] & [ Jy ] & [ \arcsec ] & [ Jy ] & [ km/s ] & & [ km/s ] & & [ kpc ] & [ kpc ] & [ 10$^{23}$ cm$^{-2}$ ] & [ M$_{\odot}$ ] & [ M$_{\odot}$ ] & \\
    75&         10.6075&         -0.3708&        272.5993&        -19.9382&            0.92&              36&           10.43&            -2.9&   0&             ---&   0&            -0.5&            17.0&            1.93&             ---&          62000.&     1\\
    76&         10.5958&         -0.3642&        272.5871&        -19.9452&            0.92&              31&            7.23&            -2.9&   0&             ---&   0&            -0.5&            17.0&            1.93&             ---&          43000.&     1\\
    80&         10.6858&         -0.3075&        272.5805&        -19.8390&            0.86&              25&            3.72&            -1.5&   1&            -1.7&   1&            -0.2&            16.7&            1.81&             ---&          21000.&     1\\
    82&         10.1842&         -0.4042&        272.4125&        -20.3249&            0.85&              21&            2.78&            10.5&   0&             9.4&   0&             1.8&            14.7&            1.79&            180.&          13000.&     1\\
    83&         10.6208&         -0.4225&        272.6543&        -19.9515&            0.83&              28&            5.64&            -2.0&   1&             ---&   0&            -0.3&            16.8&            1.74&             ---&          33000.&     1\\
    86&         10.1658&         -0.3342&        272.3378&        -20.3071&            0.81&              18&            2.18&            10.5&   0&             9.4&   0&             1.8&            14.7&            1.70&            140.&           9800.&     1\\
    87&         10.1592&         -0.3008&        272.3032&        -20.2968&            0.81&              18&            2.08&            10.5&   0&             9.4&   0&             1.8&            14.7&            1.70&            140.&           9400.&     1\\
    94&         10.6225&         -0.5092&        272.7359&        -19.9918&            0.78&              22&            3.44&            -4.1&   0&            -3.6&   0&            -0.7&            17.2&            1.64&             ---&          21000.&     1\\
    95&         10.9825&         -0.3692&        272.7899&        -19.6089&            0.78&              23&            3.33&             ---&   0&            -0.6&   1&             ---&            16.5&            1.64&             ---&          19000.&     0\\
    101&         10.1375&         -0.3575&        272.3449&        -20.3432&            0.75&              17&            2.03&            10.5&   0&             9.4&   0&             1.8&            14.7&            1.58&            130.&           9100.&     1\\
    105&         10.0675&         -0.4075&        272.3554&        -20.4286&            0.75&              20&            2.43&             ---&   0&            11.4&   1&             1.9&            14.6&            1.58&            180.&          11000.&     1\\
    106&         10.1775&         -0.4025&        272.4075&        -20.3300&            0.74&              13&            1.11&            10.5&   0&             9.4&   0&             1.8&            14.7&            1.56&             73.&           5000.&     1\\
    111&         10.1375&         -0.4108&        272.3947&        -20.3690&            0.73&              17&            1.79&            12.9&   0&             ---&   0&             2.1&            14.5&            1.53&            160.&           7700.&     1\\
    114&         10.1325&         -0.4108&        272.3921&        -20.3734&            0.72&              15&            1.41&            12.9&   0&             ---&   0&             2.1&            14.5&            1.51&            130.&           6100.&     1\\
    118&         10.1342&         -0.3475&        272.3339&        -20.3413&            0.71&              22&            3.08&            10.5&   0&             9.4&   0&             1.8&            14.7&            1.49&            200.&          14000.&     1\\
    119&         10.5758&         -0.3475&        272.5613&        -19.9547&            0.71&              17&            1.85&            -2.9&   0&            -2.6&   1&            -0.5&            17.0&            1.49&             ---&          11000.&     1\\
    121&         10.2992&         -0.1658&        272.2496&        -20.1089&            0.71&              26&            4.79&            12.8&   0&            13.6&   0&             2.1&            14.5&            1.49&            420.&          21000.&     1\\
    122&         11.0575&         -0.0958&        272.5743&        -19.4114&            0.70&              16&            1.72&            29.8&   0&            29.0&   0&             3.5&            13.0&            1.47&            420.&           6000.&     1\\
    123&         10.5775&         -0.3508&        272.5653&        -19.9548&            0.70&              19&            2.13&            -2.9&   0&            -2.6&   1&            -0.5&            17.0&            1.47&             ---&          13000.&     1\\
    140&         10.6625&          0.0825&        272.2059&        -19.6708&            0.64&              21&            2.44&             ---&   0&            21.1&   1&             2.8&            13.7&            1.35&            410.&           9400.&     0\\
    143&         10.1958&         -0.2892&        272.3112&        -20.2591&            0.62&              22&            2.64&            10.5&   0&             9.4&   0&             1.8&            14.8&            1.30&            170.&          12000.&     1\\
    155&         10.6325&         -0.4225&        272.6603&        -19.9412&            0.59&              18&            1.60&            -2.9&   0&             ---&   0&            -0.5&            17.0&            1.24&             ---&           9500.&     1\\
    156&         10.2475&         -0.3358&        272.3814&        -20.2365&            0.59&              11&            0.72&             ---&   0&            11.4&   1&             1.9&            14.7&            1.24&             53.&           3200.&     0\\
    161&         10.2542&         -0.3392&        272.3880&        -20.2322&            0.59&              12&            0.86&             ---&   0&            11.4&   1&             1.9&            14.7&            1.24&             63.&           3800.&     1\\
    162&         11.9139&          0.7356&        272.2431&        -18.2597&            0.59&              21&            2.25&            24.0&   0&             ---&   0&             2.9&            13.5&            1.24&            390.&           8500.&     1\\
    167&         11.0541&         -0.0792&        272.5571&        -19.4062&            0.58&              19&            2.05&            29.8&   0&            29.0&   0&             3.5&            13.0&            1.22&            510.&           7200.&     1\\
    168&         11.9005&          0.7206&        272.2501&        -18.2786&            0.58&              19&            1.94&             ---&   0&             ---&   0&             ---&             ---&            1.22&             ---&             ---&     1\\
    169&         11.3041&         -0.0608&        272.6679&        -19.1785&            0.58&              14&            1.17&             ---&   0&            31.6&   1&             3.5&            12.9&            1.22&            300.&           4000.&     1\\
    170&         10.5742&         -0.7891&        272.9724&        -20.1689&            0.57&              25&            3.18&             ---&   0&             ---&   0&             ---&             ---&            1.20&             ---&             ---&     1\\
    173&         10.2542&         -0.1225&        272.1860&        -20.1273&            0.57&              23&            2.94&            12.8&   0&            13.6&   0&             2.1&            14.5&            1.20&            260.&          13000.&     1\\
    179&         10.6192&         -0.0325&        272.2904&        -19.7644&            0.56&              14&            1.16&             ---&   0&            64.0&   1&             5.2&            11.3&            1.18&            650.&           3100.&     0\\
    181&         10.3475&         -0.1825&        272.2901&        -20.0747&            0.55&              20&            2.05&            12.8&   0&            13.6&   0&             2.0&            14.5&            1.16&            180.&           8900.&     1\\
    184&         10.0275&         -0.3525&        272.2834&        -20.4370&            0.55&              16&            1.35&             ---&   0&             ---&   1&             ---&             ---&            1.16&             ---&             ---&     0\\
    185&         10.6858&         -0.2125&        272.4921&        -19.7931&            0.55&              11&            0.69&            29.0&   0&            29.5&   0&             3.5&            13.0&            1.16&            170.&           2400.&     1\\
    188&         10.3342&         -0.1792&        272.2801&        -20.0848&            0.54&              23&            2.95&            12.8&   0&            13.6&   0&             2.0&            14.5&            1.14&            260.&          13000.&     1\\
    189&         11.3508&          0.7957&        271.8995&        -18.7228&            0.54&              19&            1.82&             ---&   0&             ---&   0&             ---&             ---&            1.14&             ---&             ---&     1\\
    191&         11.9139&          0.7189&        272.2584&        -18.2678&            0.54&              11&            0.63&             ---&   0&             ---&   0&             ---&             ---&            1.14&             ---&             ---&     1\\
    193&         11.0575&         -0.0875&        272.5665&        -19.4074&            0.54&              13&            1.00&            29.8&   0&            29.0&   0&             3.5&            13.0&            1.14&            250.&           3500.&     1\\
%%    197&         10.2375&         -0.1175&        272.1728&        -20.1395&            0.54&              16&            1.32&            12.8&   0&            13.6&   0&             2.1&            14.5&            1.14&            120.&           5700.&     1\\
%%    206&         10.1675&         -0.4042&        272.4039&        -20.3395&            0.52&              10&            0.56&            12.9&   0&             ---&   0&             2.1&            14.5&            1.09&             50.&           2400.&     1\\
%%    208&         10.1825&         -0.2925&        272.3075&        -20.2724&            0.52&              14&            0.99&            10.5&   0&             9.4&   0&             1.8&            14.7&            1.09&             65.&           4400.&     1\\
  \end{tabular}
  \footnotesize{Notes: Columns are identifier, galactic longitude, galactic latitude, right ascension, declination, peak flux, radius as calculated by
    clumpfind, integrated flux, NH$_3$ velocity from Wienen et al. (submitted), flag indicating presence of direct NH$_3$ observation, HCO$^+$
    velocity from \citet{Schlingman2011}, flag indicating presence of direct HCO$^+$ observation, calculated near distance, calculated far distance,
    peak column density, mass calculated for the near distance, mass calculated for the far distance, flag indicating the connection to an
    IRDC. Negative near distances are meaningless; the velocities of those sources place them in the outer Galaxy, while we are looking towards the inner Galaxy. However, their galactic latitude and velocities are also consistent with the near 3 kpc arm at a distance of 5.2 kpc \citep{Dame2008}. The NH$_3$ flag means that Wienen et al. (submitted) have observed a position within the clump's boundary definition. If the NH$_3$ flag is absent but a NH$_3$ velocity is given, the velocity is derived from neighboring clumps (see Sec. \ref{sec:dist_col_mass} for details). The same yields for the HCO$^+$ flag and HCO$^+$ observations in \citet{Schlingman2011}.}
  \label{tab:list}
\end{table}

%%  \input{./HMSC_table_latex}
  %% \twocolumn
\end{landscape}

\bibliographystyle{aa}
\bibliography{lib}

\begin{thebibliography}{84}
\expandafter\ifx\csname natexlab\endcsname\relax\def\natexlab#1{#1}\fi

\bibitem[{{Aguirre} {et~al.}(2011){Aguirre}, {Ginsburg}, {Dunham}, {Drosback},
  {Bally}, {Battersby}, {Bradley}, {Cyganowski}, {Dowell}, {Evans}, {Glenn},
  {Harvey}, {Rosolowsky}, {Stringfellow}, {Walawender}, \&
  {Williams}}]{Aguirre2011}
{Aguirre}, J.~E., {Ginsburg}, A.~G., {Dunham}, M.~K., {et~al.} 2011, \apjs,
  192, 4

\bibitem[{{Alves} {et~al.}(2007){Alves}, {Lombardi}, \& {Lada}}]{Alves2007}
{Alves}, J., {Lombardi}, M., \& {Lada}, C.~J. 2007, \aap, 462, L17

\bibitem[{{Bania} {et~al.}(2010){Bania}, {Anderson}, {Balser}, \&
  {Rood}}]{Bania2010}
{Bania}, T.~M., {Anderson}, L.~D., {Balser}, D.~S., \& {Rood}, R.~T. 2010,
  \apjl, 718, L106

\bibitem[{{Beltr{\'a}n} {et~al.}(2006){Beltr{\'a}n}, {Brand}, {Cesaroni},
  {Fontani}, {Pezzuto}, {Testi}, \& {Molinari}}]{Beltran2006}
{Beltr{\'a}n}, M.~T., {Brand}, J., {Cesaroni}, R., {et~al.} 2006, \aap, 447,
  221

\bibitem[{{Benjamin} {et~al.}(2003){Benjamin}, {Churchwell}, {Babler}, {Bania},
  {Clemens}, {Cohen}, {Dickey}, {Indebetouw}, {Jackson}, {Kobulnicky},
  {Lazarian}, {Marston}, {Mathis}, {Meade}, {Seager}, {Stolovy}, {Watson},
  {Whitney}, {Wolff}, \& {Wolfire}}]{Benjamin2003}
{Benjamin}, R.~A., {Churchwell}, E., {Babler}, B.~L., {et~al.} 2003, \pasp,
  115, 953

\bibitem[{{Beuther} {et~al.}(2007){Beuther}, {Churchwell}, {McKee}, \&
  {Tan}}]{Beuther2007}
{Beuther}, H., {Churchwell}, E.~B., {McKee}, C.~F., \& {Tan}, J.~C. 2007,
  Protostars and Planets V, 165

\bibitem[{{Beuther} {et~al.}(2010){Beuther}, {Henning}, {Linz}, {Krause},
  {Nielbock}, \& {Steinacker}}]{Beuther2010}
{Beuther}, H., {Henning}, T., {Linz}, H., {et~al.} 2010, \aap, 518, L78+

\bibitem[{{Beuther} \& {Sridharan}(2007)}]{Beuther2007a}
{Beuther}, H. \& {Sridharan}, T.~K. 2007, \apj, 668, 348

\bibitem[{{Beuther} \& {Steinacker}(2007)}]{Beuther2007b}
{Beuther}, H. \& {Steinacker}, J. 2007, \apjl, 656, L85

\bibitem[{{Bonnell} \& {Bate}(2005)}]{bonnell2005}
{Bonnell}, I.~A. \& {Bate}, M.~R. 2005, \mnras, 362, 915

\bibitem[{{Bontemps} {et~al.}(2010{\natexlab{a}}){Bontemps}, {Andr{\'e}},
  {K{\"o}nyves}, {Men'shchikov}, {Schneider}, {Maury}, {Peretto},
  {Arzoumanian}, {Attard}, {Motte}, {Minier}, {Didelon}, {Saraceno}, {Abergel},
  {Baluteau}, {Bernard}, {Cambr{\'e}sy}, {Cox}, {di Francesco}, {di Giorgo},
  {Griffin}, {Hargrave}, {Huang}, {Kirk}, {Li}, {Martin}, {Mer{\'{\i}}n},
  {Molinari}, {Olofsson}, {Pezzuto}, {Prusti}, {Roussel}, {Russeil}, {Sauvage},
  {Sibthorpe}, {Spinoglio}, {Testi}, {Vavrek}, {Ward-Thompson}, {White},
  {Wilson}, {Woodcraft}, \& {Zavagno}}]{Bontemps2010a}
{Bontemps}, S., {Andr{\'e}}, P., {K{\"o}nyves}, V., {et~al.}
  2010{\natexlab{a}}, \aap, 518, L85

\bibitem[{{Bontemps} {et~al.}(2010{\natexlab{b}}){Bontemps}, {Motte},
  {Csengeri}, \& {Schneider}}]{Bontemps2010}
{Bontemps}, S., {Motte}, F., {Csengeri}, T., \& {Schneider}, N.
  2010{\natexlab{b}}, \aap, 524, A18+

\bibitem[{{Bronfman} {et~al.}(2000){Bronfman}, {Casassus}, {May}, \&
  {Nyman}}]{Bronfman2000}
{Bronfman}, L., {Casassus}, S., {May}, J., \& {Nyman}, L. 2000, \aap, 358, 521

\bibitem[{{Carey} {et~al.}(1998){Carey}, {Clark}, {Egan}, {Price}, {Shipman},
  \& {Kuchar}}]{Carey1998}
{Carey}, S.~J., {Clark}, F.~O., {Egan}, M.~P., {et~al.} 1998, \apj, 508, 721

\bibitem[{{Carey} {et~al.}(2009){Carey}, {Noriega-Crespo}, {Mizuno}, {Shenoy},
  {Paladini}, {Kraemer}, {Price}, {Flagey}, {Ryan}, {Ingalls}, {Kuchar},
  {Pinheiro Gon{\c c}alves}, {Indebetouw}, {Billot}, {Marleau}, {Padgett},
  {Rebull}, {Bressert}, {Ali}, {Molinari}, {Martin}, {Berriman}, {Boulanger},
  {Latter}, {Miville-Deschenes}, {Shipman}, \& {Testi}}]{Carey2009}
{Carey}, S.~J., {Noriega-Crespo}, A., {Mizuno}, D.~R., {et~al.} 2009, \pasp,
  121, 76

\bibitem[{{Chabrier} \& {Hennebelle}(2010)}]{Chabrier2010}
{Chabrier}, G. \& {Hennebelle}, P. 2010, \apjl, 725, L79

\bibitem[{{Commer{\c c}on} {et~al.}(2011){Commer{\c c}on}, {Hennebelle}, \&
  {Henning}}]{commercon2011}
{Commer{\c c}on}, B., {Hennebelle}, P., \& {Henning}, T. 2011, \apjl, 742, L9

\bibitem[{{Cyganowski} {et~al.}(2008){Cyganowski}, {Whitney}, {Holden},
  {Braden}, {Brogan}, {Churchwell}, {Indebetouw}, {Watson}, {Babler},
  {Benjamin}, {Gomez}, {Meade}, {Povich}, {Robitaille}, \&
  {Watson}}]{Cyganowski2008}
{Cyganowski}, C.~J., {Whitney}, B.~A., {Holden}, E., {et~al.} 2008, \aj, 136,
  2391

\bibitem[{{Dame} {et~al.}(2001){Dame}, {Hartmann}, \& {Thaddeus}}]{Dame2001}
{Dame}, T.~M., {Hartmann}, D., \& {Thaddeus}, P. 2001, \apj, 547, 792

\bibitem[{{Dame} \& {Thaddeus}(2008)}]{Dame2008}
{Dame}, T.~M. \& {Thaddeus}, P. 2008, \apjl, 683, L143

\bibitem[{{Diehl} {et~al.}(2006){Diehl}, {Halloin}, {Kretschmer}, {Lichti},
  {Sch{\"o}nfelder}, {Strong}, {von Kienlin}, {Wang}, {Jean}, {Kn{\"o}dlseder},
  {Roques}, {Weidenspointner}, {Schanne}, {Hartmann}, {Winkler}, \&
  {Wunderer}}]{Diehl2006}
{Diehl}, R., {Halloin}, H., {Kretschmer}, K., {et~al.} 2006, \nat, 439, 45

\bibitem[{{Diolaiti} {et~al.}(2000){Diolaiti}, {Bendinelli}, {Bonaccini},
  {Close}, {Currie}, \& {Parmeggiani}}]{Diolaiti2000}
{Diolaiti}, E., {Bendinelli}, O., {Bonaccini}, D., {et~al.} 2000, \aaps, 147,
  335

\bibitem[{{Draine} \& {Lee}(1984)}]{Draine1984}
{Draine}, B.~T. \& {Lee}, H.~M. 1984, \apj, 285, 89

\bibitem[{{Dunham} {et~al.}(2011){Dunham}, {Robitaille}, {Evans}, {Schlingman},
  {Cyganowski}, \& {Urquhart}}]{Dunham2011}
{Dunham}, M.~K., {Robitaille}, T.~P., {Evans}, II, N.~J., {et~al.} 2011, \apj,
  731, 90

\bibitem[{{Enoch} {et~al.}(2008){Enoch}, {Evans}, {Sargent}, {Glenn},
  {Rosolowsky}, \& {Myers}}]{Enoch2008}
{Enoch}, M.~L., {Evans}, II, N.~J., {Sargent}, A.~I., {et~al.} 2008, \apj, 684,
  1240

\bibitem[{{Foster} {et~al.}(2011){Foster}, {Jackson}, {Barris}, {Brooks},
  {Cunningham}, {Finn}, {Fuller}, {Longmore}, {Mascoop}, {Peretto},
  {Rathborne}, {Sanhueza}, {Schuller}, \& {Wyrowski}}]{Foster2011}
{Foster}, J.~B., {Jackson}, J.~M., {Barris}, E., {et~al.} 2011, ArXiv e-prints

\bibitem[{{Green} {et~al.}(2011){Green}, {Caswell}, {McClure-Griffiths},
  {Avison}, {Breen}, {Burton}, {Ellingsen}, {Fuller}, {Gray}, {Pestalozzi},
  {Thompson}, \& {Voronkov}}]{Green2011}
{Green}, J.~A., {Caswell}, J.~L., {McClure-Griffiths}, N.~M., {et~al.} 2011,
  \apj, 733, 27

\bibitem[{{Gutermuth} {et~al.}(2008){Gutermuth}, {Myers}, {Megeath}, {Allen},
  {Pipher}, {Muzerolle}, {Porras}, {Winston}, \& {Fazio}}]{Gutermuth2008}
{Gutermuth}, R.~A., {Myers}, P.~C., {Megeath}, S.~T., {et~al.} 2008, \apj, 674,
  336

\bibitem[{{Hatchell} \& {Fuller}(2008)}]{Hatchell2008}
{Hatchell}, J. \& {Fuller}, G.~A. 2008, \aap, 482, 855

\bibitem[{{Henning} {et~al.}(2010){Henning}, {Linz}, {Krause}, {Ragan},
  {Beuther}, {Launhardt}, {Nielbock}, \& {Vasyunina}}]{Henning2010}
{Henning}, T., {Linz}, H., {Krause}, O., {et~al.} 2010, \aap, 518, L95+

\bibitem[{{Hildebrand}(1983)}]{Hildebrand1983}
{Hildebrand}, R.~H. 1983, \qjras, 24, 267

\bibitem[{{Hillenbrand} \& {Hartmann}(1998)}]{Hillenbrand1998}
{Hillenbrand}, L.~A. \& {Hartmann}, L.~W. 1998, \apj, 492, 540

\bibitem[{{Hoare} {et~al.}(2004){Hoare}, {Lumsden}, {Oudmaijer}, {Busfield},
  {King}, \& {Moore}}]{Hoare2004}
{Hoare}, M.~G., {Lumsden}, S.~L., {Oudmaijer}, R.~D., {et~al.} 2004, in
  Astronomical Society of the Pacific Conference Series, Vol. 317, Milky Way
  Surveys: The Structure and Evolution of our Galaxy, ed. {D.~Clemens, R.~Shah,
  \& T.~Brainerd}, 156--+

\bibitem[{{Johnston} {et~al.}(2009){Johnston}, {Shepherd}, {Aguirre}, {Dunham},
  {Rosolowsky}, \& {Wood}}]{Johnston2009}
{Johnston}, K.~G., {Shepherd}, D.~S., {Aguirre}, J.~E., {et~al.} 2009, \apj,
  707, 283

\bibitem[{{Johnstone} {et~al.}(2000){Johnstone}, {Wilson}, {Moriarty-Schieven},
  {Joncas}, {Smith}, {Gregersen}, \& {Fich}}]{Johnstone2000}
{Johnstone}, D., {Wilson}, C.~D., {Moriarty-Schieven}, G., {et~al.} 2000, \apj,
  545, 327

\bibitem[{{Kainulainen} {et~al.}(2011){Kainulainen}, {Alves}, {Beuther},
  {Henning}, \& {Schuller}}]{Kainulainen2011}
{Kainulainen}, J., {Alves}, J., {Beuther}, H., {Henning}, T., \& {Schuller}, F.
  2011, ArXiv e-prints

\bibitem[{{Kainulainen} {et~al.}(2009){Kainulainen}, {Lada}, {Rathborne}, \&
  {Alves}}]{Kainulainen2009}
{Kainulainen}, J., {Lada}, C.~J., {Rathborne}, J.~M., \& {Alves}, J.~F. 2009,
  \aap, 497, 399

\bibitem[{{Keto}(2003)}]{keto2003}
{Keto}, E. 2003, \apj, 599, 1196

\bibitem[{{Kirk} {et~al.}(2005){Kirk}, {Ward-Thompson}, \&
  {Andr{\'e}}}]{Kirk2005}
{Kirk}, J.~M., {Ward-Thompson}, D., \& {Andr{\'e}}, P. 2005, \mnras, 360, 1506

\bibitem[{{Kroupa}(2001)}]{Kroupa2001}
{Kroupa}, P. 2001, \mnras, 322, 231

\bibitem[{{Krumholz} {et~al.}(2007){Krumholz}, {Klein}, \&
  {McKee}}]{Krumholz2007}
{Krumholz}, M.~R., {Klein}, R.~I., \& {McKee}, C.~F. 2007, \apj, 656, 959

\bibitem[{{Krumholz} \& {McKee}(2008)}]{Krumholz2008}
{Krumholz}, M.~R. \& {McKee}, C.~F. 2008, \nat, 451, 1082

\bibitem[{{Kuiper} {et~al.}(2010){Kuiper}, {Klahr}, {Beuther}, \&
  {Henning}}]{Kuiper2010}
{Kuiper}, R., {Klahr}, H., {Beuther}, H., \& {Henning}, T. 2010, \apj, 722,
  1556

\bibitem[{{Lada} \& {Lada}(2003)}]{Lada2003}
{Lada}, C.~J. \& {Lada}, E.~A. 2003, \araa, 41, 57

\bibitem[{{Liszt} {et~al.}(1981){Liszt}, {Burton}, \& {Bania}}]{Liszt1981}
{Liszt}, H.~S., {Burton}, W.~B., \& {Bania}, T.~M. 1981, \apj, 246, 74

\bibitem[{{McKee} \& {Tan}(2003)}]{mckee2003}
{McKee}, C.~F. \& {Tan}, J.~C. 2003, \apj, 585, 850

\bibitem[{{Molinari} {et~al.}(2010){Molinari}, {Swinyard}, {Bally}, {Barlow},
  {Bernard}, {Martin}, {Moore}, {Noriega-Crespo}, {Plume}, {Testi}, {Zavagno},
  {Abergel}, {Ali}, {Andr{\'e}}, {Baluteau}, {Benedettini}, {Bern{\'e}},
  {Billot}, {Blommaert}, {Bontemps}, {Boulanger}, {Brand}, {Brunt}, {Burton},
  {Campeggio}, {Carey}, {Caselli}, {Cesaroni}, {Cernicharo}, {Chakrabarti},
  {Chrysostomou}, {Codella}, {Cohen}, {Compiegne}, {Davis}, {de Bernardis}, {de
  Gasperis}, {Di Francesco}, {di Giorgio}, {Elia}, {Faustini}, {Fischera},
  {Fukui}, {Fuller}, {Ganga}, {Garcia-Lario}, {Giard}, {Giardino}, {Glenn},
  {Goldsmith}, {Griffin}, {Hoare}, {Huang}, {Jiang}, {Joblin}, {Joncas},
  {Juvela}, {Kirk}, {Lagache}, {Li}, {Lim}, {Lord}, {Lucas}, {Maiolo},
  {Marengo}, {Marshall}, {Masi}, {Massi}, {Matsuura}, {Meny}, {Minier},
  {Miville-Desch{\^e}nes}, {Montier}, {Motte}, {M{\"u}ller}, {Natoli}, {Neves},
  {Olmi}, {Paladini}, {Paradis}, {Pestalozzi}, {Pezzuto}, {Piacentini},
  {Pomar{\`e}s}, {Popescu}, {Reach}, {Richer}, {Ristorcelli}, {Roy}, {Royer},
  {Russeil}, {Saraceno}, {Sauvage}, {Schilke}, {Schneider-Bontemps},
  {Schuller}, {Schultz}, {Shepherd}, {Sibthorpe}, {Smith}, {Smith},
  {Spinoglio}, {Stamatellos}, {Strafella}, {Stringfellow}, {Sturm}, {Taylor},
  {Thompson}, {Tuffs}, {Umana}, {Valenziano}, {Vavrek}, {Viti}, {Waelkens},
  {Ward-Thompson}, {White}, {Wyrowski}, {Yorke}, \& {Zhang}}]{Molinari2010}
{Molinari}, S., {Swinyard}, B., {Bally}, J., {et~al.} 2010, \pasp, 122, 314

\bibitem[{{Motte} {et~al.}(1998){Motte}, {Andre}, \& {Neri}}]{Motte1998}
{Motte}, F., {Andre}, P., \& {Neri}, R. 1998, \aap, 336, 150

\bibitem[{{Motte} {et~al.}(2007){Motte}, {Bontemps}, {Schilke}, {Schneider},
  {Menten}, \& {Brogui{\`e}re}}]{Motte2007}
{Motte}, F., {Bontemps}, S., {Schilke}, P., {et~al.} 2007, \aap, 476, 1243

\bibitem[{{Motte} {et~al.}(2010){Motte}, {Zavagno}, {Bontemps}, {Schneider},
  {Hennemann}, {di Francesco}, {Andr{\'e}}, {Saraceno}, {Griffin}, {Marston},
  {Ward-Thompson}, {White}, {Minier}, {Men'shchikov}, {Hill}, {Abergel},
  {Anderson}, {Aussel}, {Balog}, {Baluteau}, {Bernard}, {Cox}, {Csengeri},
  {Deharveng}, {Didelon}, {di Giorgio}, {Hargrave}, {Huang}, {Kirk}, {Leeks},
  {Li}, {Martin}, {Molinari}, {Nguyen-Luong}, {Olofsson}, {Persi}, {Peretto},
  {Pezzuto}, {Roussel}, {Russeil}, {Sadavoy}, {Sauvage}, {Sibthorpe},
  {Spinoglio}, {Testi}, {Teyssier}, {Vavrek}, {Wilson}, \&
  {Woodcraft}}]{Motte2010}
{Motte}, F., {Zavagno}, A., {Bontemps}, S., {et~al.} 2010, \aap, 518, L77+

\bibitem[{{Mottram} {et~al.}(2011){Mottram}, {Hoare}, {Urquhart}, {Lumsden},
  {Oudmaijer}, {Robitaille}, {Moore}, {Davies}, \& {Stead}}]{Mottram2011}
{Mottram}, J.~C., {Hoare}, M.~G., {Urquhart}, J.~S., {et~al.} 2011, \aap, 525,
  A149+

\bibitem[{{Ossenkopf} \& {Henning}(1994)}]{Ossenkopf1994}
{Ossenkopf}, V. \& {Henning}, T. 1994, \aap, 291, 943

\bibitem[{{Perault} {et~al.}(1996){Perault}, {Omont}, {Simon}, {Seguin},
  {Ojha}, {Blommaert}, {Felli}, {Gilmore}, {Guglielmo}, {Habing}, {Price},
  {Robin}, {de Batz}, {Cesarsky}, {Elbaz}, {Epchtein}, {Fouque}, {Guest},
  {Levine}, {Pollock}, {Prusti}, {Siebenmorgen}, {Testi}, \&
  {Tiphene}}]{Perault1996}
{Perault}, M., {Omont}, A., {Simon}, G., {et~al.} 1996, \aap, 315, L165

\bibitem[{{Peretto} \& {Fuller}(2009)}]{Peretto2009}
{Peretto}, N. \& {Fuller}, G.~A. 2009, \aap, 505, 405

\bibitem[{{Peretto} \& {Fuller}(2010)}]{Peretto2010}
{Peretto}, N. \& {Fuller}, G.~A. 2010, \apj, 723, 555

\bibitem[{{Peretto} {et~al.}(2010){Peretto}, {Fuller}, {Plume}, {Anderson},
  {Bally}, {Battersby}, {Beltran}, {Bernard}, {Calzoletti}, {Digiorgio},
  {Faustini}, {Kirk}, {Lenfestey}, {Marshall}, {Martin}, {Molinari}, {Montier},
  {Motte}, {Ristorcelli}, {Rod{\'o}n}, {Smith}, {Traficante}, {Veneziani},
  {Ward-Thompson}, \& {Wilcock}}]{Peretto2010a}
{Peretto}, N., {Fuller}, G.~A., {Plume}, R., {et~al.} 2010, \aap, 518, L98+

\bibitem[{{Pillai} {et~al.}(2006){Pillai}, {Wyrowski}, {Carey}, \&
  {Menten}}]{Pillai2006}
{Pillai}, T., {Wyrowski}, F., {Carey}, S.~J., \& {Menten}, K.~M. 2006, \aap,
  450, 569

\bibitem[{{Pineda} {et~al.}(2009){Pineda}, {Rosolowsky}, \&
  {Goodman}}]{Pineda2009}
{Pineda}, J.~E., {Rosolowsky}, E.~W., \& {Goodman}, A.~A. 2009, \apjl, 699,
  L134

\bibitem[{{Planck Collaboration} {et~al.}(2011){Planck Collaboration}, {Ade},
  {Aghanim}, {Arnaud}, {Ashdown}, {Aumont}, {Baccigalupi}, {Balbi}, {Banday},
  {Barreiro}, \& et~al.}]{Planck2011}
{Planck Collaboration}, {Ade}, P.~A.~R., {Aghanim}, N., {et~al.} 2011, ArXiv
  e-prints

\bibitem[{{Purcell} \& {Hoare}(2010)}]{Purcell2010}
{Purcell}, C.~R. \& {Hoare}, M.~G. 2010, Highlights of Astronomy, 15, 781

\bibitem[{{Rathborne} {et~al.}(2006){Rathborne}, {Jackson}, \&
  {Simon}}]{Rathborne2006}
{Rathborne}, J.~M., {Jackson}, J.~M., \& {Simon}, R. 2006, \apj, 641, 389

\bibitem[{{Reid} {et~al.}(2010){Reid}, {Wadsley}, {Petitclerc}, \&
  {Sills}}]{Reid2010}
{Reid}, M.~A., {Wadsley}, J., {Petitclerc}, N., \& {Sills}, A. 2010, \apj, 719,
  561

\bibitem[{{Reid} \& {Wilson}(2005)}]{Reid2005}
{Reid}, M.~A. \& {Wilson}, C.~D. 2005, \apj, 625, 891

\bibitem[{{Reid} {et~al.}(2009){Reid}, {Menten}, {Zheng}, {Brunthaler},
  {Moscadelli}, {Xu}, {Zhang}, {Sato}, {Honma}, {Hirota}, {Hachisuka}, {Choi},
  {Moellenbrock}, \& {Bartkiewicz}}]{Reid2009}
{Reid}, M.~J., {Menten}, K.~M., {Zheng}, X.~W., {et~al.} 2009, \apj, 700, 137

\bibitem[{{Robitaille} {et~al.}(2008){Robitaille}, {Meade}, {Babler},
  {Whitney}, {Johnston}, {Indebetouw}, {Cohen}, {Povich}, {Sewilo}, {Benjamin},
  \& {Churchwell}}]{Robitaille2008}
{Robitaille}, T.~P., {Meade}, M.~R., {Babler}, B.~L., {et~al.} 2008, \aj, 136,
  2413

\bibitem[{{Robitaille} \& {Whitney}(2010)}]{Robitaille2010}
{Robitaille}, T.~P. \& {Whitney}, B.~A. 2010, \apjl, 710, L11

\bibitem[{{Rosolowsky} {et~al.}(2010){Rosolowsky}, {Dunham}, {Ginsburg},
  {Bradley}, {Aguirre}, {Bally}, {Battersby}, {Cyganowski}, {Dowell},
  {Drosback}, {Evans}, {Glenn}, {Harvey}, {Stringfellow}, {Walawender}, \&
  {Williams}}]{Rosolowsky2010}
{Rosolowsky}, E., {Dunham}, M.~K., {Ginsburg}, A., {et~al.} 2010, \apjs, 188,
  123

\bibitem[{{Russeil} {et~al.}(2010){Russeil}, {Zavagno}, {Motte}, {Schneider},
  {Bontemps}, \& {Walsh}}]{Russeil2010}
{Russeil}, D., {Zavagno}, A., {Motte}, F., {et~al.} 2010, \aap, 515, A55+

\bibitem[{{Salpeter}(1955)}]{Salpeter1955}
{Salpeter}, E.~E. 1955, \apj, 121, 161

\bibitem[{{Schlingman} {et~al.}(2011){Schlingman}, {Shirley}, {Schenk},
  {Rosolowsky}, {Bally}, {Battersby}, {Dunham}, {Ellsworth-Bowers}, {Evans},
  {Ginsburg}, \& {Stringfellow}}]{Schlingman2011}
{Schlingman}, W.~M., {Shirley}, Y.~L., {Schenk}, D.~E., {et~al.} 2011, \apjs,
  195, 14

\bibitem[{{Schuller} {et~al.}(2009){Schuller}, {Menten}, {Contreras},
  {Wyrowski}, {Schilke}, {Bronfman}, {Henning}, {Walmsley}, {Beuther},
  {Bontemps}, {Cesaroni}, {Deharveng}, {Garay}, {Herpin}, {Lefloch}, {Linz},
  {Mardones}, {Minier}, {Molinari}, {Motte}, {Nyman}, {Reveret}, {Risacher},
  {Russeil}, {Schneider}, {Testi}, {Troost}, {Vasyunina}, {Wienen}, {Zavagno},
  {Kovacs}, {Kreysa}, {Siringo}, \& {Wei{\ss}}}]{Schuller2009}
{Schuller}, F., {Menten}, K.~M., {Contreras}, Y., {et~al.} 2009, \aap, 504, 415

\bibitem[{{Siess} {et~al.}(2000){Siess}, {Dufour}, \& {Forestini}}]{Siess2000}
{Siess}, L., {Dufour}, E., \& {Forestini}, M. 2000, \aap, 358, 593

\bibitem[{{Simon} {et~al.}(2006){Simon}, {Jackson}, {Rathborne}, \&
  {Chambers}}]{Simon2006}
{Simon}, R., {Jackson}, J.~M., {Rathborne}, J.~M., \& {Chambers}, E.~T. 2006,
  \apj, 639, 227

\bibitem[{{Siringo} {et~al.}(2009){Siringo}, {Kreysa}, {Kov{\'a}cs},
  {Schuller}, {Wei{\ss}}, {Esch}, {Gem{\"u}nd}, {Jethava}, {Lundershausen},
  {Colin}, {G{\"u}sten}, {Menten}, {Beelen}, {Bertoldi}, {Beeman}, \&
  {Haller}}]{Siringo2009}
{Siringo}, G., {Kreysa}, E., {Kov{\'a}cs}, A., {et~al.} 2009, \aap, 497, 945

\bibitem[{{Smith} {et~al.}(2008){Smith}, {Clark}, \& {Bonnell}}]{Smith2008}
{Smith}, R.~J., {Clark}, P.~C., \& {Bonnell}, I.~A. 2008, \mnras, 391, 1091

\bibitem[{{Solomon} \& {Rivolo}(1989)}]{Solomon1989}
{Solomon}, P.~M. \& {Rivolo}, A.~R. 1989, \apj, 339, 919

\bibitem[{{Sridharan} {et~al.}(2005){Sridharan}, {Beuther}, {Saito},
  {Wyrowski}, \& {Schilke}}]{Sridharan2005}
{Sridharan}, T.~K., {Beuther}, H., {Saito}, M., {Wyrowski}, F., \& {Schilke},
  P. 2005, \apjl, 634, L57

\bibitem[{{Vasyunina} {et~al.}(2009){Vasyunina}, {Linz}, {Henning}, {Stecklum},
  {Klose}, \& {Nyman}}]{Vasyunina2009}
{Vasyunina}, T., {Linz}, H., {Henning}, T., {et~al.} 2009, \aap, 499, 149

\bibitem[{{Vasyunina} {et~al.}(2010){Vasyunina}, {Linz}, {Henning},
  {Zinchenko}, {Beuther}, \& {Voronkov}}]{Vasyunina2010}
{Vasyunina}, T., {Linz}, H., {Henning}, T., {et~al.} 2010, ArXiv e-prints

\bibitem[{{Wilcock} {et~al.}(2011){Wilcock}, {Kirk}, {Stamatellos},
  {Ward-Thompson}, {Whitworth}, {Battersby}, {Brunt}, {Fuller}, {Griffin},
  {Molinari}, {Martin}, {Mottram}, {Peretto}, {Plume}, {Smith}, \&
  {Thompson}}]{Wilcock2011}
{Wilcock}, L.~A., {Kirk}, J.~M., {Stamatellos}, D., {et~al.} 2011, \aap, 526,
  A159+

\bibitem[{{Williams} {et~al.}(2000){Williams}, {Blitz}, \&
  {McKee}}]{Williams2000}
{Williams}, J.~P., {Blitz}, L., \& {McKee}, C.~F. 2000, Protostars and Planets
  IV, 97

\bibitem[{{Williams} {et~al.}(1994){Williams}, {de Geus}, \&
  {Blitz}}]{Williams1994}
{Williams}, J.~P., {de Geus}, E.~J., \& {Blitz}, L. 1994, \apj, 428, 693

\bibitem[{{Williams} {et~al.}(2004){Williams}, {Fuller}, \&
  {Sridharan}}]{Williams2004}
{Williams}, S.~J., {Fuller}, G.~A., \& {Sridharan}, T.~K. 2004, \aap, 417, 115

\bibitem[{{Zinnecker} \& {Yorke}(2007)}]{Zinnecker2007}
{Zinnecker}, H. \& {Yorke}, H.~W. 2007, \araa, 45, 481

\end{thebibliography}

\begin{appendix}
\section{Errors and uncertainties}
\label{uncertainties}
As conventional error propagation breaks down when the uncertainties become larger than a few percent, one can only point out the individual sources of errors and estimate the final uncertainties.

The distance error mainly stems from uncertainties in the Galactic rotation curve and errors in the gas velocities can be neglected. Error propagation including the velocity uncertainties following \citet{Reid2009} suggests uncertainties of smaller than 0.1 kpc. However, owing to intrinsic errors and deviations from the global Galactic rotation we estimate the distance to be uncertain to within 0.5 kpc. This leads to a contribution to the final mass uncertainties ranging from 10\% to 50\%, that depends on the absolute distance. In addition, individual objects close to the Galactic center that have non-circular orbits may be placed at random distances and contaminate the sample.

In the literature, temperatures of starless cores range from 10 K to 20 K with the bulk at 15 K \citep{Sridharan2005, Pillai2006, Vasyunina2010, Peretto2010a}. The temperature estimate here is based on direct observations of 15 out of 210 starless clumps and is in good agreement with earlier
studies. A temperature uncertainty of $\pm$ 5 K at 15 K may introduce mass uncertainties of about a factor of two.

The dust properties and the gas-to-dust ratio are very uncertain as well and might contribute another factor of two to the errors. 
The flux uncertainties are dominated by the calibration uncertainties, which are $\sim$ 15\% \citep{Schuller2009}.

When calculating the column density as well as the masses, the predominant uncertainties are those of the dust properties and temperatures. For the mass, the uncertainty in the distance is equally important.
Altogether, the total uncertainties in the mass may be as large as a factor of five. 
\end{appendix}
\begin{appendix}
\section{Omitted regions}
\begin{table*}[h]
\caption{Regions where MIPSGAL 24 $\mu$m images are to saturated to do a classification. Regions listed below were omitted during the classification.}
\begin{tabular}{ p{2.2cm} *{4}{l} *{1}{p{1.8cm}} *{2}{p{0.7cm}} }
global identifier & Gal Lon & Gal Lat & Ra & Dec & Peak Flux & Total Flux & Radius \\
 & [ $^{\circ}$ ] & [ $^{\circ}$ ] & [ $^{\circ}$ ] & [ $^{\circ}$ ] & [ Jy / beam ] & [ Jy ] & [ \arcsec ] \\
     8&         10.3225&         -0.1608&        272.2570&        -20.0861&            4.40&           45.36&             50.\\
    45&         10.1325&         -0.3775&        272.3610&        -20.3573&            1.30&            6.38&             28.\\
    59&         10.1392&         -0.3658&        272.3535&        -20.3458&            1.06&            5.59&             26.\\
    65&         10.1458&         -0.3158&        272.3103&        -20.3157&            1.00&            7.15&             31.\\
    70&         10.1808&         -0.3692&        272.3781&        -20.3109&            0.94&            6.48&             29.\\
    79&         10.1575&         -0.3775&        272.3739&        -20.3354&            0.88&            3.34&             23.\\
    96&         10.1442&         -0.3558&        272.3468&        -20.3365&            0.78&            2.34&             19.\\
   465&         14.9838&         -0.6956&        275.0954&        -16.2465&            6.60&          166.25&             74.\\
   467&         15.0121&         -0.7056&        275.1185&        -16.2262&            5.24&          118.69&             64.\\
   502&         14.9755&         -0.7139&        275.1082&        -16.2624&            1.90&           27.56&             49.\\
   514&         15.0022&         -0.7255&        275.1320&        -16.2444&            1.45&           17.96&             44.\\
   519&         14.9938&         -0.7305&        275.1325&        -16.2541&            1.34&           12.58&             36.\\
   554&         14.9738&         -0.7405&        275.1318&        -16.2765&            0.92&            4.37&             25.\\
   561&         14.9755&         -0.7372&        275.1296&        -16.2734&            0.91&            6.55&             28.\\
   617&         14.9688&         -0.7389&        275.1279&        -16.2801&            0.72&            2.42&             20.\\
   656&         13.9908&         -0.1208&        274.0768&        -16.8488&            0.61&            1.40&             16.\\
   687&         12.7975&         -0.2275&        273.5786&        -17.9485&            0.56&            2.69&             23.\\
   707&         13.9892&         -0.1358&        274.0897&        -16.8574&            0.54&            1.52&             18.\\
   786&         14.9738&         -0.7539&        275.1441&        -16.2827&            0.50&            1.33&             17.\\
  1114&         15.0545&         -0.6256&        275.0659&        -16.1512&            3.59&           28.11&             40.\\
  1116&         15.0511&         -0.6423&        275.0795&        -16.1620&            2.98&           46.09&             50.\\
  1119&         15.0678&         -0.6140&        275.0617&        -16.1339&            2.61&           22.05&             37.\\
  1122&         15.0995&         -0.6889&        275.1461&        -16.1414&            2.25&           22.50&             38.\\
  1125&         15.0778&         -0.6073&        275.0605&        -16.1220&            2.02&           21.50&             36.\\
  1126&         15.1844&         -0.6223&        275.1266&        -16.0351&            1.96&           37.36&             58.\\
  1131&         15.1311&         -0.6706&        275.1448&        -16.1048&            1.74&           21.13&             46.\\
  1137&         15.1028&         -0.6573&        275.1187&        -16.1235&            1.40&           31.14&             56.\\
  1139&         15.0028&         -0.7206&        275.1277&        -16.2415&            1.36&            9.88&             30.\\
  1140&         14.9929&         -0.7322&        275.1335&        -16.2558&            1.36&           11.98&             35.\\
  1141&         15.1061&         -0.6939&        275.1539&        -16.1378&            1.36&           12.79&             36.\\
  1152&         15.1111&         -0.7122&        275.1732&        -16.1421&            1.16&           11.03&             36.\\
  1153&         15.1161&         -0.6340&        275.1038&        -16.1008&            1.16&            9.14&             32.\\
  1157&         15.0961&         -0.7106&        275.1643&        -16.1545&            1.10&            5.29&             26.\\
  1168&         15.0944&         -0.7139&        275.1666&        -16.1576&            0.95&            3.61&             22.\\
  1174&         15.0944&         -0.6073&        275.0687&        -16.1073&            0.92&            4.57&             23.\\
  1180&         15.1078&         -0.6273&        275.0936&        -16.1050&            0.91&            5.14&             25.\\
  1181&         15.1061&         -0.5923&        275.0606&        -16.0900&            0.90&            5.60&             29.\\
  1187&         15.0944&         -0.7322&        275.1834&        -16.1662&            0.85&            4.65&             27.\\
  1193&         15.0412&         -0.6189&        275.0532&        -16.1598&            0.81&            5.39&             28.\\
  1202&         15.1278&         -0.6923&        275.1630&        -16.1180&            0.76&            3.21&             23.\\
  1207&         14.9845&         -0.7455&        275.1417&        -16.2694&            0.74&            2.95&             22.\\
  1214&         15.0995&         -0.7322&        275.1859&        -16.1618&            0.73&            2.98&             22.\\
  1215&         15.0778&         -0.6823&        275.1293&        -16.1573&            0.72&            4.79&             27.\\
  1223&         15.1028&         -0.6223&        275.0865&        -16.1070&            0.71&            3.04&             22.\\
  1224&         14.9845&         -0.6589&        275.0621&        -16.2286&            0.71&            4.20&             26.\\
  1231&         15.1278&         -0.7023&        275.1722&        -16.1227&            0.68&            1.34&             16.\\
  1233&         15.0861&         -0.6273&        275.0829&        -16.1241&            0.68&            2.48&             21.\\
  1285&         15.1228&         -0.7106&        275.1774&        -16.1310&            0.54&            1.19&             15.\\
  1291&         15.0561&         -0.6989&        275.1340&        -16.1843&            0.53&            1.55&             18.\\
  1294&         14.9879&         -0.7555&        275.1525&        -16.2712&            0.53&            0.87&             14.\\
  1297&         15.0328&         -0.7372&        275.1577&        -16.2229&            0.52&            1.78&             19.\\
  1301&         15.1094&         -0.5823&        275.0531&        -16.0823&            0.52&            1.33&             17.\\
  1302&         15.1244&         -0.7023&        275.1706&        -16.1256&            0.52&            1.18&             15.\\
  1303&         15.0245&         -0.7389&        275.1552&        -16.2310&            0.52&            0.68&             12.\\
  1315&         15.1311&         -0.6856&        275.1585&        -16.1119&            0.52&            2.16&             21.\\
  1324&         15.1011&         -0.6323&        275.0949&        -16.1132&            0.51&            1.64&             18.\\
  1327&         15.0628&         -0.6023&        275.0486&        -16.1328&            0.51&            1.82&             19.\\
  1331&         15.0278&         -0.7339&        275.1522&        -16.2257&            0.51&            1.68&             18.\\
  1737&         19.0742&         -0.2742&        276.6897&        -12.4356&            0.52&            2.73&             23.\\
\end{tabular}
\footnotesize{\\Notes: Columns are identifier, galactic longitude, galactic latitude, right ascension, declination, peak flux, integrated flux and radius as calculated by CLUMPFIND.}
\label{tab:omitted}
\end{table*}

\end{appendix}
\Online
\begin{appendix}
\section{Stamps of starless regions}
\begin{figure}[p]
\includegraphics[angle=-90.,width=.5\textwidth]{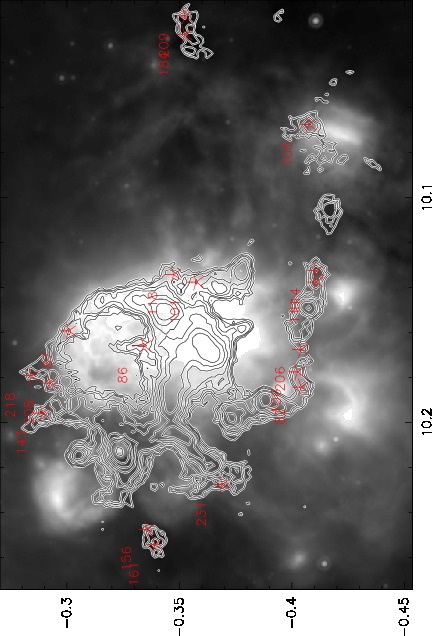}
\caption{MIPSGAL 24 $\mu$m image with ATLASGAL contours on top. Starless clumps are marked with a red asterisk. The numbers correspond to the global identifier given in Table \ref{tab:list}.}
\label{app:stamps}
\end{figure}
\begin{figure}[p]
\includegraphics[angle=-90.,width=.5\textwidth]{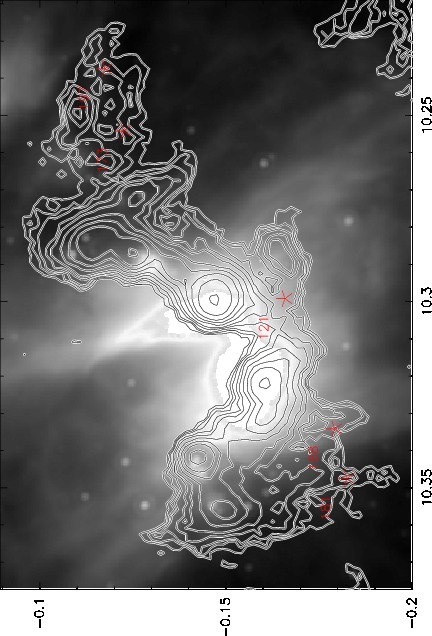}
\end{figure}
\begin{figure}[p]
\includegraphics[angle=-90.,width=.5\textwidth]{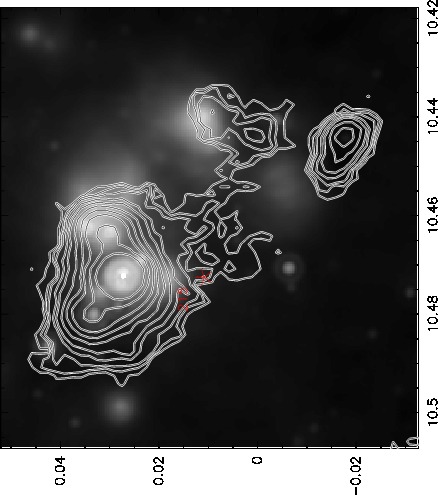}
\end{figure}
\begin{figure}[p]
\includegraphics[angle=-90.,width=.5\textwidth]{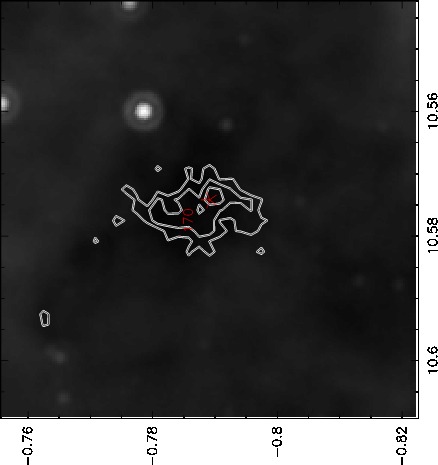}
\end{figure}
\begin{figure}[p]
\includegraphics[angle=-90.,width=.5\textwidth]{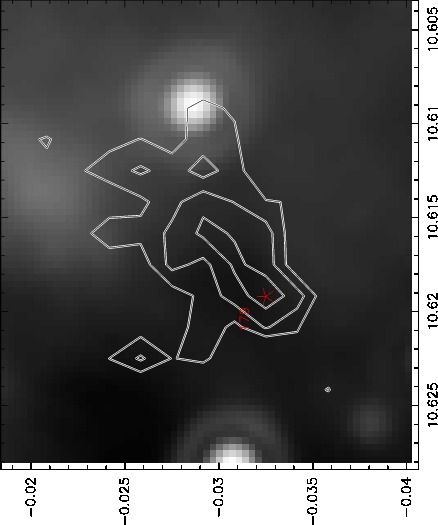}
\end{figure}
\begin{figure}[p]
\includegraphics[angle=-90.,width=.5\textwidth]{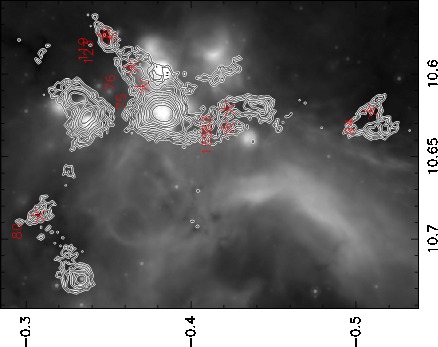}
\end{figure}
\begin{figure}[p]
\includegraphics[angle=-90.,width=.5\textwidth]{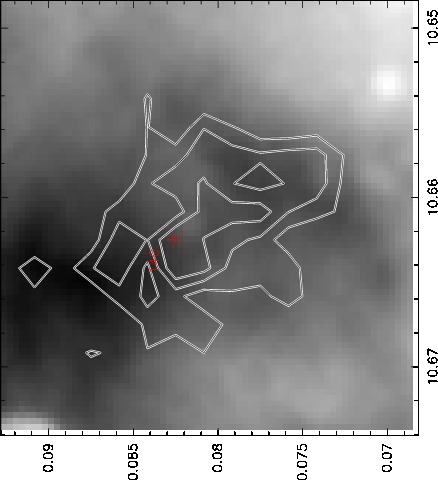}
\end{figure}
\begin{figure}[p]
\includegraphics[angle=-90.,width=.5\textwidth]{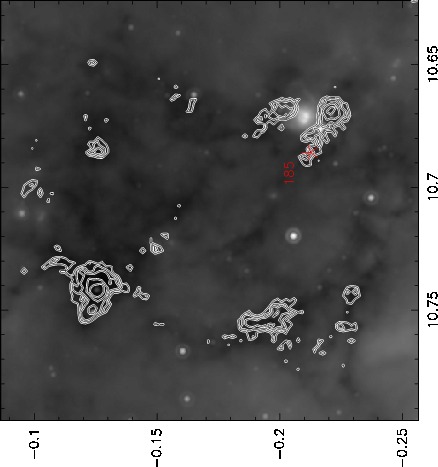}
\end{figure}
\begin{figure}[p]
\includegraphics[angle=-90.,width=.5\textwidth]{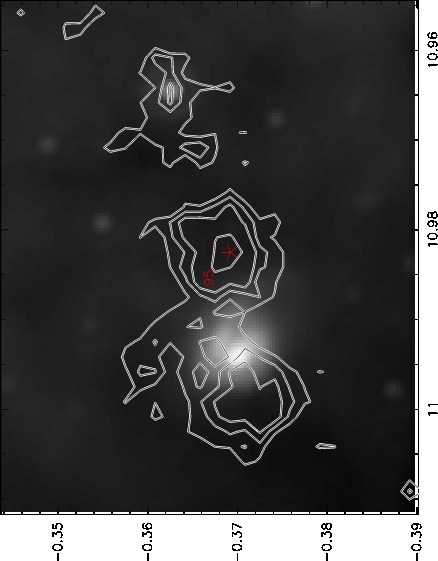}
\end{figure}
\begin{figure}[p]
\includegraphics[angle=-90.,width=.5\textwidth]{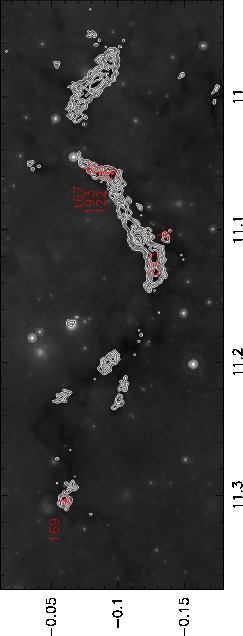}
\end{figure}
\begin{figure}[p]
\includegraphics[angle=-90.,width=.5\textwidth]{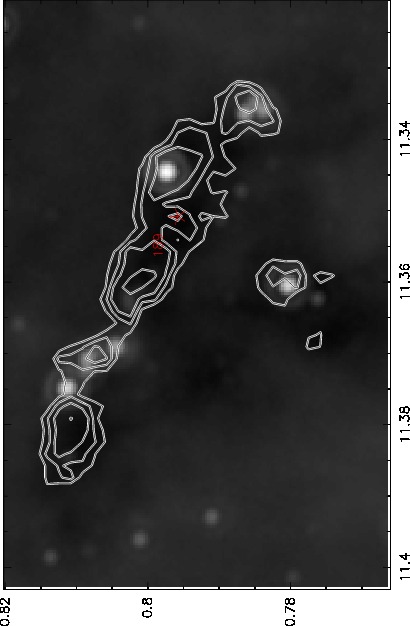}
\end{figure}
\clearpage
\begin{figure}[p]
\includegraphics[angle=-90.,width=.5\textwidth]{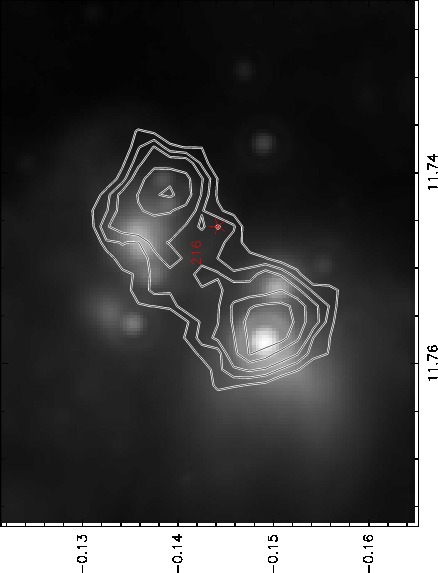}
\end{figure}
\begin{figure}[p]
\includegraphics[angle=-90.,width=.5\textwidth]{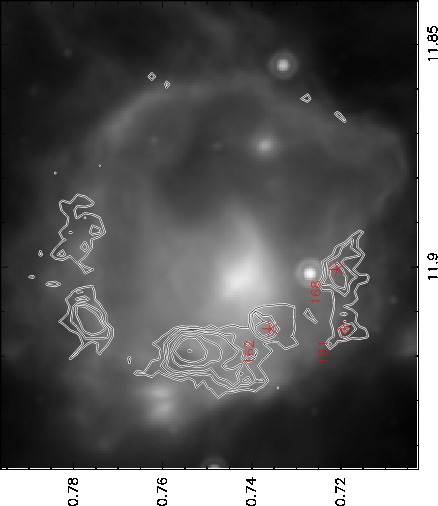}
\end{figure}
\begin{figure}[p]
\includegraphics[angle=-90.,width=.5\textwidth]{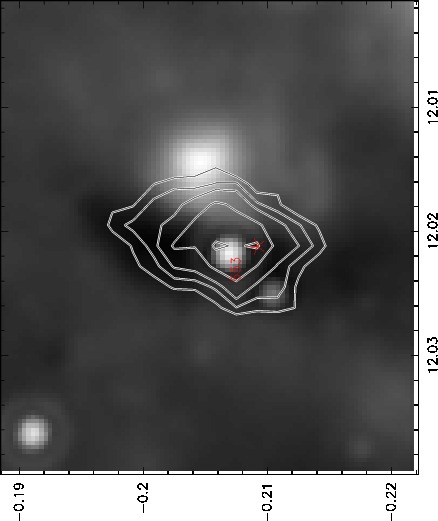}
\end{figure}
\begin{figure}[p]
\includegraphics[angle=-90.,width=.5\textwidth]{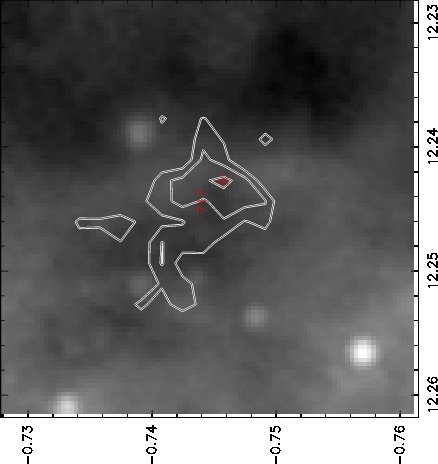}
\end{figure}
\begin{figure}[p]
\includegraphics[angle=-90.,width=.5\textwidth]{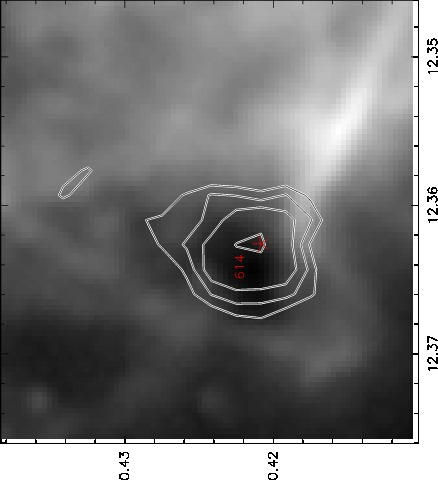}
\end{figure}
\begin{figure}[p]
\includegraphics[angle=-90.,width=.5\textwidth]{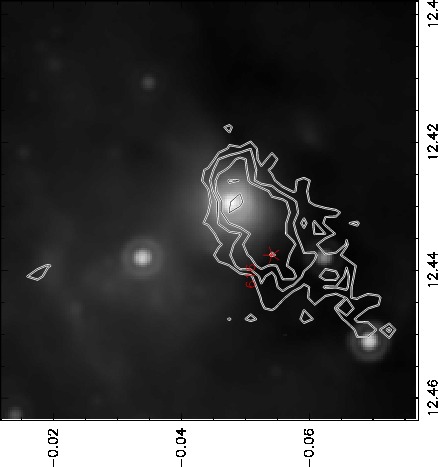}
\end{figure}
\begin{figure}[p]
\includegraphics[angle=-90.,width=.5\textwidth]{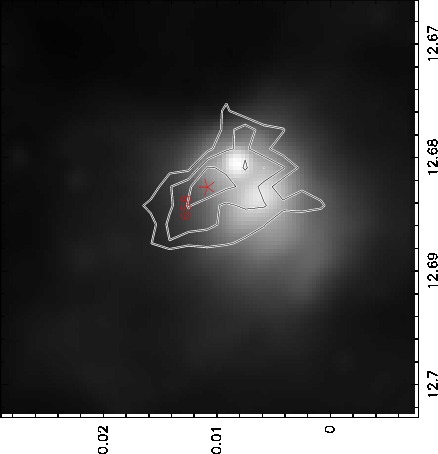}
\end{figure}
\begin{figure}[p]
\includegraphics[angle=-90.,width=.5\textwidth]{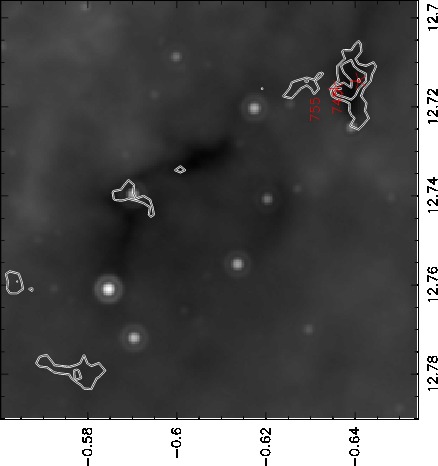}
\end{figure}
\begin{figure}[p]
\includegraphics[angle=-90.,width=.5\textwidth]{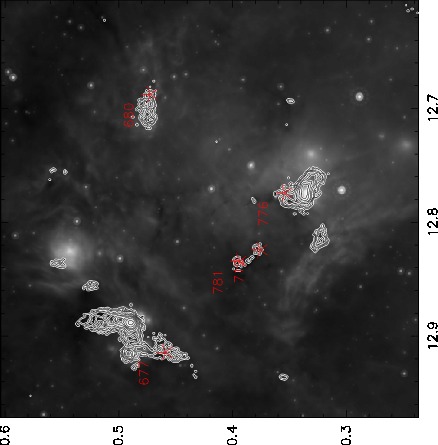}
\end{figure}
\begin{figure}[p]
\includegraphics[angle=-90.,width=.5\textwidth]{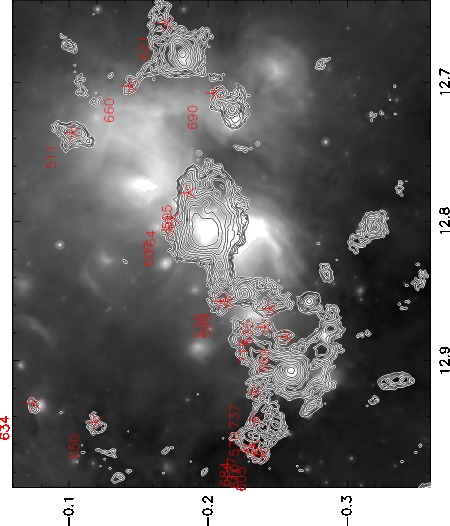}
\end{figure}
\clearpage
\begin{figure}[p]
\includegraphics[angle=-90.,width=.5\textwidth]{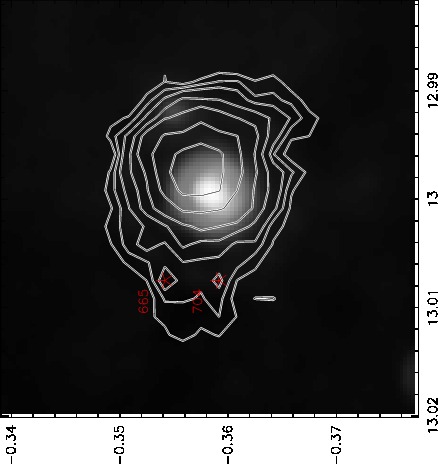}
\end{figure}
\begin{figure}[p]
\includegraphics[angle=-90.,width=.5\textwidth]{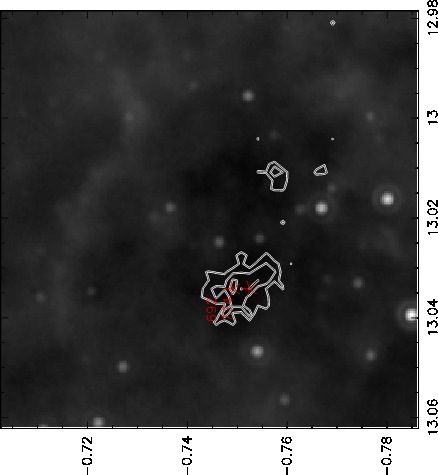}
\end{figure}
\begin{figure}[p]
\includegraphics[angle=-90.,width=.5\textwidth]{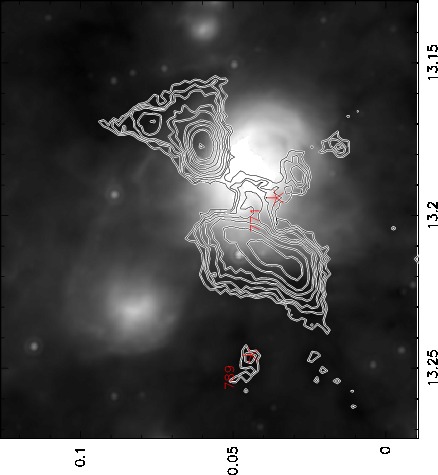}
\end{figure}
\begin{figure}[p]
\includegraphics[angle=-90.,width=.5\textwidth]{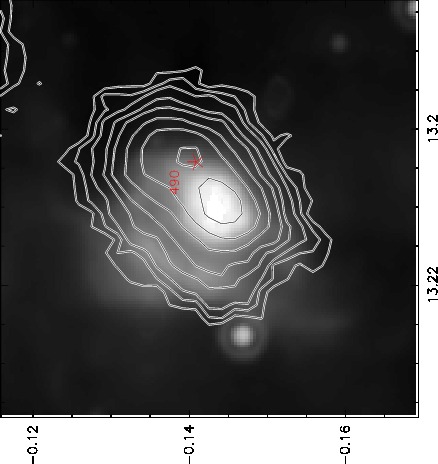}
\end{figure}
\begin{figure}[p]
\includegraphics[angle=-90.,width=.5\textwidth]{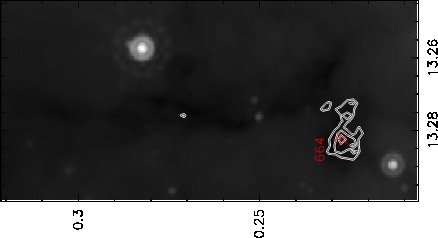}
\end{figure}
\begin{figure}[p]
\includegraphics[angle=-90.,width=.5\textwidth]{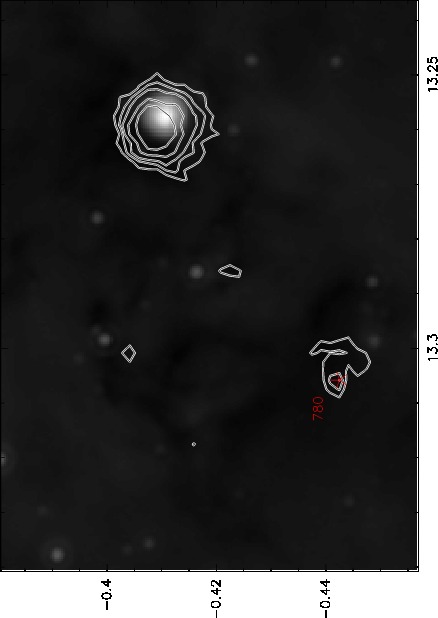}
\end{figure}
\begin{figure}[p]
\includegraphics[angle=-90.,width=.5\textwidth]{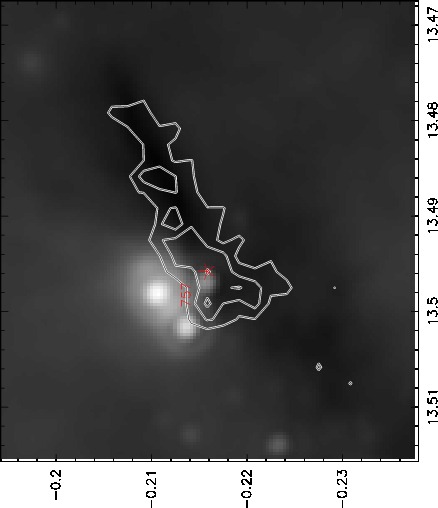}
\end{figure}
\begin{figure}[p]
\includegraphics[angle=-90.,width=.5\textwidth]{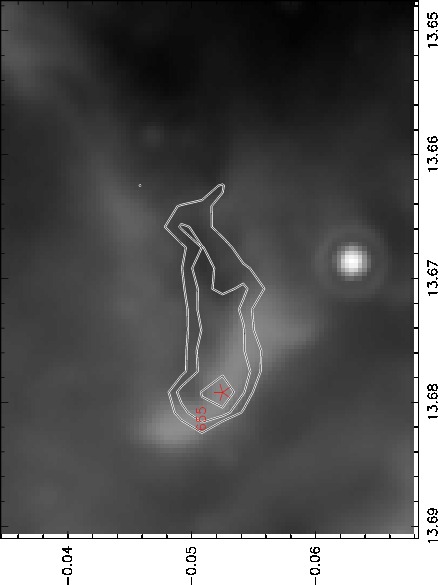}
\end{figure}
\begin{figure}[p]
\includegraphics[angle=-90.,width=.5\textwidth]{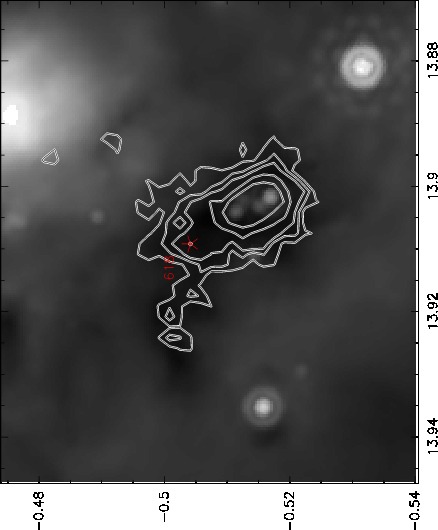}
\end{figure}
\begin{figure}[p]
\includegraphics[angle=-90.,width=.5\textwidth]{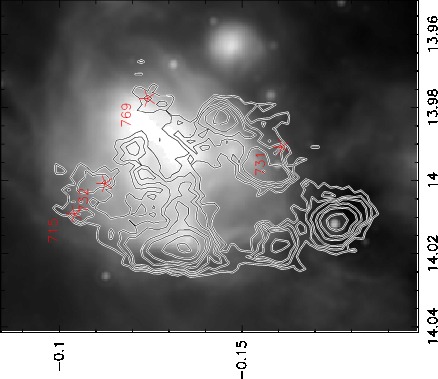}
\end{figure}
\clearpage
\begin{figure}[p]
\includegraphics[angle=-90.,width=.5\textwidth]{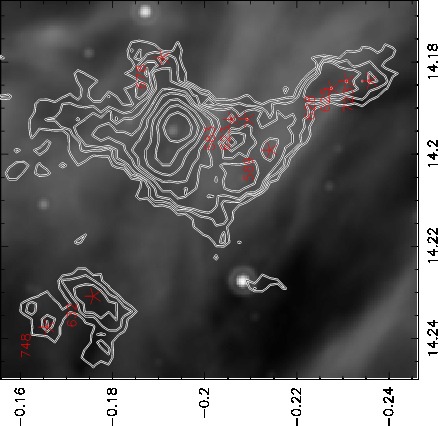}
\end{figure}
\begin{figure}[p]
\includegraphics[angle=-90.,width=.5\textwidth]{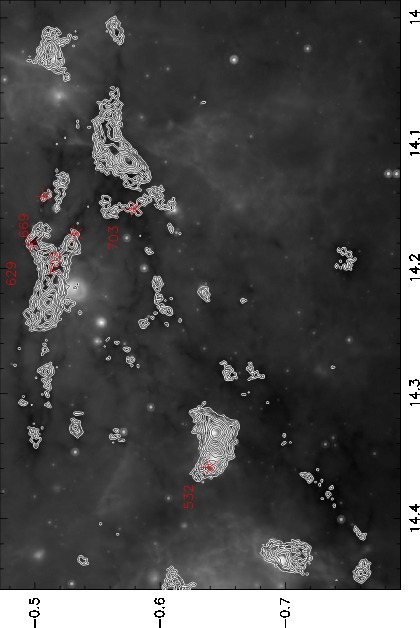}
\end{figure}
\begin{figure}[p]
\includegraphics[angle=-90.,width=.5\textwidth]{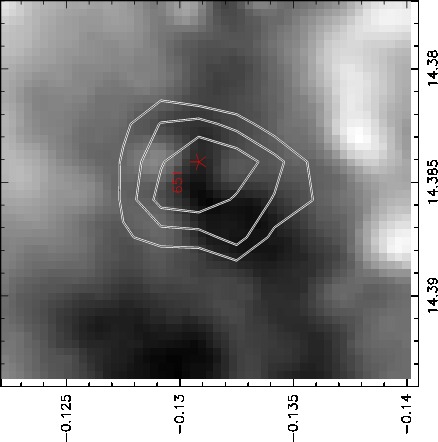}
\end{figure}
\begin{figure}[p]
\includegraphics[angle=-90.,width=.5\textwidth]{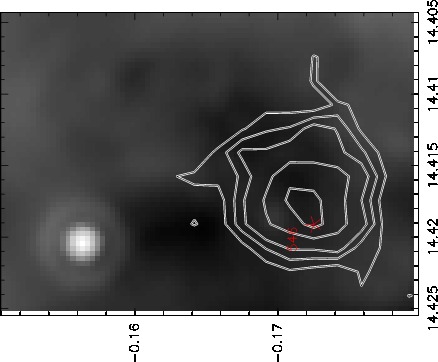}
\end{figure}
\begin{figure}[p]
\includegraphics[angle=-90.,width=.5\textwidth]{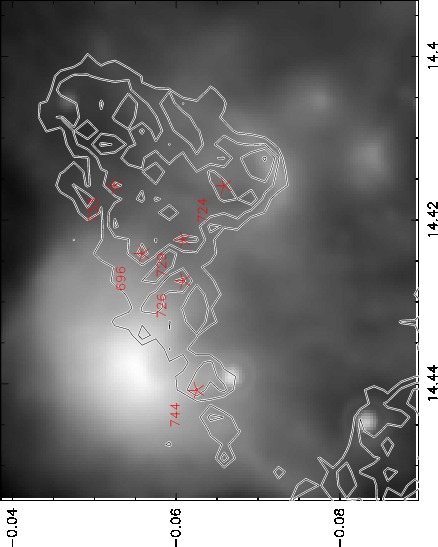}
\end{figure}
\begin{figure}[p]
\includegraphics[angle=-90.,width=.5\textwidth]{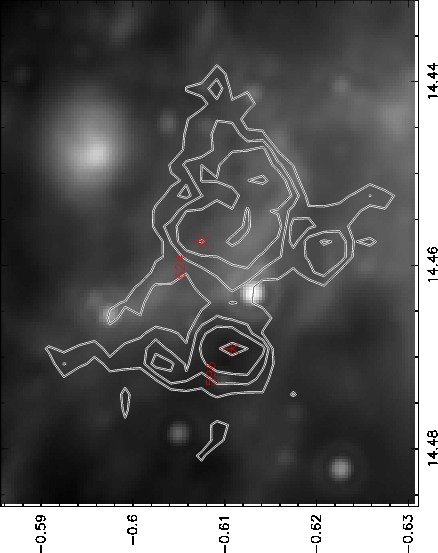}
\end{figure}
\begin{figure}[p]
\includegraphics[angle=-90.,width=.5\textwidth]{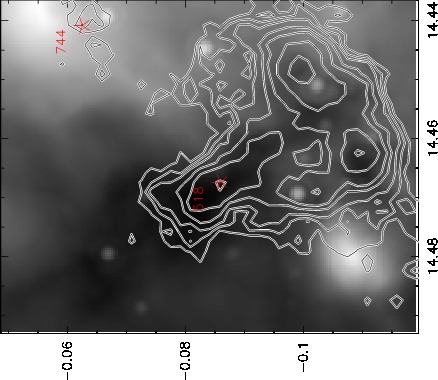}
\end{figure}
\begin{figure}[p]
\includegraphics[angle=-90.,width=.5\textwidth]{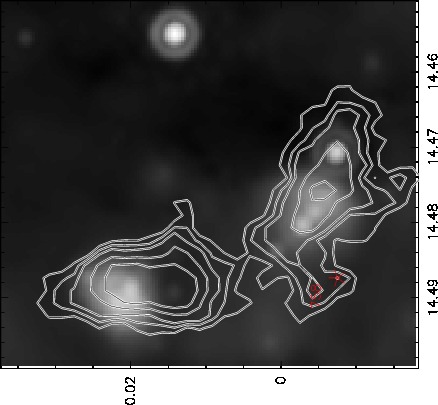}
\end{figure}
\begin{figure}[p]
\includegraphics[angle=-90.,width=.5\textwidth]{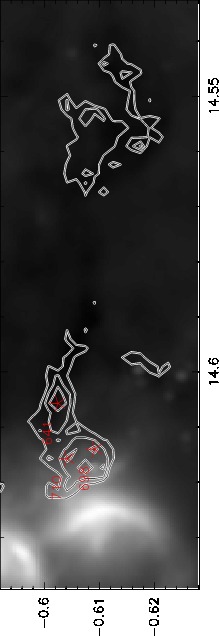}
\end{figure}
\begin{figure}[p]
\includegraphics[angle=-90.,width=.5\textwidth]{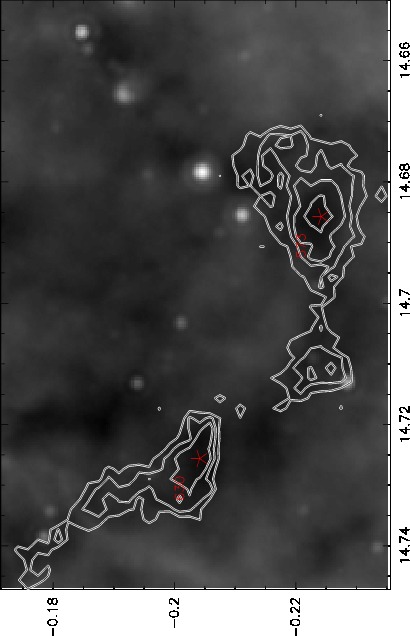}
\end{figure}
\clearpage
\begin{figure}[p]
\includegraphics[angle=-90.,width=.5\textwidth]{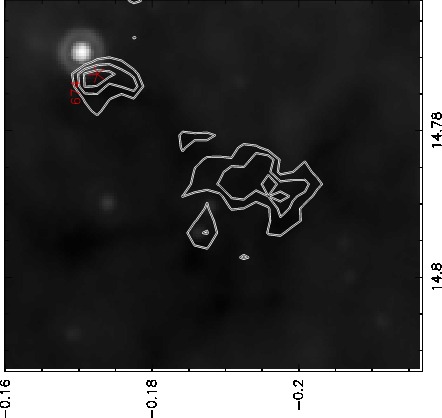}
\end{figure}
\begin{figure}[p]
\includegraphics[angle=-90.,width=.5\textwidth]{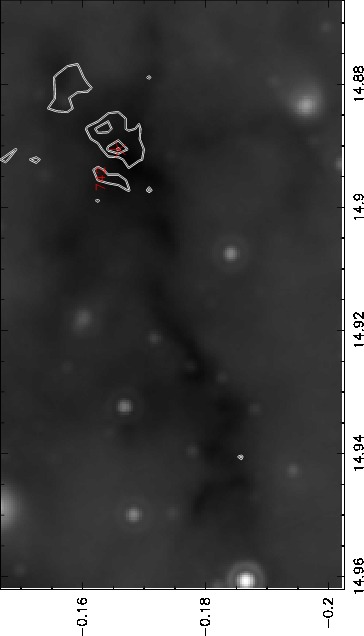}
\end{figure}
\begin{figure}[p]
\includegraphics[angle=-90.,width=.5\textwidth]{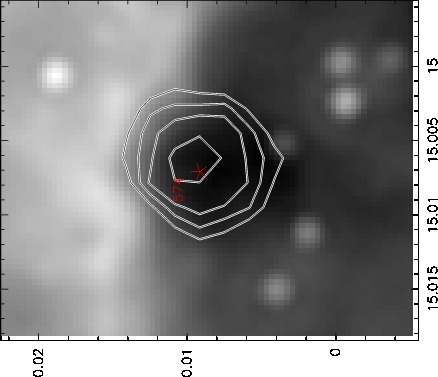}
\end{figure}
\begin{figure}[p]
\includegraphics[angle=-90.,width=.5\textwidth]{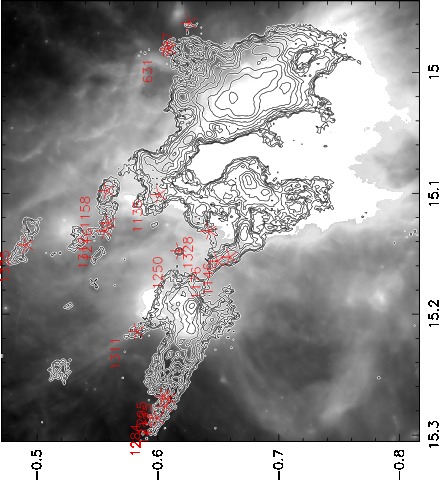}
\end{figure}
\begin{figure}[p]
\includegraphics[angle=-90.,width=.5\textwidth]{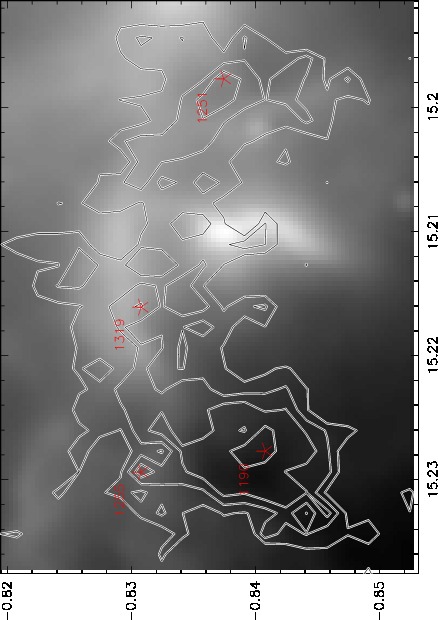}
\end{figure}
\begin{figure}[p]
\includegraphics[angle=-90.,width=.5\textwidth]{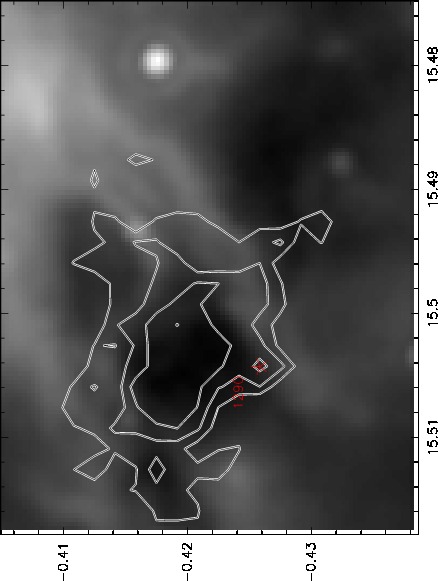}
\end{figure}
\begin{figure}[p]
\includegraphics[angle=-90.,width=.5\textwidth]{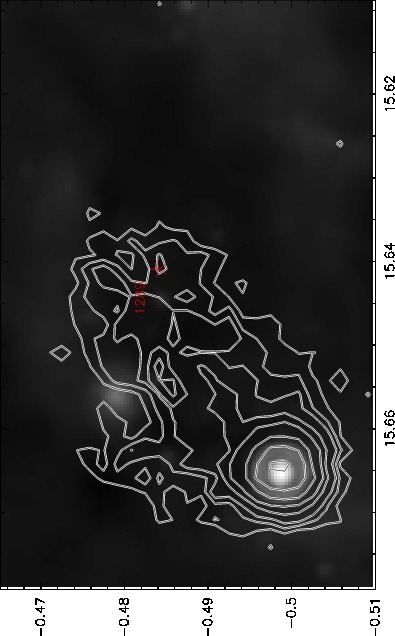}
\end{figure}
\begin{figure}[p]
\includegraphics[angle=-90.,width=.5\textwidth]{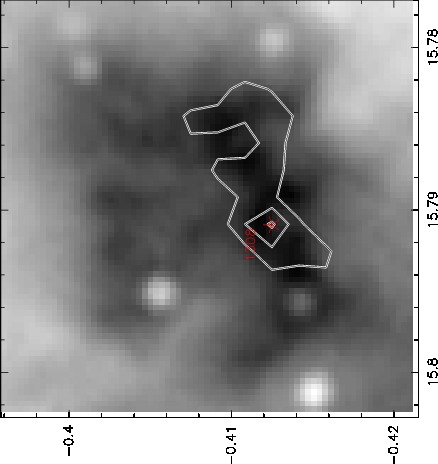}
\end{figure}
\begin{figure}[p]
\includegraphics[angle=-90.,width=.5\textwidth]{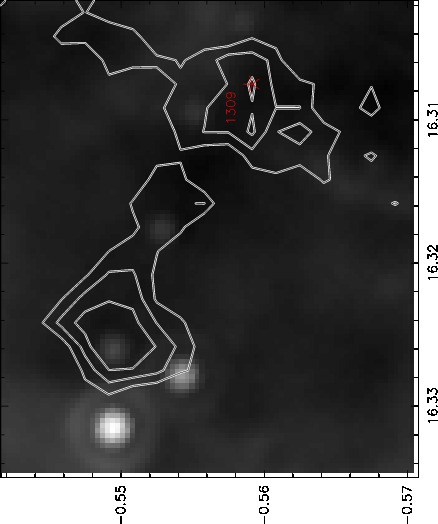}
\end{figure}
\begin{figure}[p]
\includegraphics[angle=-90.,width=.5\textwidth]{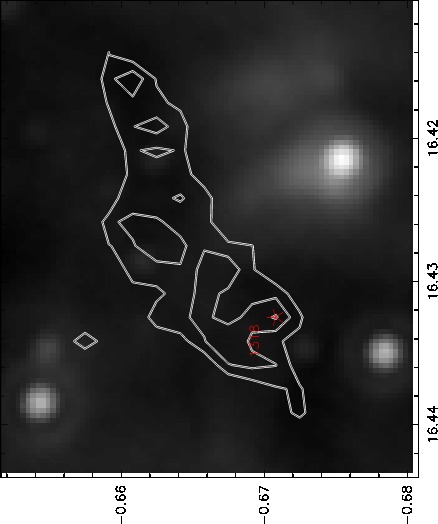}
\end{figure}
\clearpage
\begin{figure}[p]
\includegraphics[angle=-90.,width=.5\textwidth]{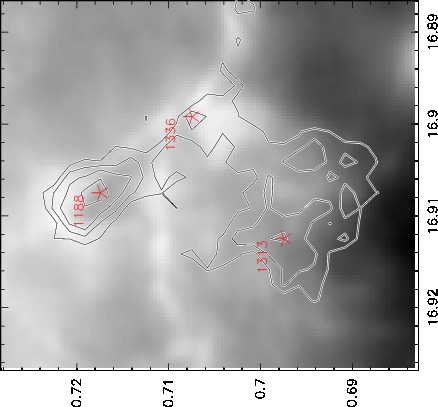}
\end{figure}
\begin{figure}[p]
\includegraphics[angle=-90.,width=.5\textwidth]{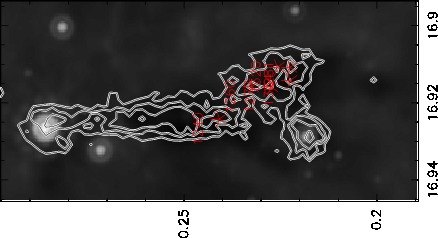}
\end{figure}
\begin{figure}[p]
\includegraphics[angle=-90.,width=.5\textwidth]{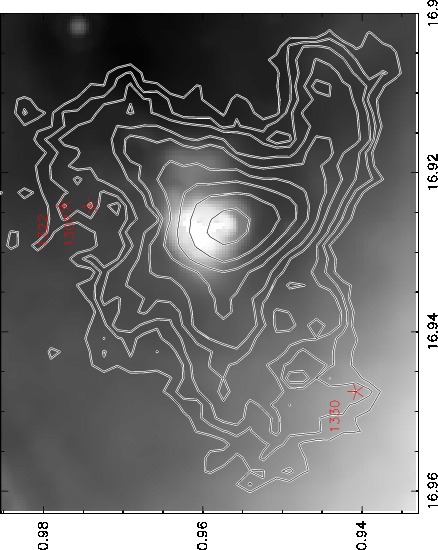}
\end{figure}
\begin{figure}[p]
\includegraphics[angle=-90.,width=.5\textwidth]{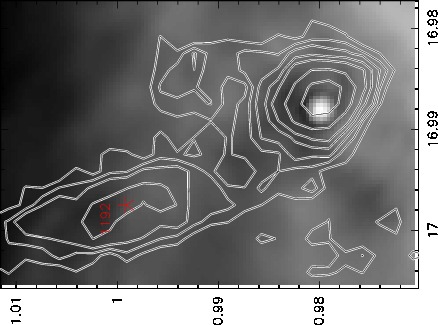}
\end{figure}
\begin{figure}[p]
\includegraphics[angle=-90.,width=.5\textwidth]{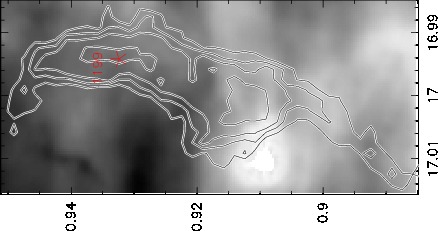}
\end{figure}
\begin{figure}[p]
\includegraphics[angle=-90.,width=.5\textwidth]{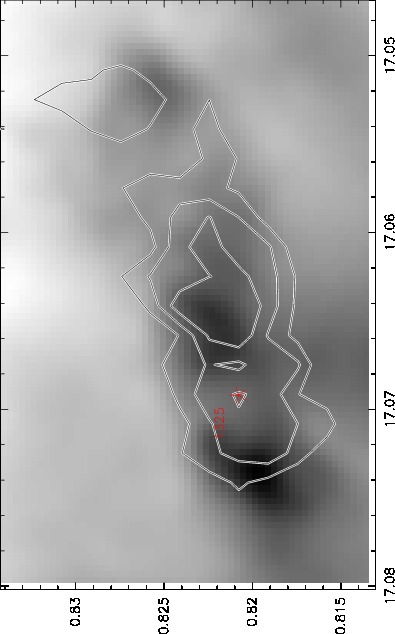}
\end{figure}
\begin{figure}[p]
\includegraphics[angle=-90.,width=.5\textwidth]{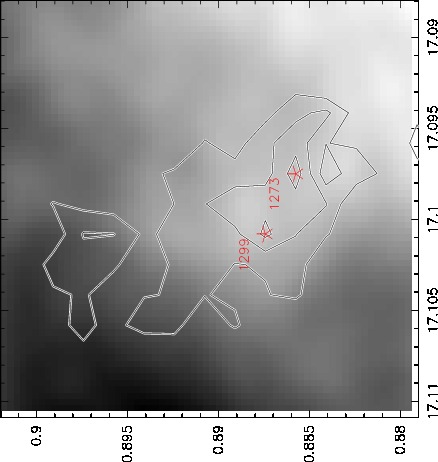}
\end{figure}
\begin{figure}[p]
\includegraphics[angle=-90.,width=.5\textwidth]{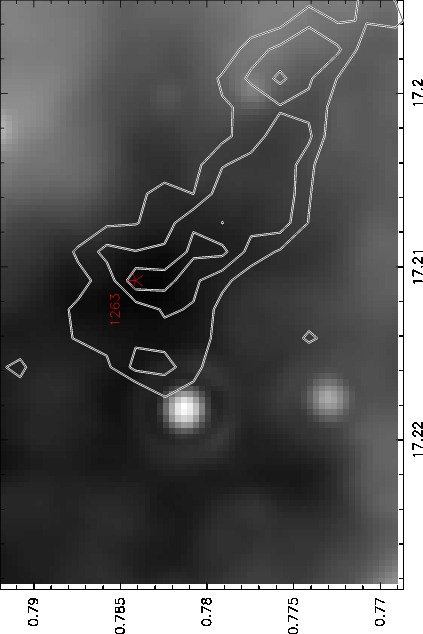}
\end{figure}
\begin{figure}[p]
\includegraphics[angle=-90.,width=.5\textwidth]{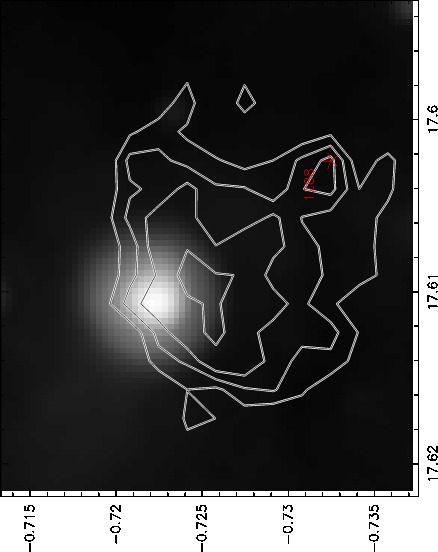}
\end{figure}
\begin{figure}[p]
\includegraphics[angle=-90.,width=.5\textwidth]{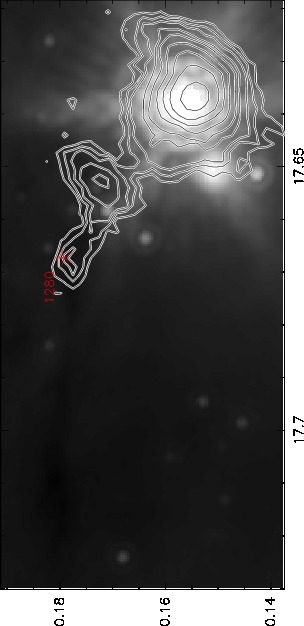}
\end{figure}
\clearpage
\begin{figure}[p]
\includegraphics[angle=-90.,width=.5\textwidth]{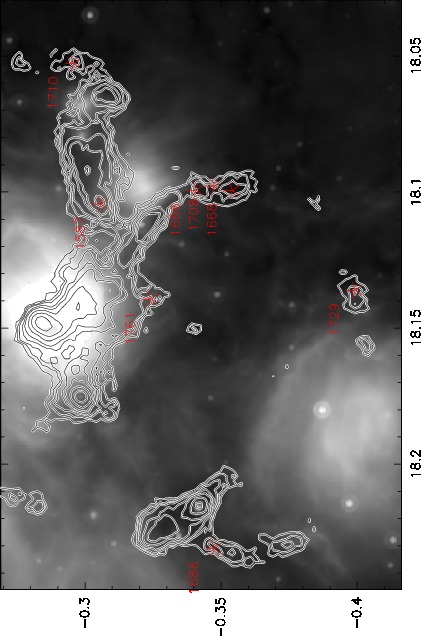}
\end{figure}
\begin{figure}[p]
\includegraphics[angle=-90.,width=.5\textwidth]{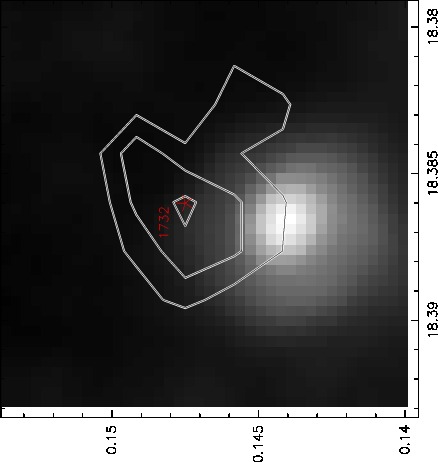}
\end{figure}
\begin{figure}[p]
\includegraphics[angle=-90.,width=.5\textwidth]{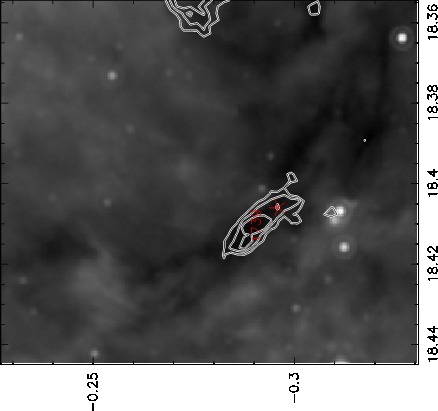}
\end{figure}
\begin{figure}[p]
\includegraphics[angle=-90.,width=.5\textwidth]{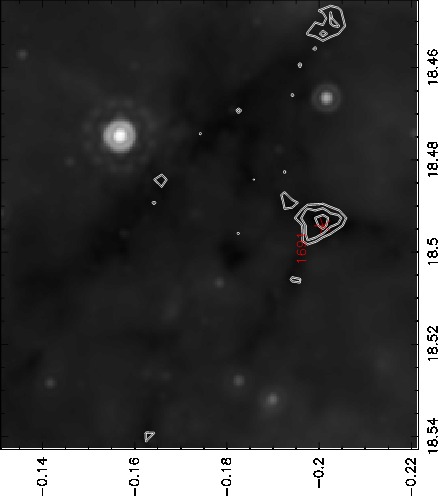}
\end{figure}
\begin{figure}[p]
\includegraphics[angle=-90.,width=.5\textwidth]{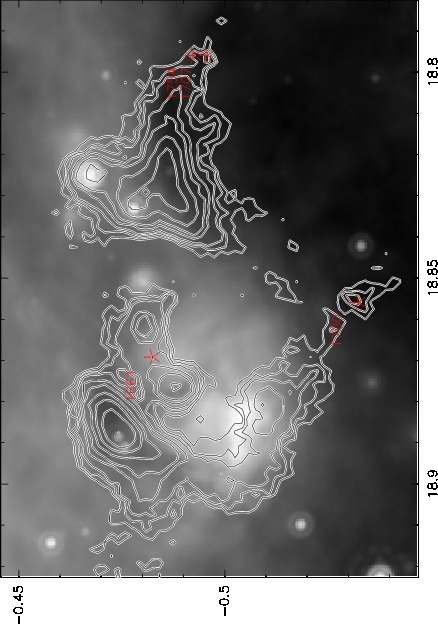}
\end{figure}
\begin{figure}[p]
\includegraphics[angle=-90.,width=.5\textwidth]{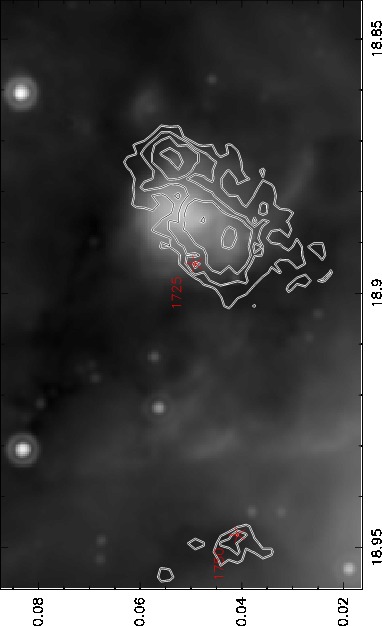}
\end{figure}
\begin{figure}[p]
\includegraphics[angle=-90.,width=.5\textwidth]{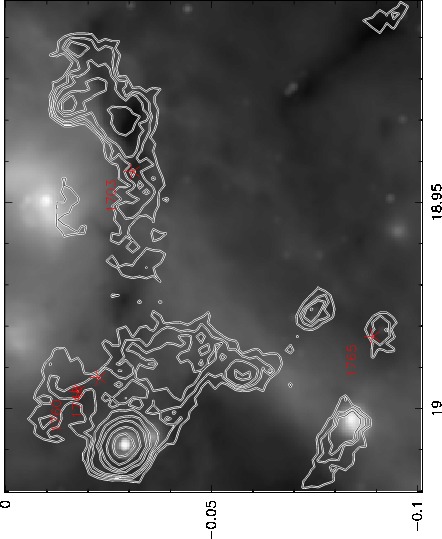}
\end{figure}
\begin{figure}[p]
\includegraphics[angle=-90.,width=.5\textwidth]{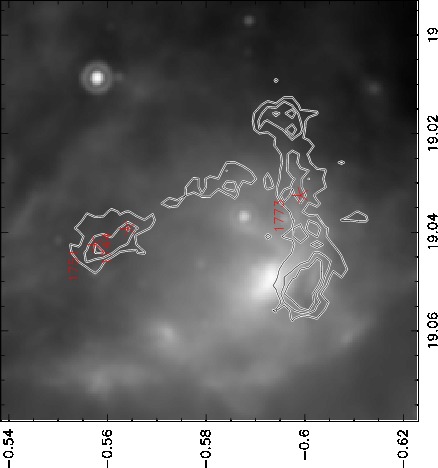}
\end{figure}
\begin{figure}[p]
\includegraphics[angle=-90.,width=.5\textwidth]{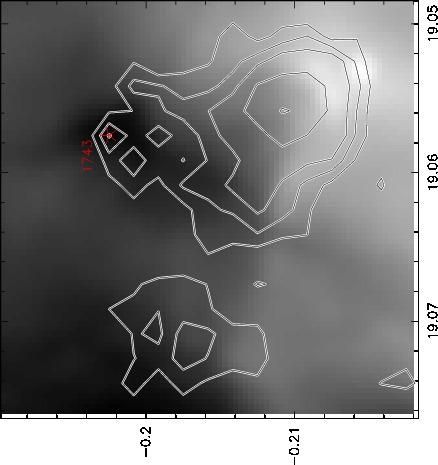}
\end{figure}
\begin{figure}[p]
\includegraphics[angle=-90.,width=.5\textwidth]{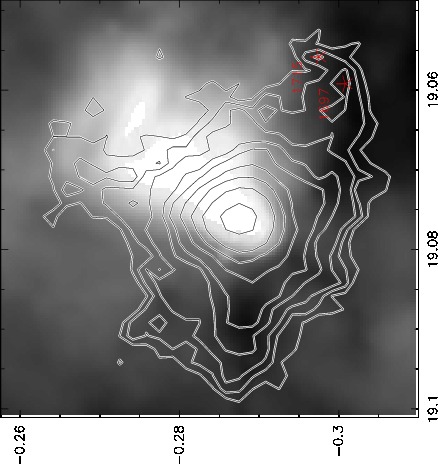}
\end{figure}
\clearpage
\begin{figure}[p]
\includegraphics[angle=-90.,width=.5\textwidth]{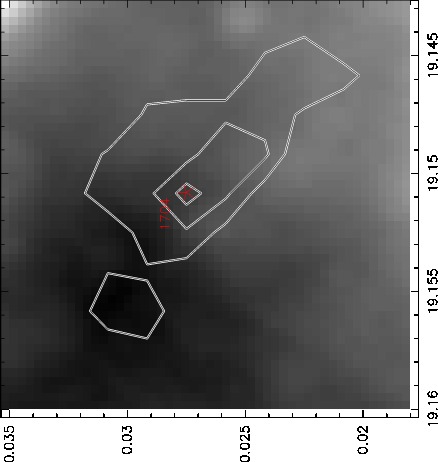}
\end{figure}
\begin{figure}[p]
\includegraphics[angle=-90.,width=.5\textwidth]{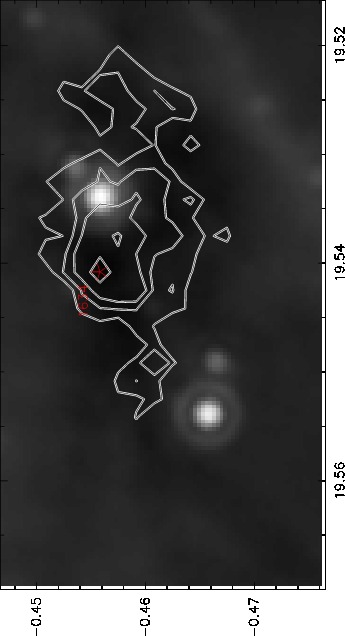}
\end{figure}
\begin{figure}[p]
\includegraphics[angle=-90.,width=.5\textwidth]{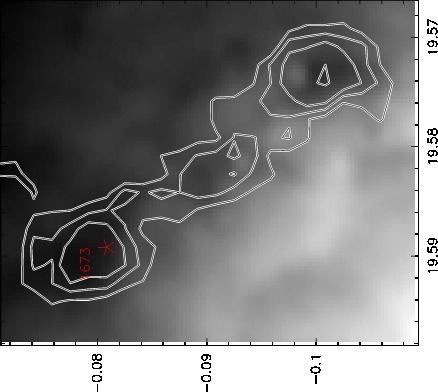}
\end{figure}
\begin{figure}[p]
\includegraphics[angle=-90.,width=.5\textwidth]{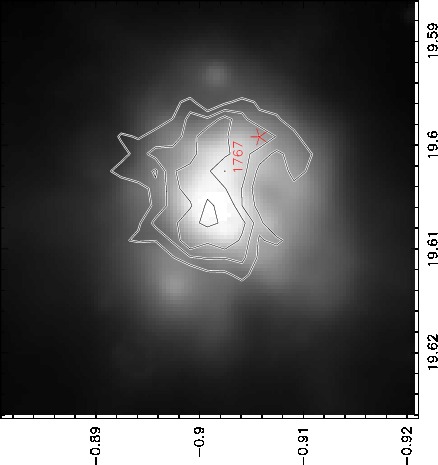}
\end{figure}
\begin{figure}[p]
\includegraphics[angle=-90.,width=.5\textwidth]{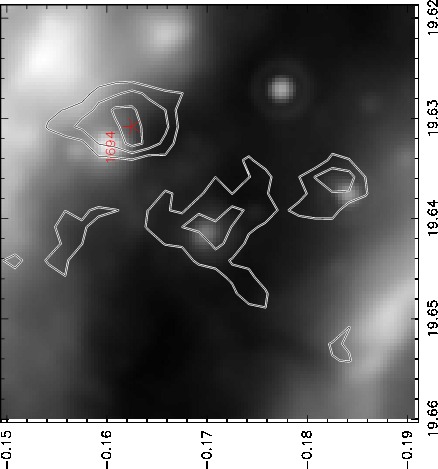}
\end{figure}
\begin{figure}[p]
\includegraphics[angle=-90.,width=.5\textwidth]{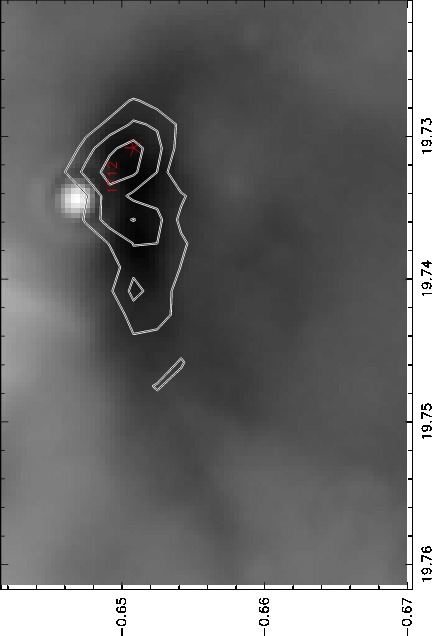}
\end{figure}
\begin{figure}[p]
\includegraphics[angle=-90.,width=.5\textwidth]{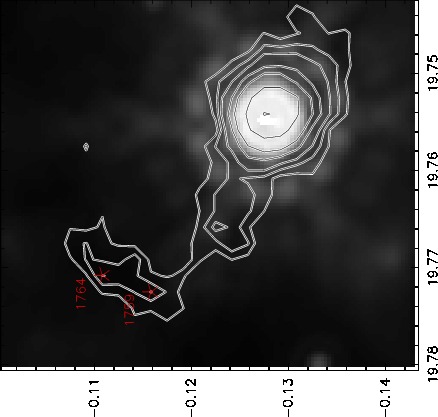}
\end{figure}
\begin{figure}[p]
\includegraphics[angle=-90.,width=.5\textwidth]{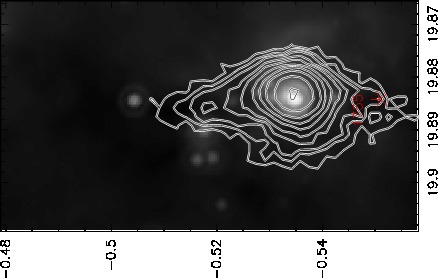}
\end{figure}

\end{appendix}

\end{document}